\documentclass[twocolumn]{aastex631}

\shorttitle{AGN properties of HSC galaxy groups and clusters at $z < 1.4$}
\shortauthors{Toba et al.}
\graphicspath{{./}{figures/}}
\begin{document}

\title{AGN properties of $\sim$1 million member galaxies of galaxy groups and clusters at $z < 1.4$ based on the Subaru Hyper Suprime-Cam survey}

\correspondingauthor{Yoshiki Toba}
\email{yoshiki.toba@nao.ac.jp}

\author[0000-0002-3531-7863]{Yoshiki Toba}
\altaffiliation{NAOJ fellow}
\affiliation{National Astronomical Observatory of Japan, 2-21-1 Osawa, Mitaka, Tokyo 181-8588, Japan}
\affiliation{Department of Physics, Nara Women's University, Kitauoyanishi-machi, Nara, Nara 630-8506, Japan}
\affiliation{Academia Sinica Institute of Astronomy and Astrophysics, 11F of Astronomy-Mathematics Building, AS/NTU, No.1, Section 4, Roosevelt Road, Taipei 10617, Taiwan}
\affiliation{Research Center for Space and Cosmic Evolution, Ehime University, 2-5 Bunkyo-cho, Matsuyama, Ehime 790-8577, Japan}

\author{Aoi Hashiguchi}
\affiliation{Department of Physics, Nara Women's University, Kitauoyanishi-machi, Nara, Nara 630-8506, Japan}

\author[0000-0002-2784-3652]{Naomi Ota}
\affiliation{Department of Physics, Nara Women's University, Kitauoyanishi-machi, Nara, Nara 630-8506, Japan}

\author[0000-0003-3484-399X]{Masamune Oguri}
\affiliation{Center for Frontier Science, Chiba University, 1-33 Yayoi-cho, Inage-ku, Chiba 263-8522, Japan} 
\affiliation{Department of Physics, Graduate School of Science, Chiba University, 1-33 Yayoi-Cho, Inage-Ku, Chiba 263-8522, Japan} 

\author[0000-0003-2898-0728]{Nobuhiro Okabe}
\affiliation{Department of Physical Science, Hiroshima University, 1-3-1 Kagamiyama, Higashi-Hiroshima, Hiroshima 739-8526, Japan}
\affiliation{Hiroshima Astrophysical Science Center, Hiroshima University, 1-3-1, Kagamiyama, Higashi-Hiroshima, Hiroshima 739-8526, Japan}
\affiliation{Core Research for Energetic Universe, Hiroshima University, 1-3-1, Kagamiyama, Higashi-Hiroshima, Hiroshima 739-8526, Japan}

\author[0000-0001-7821-6715]{Yoshihiro Ueda}
\affiliation{Department of Astronomy, Kyoto University, Kitashirakawa-Oiwake-cho, Sakyo-ku, Kyoto 606-8502, Japan}

\author[0000-0001-6186-8792]{Masatoshi Imanishi}
\affiliation{National Astronomical Observatory of Japan, 2-21-1 Osawa, Mitaka, Tokyo 181-8588, Japan}

\author[0000-0002-6109-2397]{Atsushi J. Nishizawa}
\affiliation{DX Center, Gifu Shotoku Gakuen University, Takakuwa-Nishi, Yanaizucho, Gifu 501-6194, Japan}
\affiliation{Institute for Advanced Research/Kobayashi Maskawa Institute, Nagoya University, Nagoya 464-8602, Japan}

\author[0000-0002-6821-8669]{Tomotsugu Goto}
\affiliation{Institute of Astronomy, National Tsing Hua University, No. 101, Section 2, Kuang-Fu Road, Hsinchu City 30013, Taiwan}

\author[0000-0001-5615-4904]{Bau-Ching Hsieh}
\affiliation{Academia Sinica Institute of Astronomy and Astrophysics, 11F of Astronomy-Mathematics Building, AS/NTU, No.1, Section 4, Roosevelt Road, Taipei 10617, Taiwan}

\author{Marie Kondo}
\affiliation{Graduate School of Science and Engineering, Saitama University, Shimo-Okubo 255, Sakura, Saitama 338-8570, Japan}

\author[0000-0002-0100-1238]{Shuhei Koyama}
\affiliation{Institute of Astronomy, Graduate School of Science, The University of Tokyo, 2-21-1 Osawa, Mitaka, Tokyo 181-0015, Japan}

\author[0000-0003-4814-0101]{Kianhong Lee}
\affiliation{Astronomical Institute, Tohoku University, Aramaki, Aoba-ku, Sendai, 980-8578, Japan}
\affiliation{Institute of Astronomy, Graduate School of Science, The University of Tokyo, 2-21-1 Osawa, Mitaka, Tokyo 181-0015, Japan}

\author[0000-0002-9901-233X]{Ikuyuki Mitsuishi}
\affiliation{Graduate School of Science, Division of Particle and Astrophysical Science, Nagoya University, Furocho, Chikusa-ku, Nagoya, Aichi 464-8602, Japan}

\author[0000-0002-7402-5441]{Tohru Nagao}
\affiliation{Research Center for Space and Cosmic Evolution, Ehime University, 2-5 Bunkyo-cho, Matsuyama, Ehime 790-8577, Japan}

\author{Taira Oogi}
\affiliation{National Institute of Technology, Asahikawa College, 2-2-1-6 Syunkodai, Asahikawa, Hokkaido 071-8142, Japan}
\affiliation{Research Center for Space and Cosmic Evolution, Ehime University, 2-5 Bunkyo-cho, Matsuyama, Ehime 790-8577, Japan}

\author{Koki Sakuta}
\affiliation{Graduate School of Science, Division of Particle and Astrophysical Science, Nagoya University, Furocho, Chikusa-ku, Nagoya, Aichi 464-8602, Japan}
   
\author[0000-0001-7825-0075]{Malte Schramm}
\affiliation{Universit\"at Potsdam, Karl-Liebknecht-Str. 24/25, D-14476 Potsdam, Germany}

\author{Anri Yanagawa}
\affiliation{Department of Physics, Nara Women's University, Kitauoyanishi-machi, Nara, Nara 630-8506, Japan}

\author{Anje Yoshimoto}
\affiliation{Department of Physics, Nara Women's University, Kitauoyanishi-machi, Nara, Nara 630-8506, Japan}




%%%%%%%%%%%%%%%%%%
%    Abstract
%%%%%%%%%%%%%%%%%%
\begin{abstract}

Herein, we present the statistical properties of active galactic nuclei (AGNs) for approximately 1 million member galaxies of galaxy groups and clusters, with 0.1 $<$ cluster redshift ($z_{\rm cl}$) $<$ 1.4, selected using Subaru Hyper Suprime-Cam, the so-called CAMIRA clusters. 
In this research, we focused on the AGN power fraction ($f_{\rm AGN}$), which is defined as the proportion of the contribution of AGNs to the total infrared (IR) luminosity, $L_{\rm IR}$ (AGN)/$L_{\rm IR}$, and examined how $f_{\rm AGN}$ depends on (i) $z_{\rm cl}$ and (ii) the distance from the cluster center. 
We compiled multiwavelength data using the ultraviolet--mid-IR range.
Moreover, we performed spectral energy distribution fits to determine $f_{\rm AGN}$ using the {\tt CIGALE} code with the {\tt SKIRTOR} AGN model.
We found that (i) the value of $f_{\rm AGN}$ in the CAMIRA clusters is positively correlated with $z_{\rm cl}$, with the correlation slope being steeper than that for field galaxies, and (ii) $f_{\rm AGN}$ exhibits a high value at the cluster outskirts. 
These results indicate that the emergence of AGN population depends on the redshift and environment and that galaxy groups and clusters at high redshifts are important in AGN evolution. 
Additionally, we demonstrated that cluster--cluster mergers may enhance AGN activity at the outskirts of particularly massive galaxy clusters. 
Our findings are consistent with a related study on the CAMIRA clusters that was based on the AGN number fraction.

\end{abstract}



%####################
%   Introduction
%####################
\section{Introduction}
\label{Intro}

Determination of the effects of active galactic nuclei (AGNs) on the formation and evolution of galaxy clusters and their member galaxies during the universe history is important.
This is because (i) almost all galaxies contain supermassive black holes (SMBHs) that can influence the host galaxies \cite[e.g.,][]{Magorrian,Ferrarese,Woo} and (ii) AGNs may affect the dynamics and energetics of galaxy clusters \citep[e.g.,][and references therein]{Fabian}.
Therefore, studies on these systems offer a unique opportunity to investigate the relation between AGNs and the host galaxies.

Notably, AGN number fraction is an important parameter for understanding the abovementioned effects. 
It is often defined as the number of AGNs within the member galaxies of a cluster. 
Many researchers have investigated the AGN number fraction for galaxy groups and clusters \citep[e.g.,][]{Krick,Martini09,Tomczak,Pentericci,Ehlert14,Magliocchetti,Koulouridis,Koulouridis24}.
In addition, this parameter has been explored using semianalytic galaxy formation models, such as those reported by \cite{Marshall} and \cite{Munoz}.

For example, for galaxy clusters, the AGN number fraction is found to increase with increasing redshift \citep[e.g.,][]{Eastman,Martini09,Pentericci,Mishra,Bhargava}. 
This is similar to the redshift evolution of blue galaxies in galaxy clusters \citep{Butcher}. 
Furthermore, some studies have reported that the AGN fraction depends on the environment---that is, the fraction is higher in denser environments than in the field \citep{Manzer} \citep[see also][for counter arguments]{Manzer,Santos}.

Recently, based on numerous galaxy groups and clusters up to $z\sim 1.4$, \cite{Hashiguchi} investigated the dependence of the AGN number fraction on the cluster redshift ($z_{\rm cl}$) and distance from the cluster center.
These galaxy groups and clusters were discovered using the Hyper Suprime-Cam \citep[HSC;][]{Miyazaki} Subaru Strategic Program \citep[HSC-SSP;][]{Aihara18b,Aihara18a,Aihara19,Aihara22}.
The HSC-SSP is an optical-imaging survey covering approximately 1,200 deg$^2$ with five broadband filters and approximately 30 deg$^2$ with five broadband and four narrowband filters \citep[see][]{Bosch,Coupon,Furusawa,Huang,Kawanomoto,Komiyama}\footnote{We refer the reader to \cite{Schlafly}, \cite{Tonry}, \cite{Magnier}, \cite{Chambers}, \cite{Juric}, and \cite{Ivezic} for relevant papers.}.
\cite{Hashiguchi} constructed an unbiased AGN sample by combining AGN selection methods with multiwavelength data. 
They reported that the AGN number fraction increases with increasing $z_{\rm cl}$; its value was higher than that of field galaxies, regardless of $z_{\rm cl}$. 
Moreover, they reported that the AGN number fraction primarily contributed by radio-selected AGNs shows considerable excess in the cluster center, while that primarily contributed by infrared (IR)--selected AGNs exhibits a small excess at the cluster outskirts.

A non-negligible issue concerning the AGN number fraction is its poor statistics owing to low AGN surface densities. 
Most previous studies have employed tens--hundreds of AGN samples, which tend to cause large Poisson errors when the AGN number fractions are distributed in subsamples, such as in redshift bins. 
In addition, optical/IR color(s) and luminosity have been utilized for identifying member galaxies that host an AGN. This always involves a trade-off between purity and completeness \citep[e.g.,][]{Toba15,Assef}.
Hence, color(s)- or luminosity-based AGN selection may miss weak AGNs or induce contamination from star-forming galaxies.

Herein, we revisit the dependence of the AGN fraction on $z_{\rm cl}$ and distance from the cluster center from an AGN energy perspective. 
We define the ratio of the contribution of the AGN IR luminosity to the total IR luminosity---i.e., $L_{\rm IR}$ (AGNs)/$L_{\rm IR} $---as the AGN IR ``power fraction'' $f_{\rm AGN}$.
We are able to determine this quantity through analysis of the spectral energy distribution (SED) of each galaxy.
The advantages of using the AGN power fraction are the following: (i) $f_{\rm AGN}$ can be determined for each member galaxy, allowing discussion of the abovementioned issues, but with small statistical errors, and (ii) we can detect signatures even from weak AGNs that may be missed by color- and luminosity-based selections, as demonstrated by \cite{Pouliasis20}.
Following \cite{Hashiguchi}, we utilized the sample of galaxy groups and clusters discovered using the HSC-SSP.

The remainder of this paper is structured as follows. 
In Section \ref{DA}, we describe the galaxy cluster sample and SED analysis used herein to calculate $f_{\rm AGN}$.
The resultant dependences of the AGN power fraction on $z_{\rm cl}$ and distance from the cluster center are presented in Section \ref{Res}. 
In Section \ref{Dis}, we discuss the possible uncertainties in these results and their consistency with results based on the AGN number fraction. 
We also explain the enhancements in AGN activity at the cluster outskirts due to cluster--cluster mergers.
Finally, we summarize the primary conclusions in Section \ref{S_summary}.
All information about galaxy clusters with member galaxy samples, such as coordinates, photometry, and derived physical quantities, is available in a catalog form (Appendix \ref{app1}). 
We also provide the best-fit SED templates of the sources (Appendix \ref{app2}).
Throughout this paper, we adopt the cosmology of a flat universe with $H_0$ = 70 km s$^{-1}$ Mpc$^{-1}$, $\Omega_{\rm M}$ = 0.28, and $\Omega_{\rm \Lambda}$ = 0.72\footnote{The assumed cosmology is the same as that employed by \cite{Oguri18} and \cite{Hashiguchi}.}.
All magnitudes are given according to the AB system, unless specified otherwise.

%%%%%%%%%%%%%%%%%%%%%%
% Data and analysis
%%%%%%%%%%%%%%%%%%%%%%
\section{Data and analysis}
\label{DA}

%=========================
% CAMIRA cluster catalog
%=========================
\subsection{CAMIRA Galaxy Cluster Catalog}
\label{s_CAMIRA}

We used the same sample of galaxy groups and clusters as those used by \cite{Hashiguchi}. 
We refer the reader to that paper for details but present the following brief summary.

\begin{figure}
\centering
\includegraphics[width=0.45\textwidth]{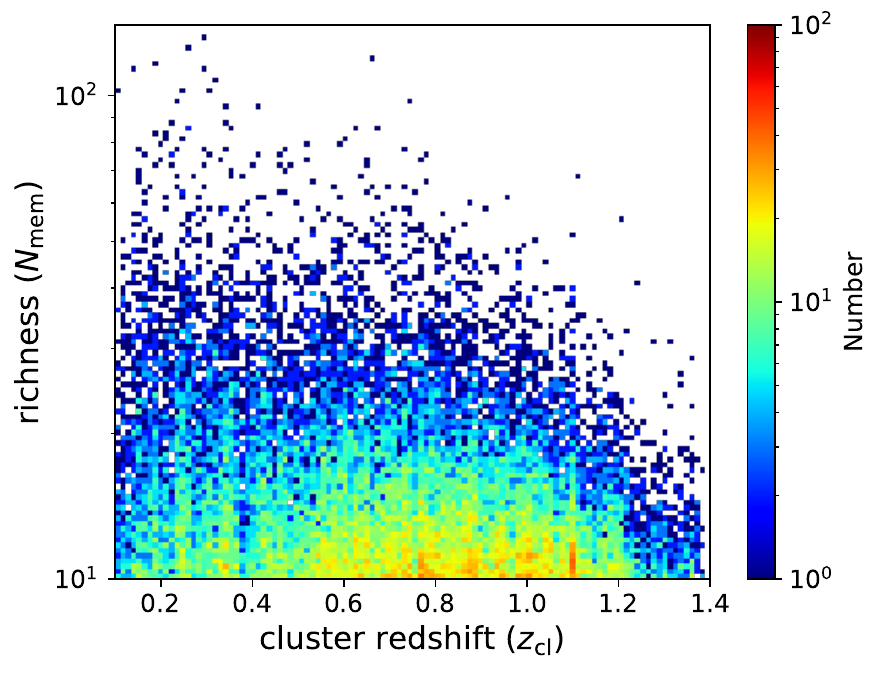}
\caption{Distributions of cluster redshift ($z_{\rm cl}$) and richness ($N_{\rm mem}$) for the CAMIRA clusters, color-coded by the number of objects per pixel.}
\label{camira}
\end{figure}

\begin{figure}
\centering
\includegraphics[width=0.45\textwidth]{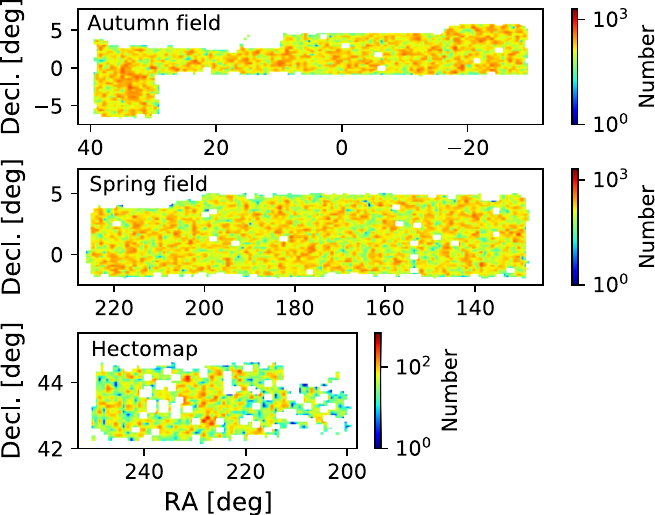}
\caption{Density maps of CAMIRA member galaxies (top: autumn field, middle: spring field, and bottom: Hectomap).}
\label{sky}
\end{figure}

We utilized the HSC-selected galaxy group and cluster catalog obtained by applying the CAMIRA (CAMIRA stands for the cluster-finding algorithm based on the multiband identification of red-sequence galaxies) \citep{Oguri} to the HSC-SSP data. 
This algorithm essentially uses the $r$, $i$, and $z$ colors for HSC sources with $z_{\rm AB}$ $<$ 24.0 to find the overdensity of galaxies \citep[see][for additional details]{Oguri18}.
We employed the latest version ({\tt s21a\_v1}) of the CAMIRA catalog with bright-star masks, which provides 27,037 galaxy groups and clusters with a richness of $N_{\rm mem} > 10$ over $\sim$ 1,027 deg$^2$. 
The $z_{\rm cl}$ of the CAMIRA groups/clusters is distributed over the range $0.1 < z_{\rm cl} < 1.4$; it could be efficiently determined, in contrast to the spectroscopic redshift ($z_{\rm spec}$) of the brightest cluster galaxies (BCGs) \citep[Figure 7 in][]{Oguri18}.
Figure \ref{camira} presents the distributions of $N_{\rm mem}$ and $z_{\rm cl}$ in the CAMIRA clusters \citep[see also Figure 1 in][]{Hashiguchi}.
The CAMIRA catalog contains 1,052,529 member galaxies and is one of the largest of such catalogs to date. The distribution of the CAMIRA member galaxies on the sky is shown in Figure \ref{sky}.
Hereafter, galaxy groups and clusters are simply referred to as galaxy clusters\footnote{Although we herein focus on galaxy clusters with $N_{\rm mem}$ $>$ 10, notably, numerous AGN-related studies only focused on galaxy groups or on galaxy pairs and compact groups \citep[e.g.,][]{Ellison,Tzanavaris,Bitsakis,Zucker,Li}.}.

%~~~~~~~~~~~~
%   Figure
%~~~~~~~~~~~~
%~~~~~~~~~~~~

%~~~~~~~~~~~~
%   Figure
%~~~~~~~~~~~~
%~~~~~~~~~~~~

The redshifts of the member galaxies originate from $z_{\rm spec}$ and photometric redshift ($z_{\rm phot}$). 
We compiled the values of $z_{\rm spec}$ from literature. such as from 
the Sloan Digital Sky Survey \citep[SDSS;][]{York} Data Release (DR) 15 \citep{Aguado},
Galaxy And Mass Assembly \citep[GAMA;][]{Driver} DR2 \citep{Baldry}, 
DEEP2/3 DR4 \citep{Cooper,Newman}, 
3D-HST v4.1.5 \citep{Skelton,Momcheva}, 
PRIMUS DR1 \citep{Coil11,Cool13}, 
VIPERS DR2 \citep{Guzzo,Scodeggio}, and
VVDS final DR \citep{LeFevre} \citep[see also][]{Tanaka,Nishizawa}.
Furthermore, we employed the Direct Empirical Photometric code \citep[{\tt DEmP}:][]{Hsieh} for member galaxies without $z_{\rm spec}$. 
{\tt DEmP} is an empirical quadratic-polynomial photometric-redshift fitting code that exhibits good performance with red-sequence galaxies \citep[see][for a full description of this redshift code]{Hsieh05,Hsieh}.

\cite{Hashiguchi} reported that the $z_{\rm phot}$ of member galaxies could be efficiently estimated with a bias ($\delta_{\rm z}$) of 0.004, a scatter ($\sigma_{\rm z}$) of 0.012, and an outlier rate ($f_{\rm out}$) of 0.008\footnote{Following \cite{Oguri18}, \cite{Hashiguchi} evaluated the accuracy of $z_{\rm cl}$ using the residual ($z_{\rm cl} - z_{\rm BCG,spec}) / (1+z_{\rm BCG,spec})$. The bias ($\delta_{\rm z}$) and scatter ($\sigma_{\rm z}$) were defined using the mean and standard deviations of the residual with 4$\sigma$ clipping. The outlier fraction ($f_{\rm out}$) was defined as the fraction of galaxies removed by 4$\sigma$ clipping.}.
Because of the involved high quality, we did not herein consider the uncertainty in $z_{\rm phot}$, which exhibits a negligible impact on the final results. 
If a member galaxy has a value of $z_{\rm spec}$, it is used; otherwise, $z_{\rm phot}$ is used. 
Hereafter, we refer to the redshift calculated in this manner as $z_{\rm mem}$. 
We narrowed down the member galaxy sample to sources with $|z_{\rm cl} - z_{\rm mem}| \leq 0.05 \times ( 1 + z_{\rm cl})$ for selecting reliable member galaxies \citep[][see also \citealt{Ando}]{Tanaka}.
This procedure left 877,642 member galaxies in the CAMIRA clusters, which were used for our subsequent analyses.

%==================
% Multi-wavelength dataset
%==================
\subsection{Multiwavelength Dataset}
\label{s_multi}

We compiled multiwavelength data from the ultraviolet (UV) to the mid-IR (MIR) range for CAMIRA member galaxies.
As described in Section \ref{cigale}, this multiwavelength dataset enables us to derive reliable AGN power fractions.

%------------
%    UV
%------------
\subsubsection{UV Data}
\label{s_UV}

For the far-UV (FUV) and near-UV (NUV) data, we utilized the revised catalog \citep[{\tt GUVcat\_AIS\footnote{\url{http://dolomiti.pha.jhu.edu/uvsky/GUVcat/GUVcat_AIS.html}.}};][]{Bianchi} from the Galaxy Evolution Explorer \citep[GALEX;][]{Martin}. This catalog contains 82,992,062 sources with 5$\sigma$ limiting magnitudes of 19.9 and 20.8 for the FUV and NUV, respectively. 

We extracted 79,750,759 sources with (i) {\tt GRANK} $\leq$ 1 and (ii) {\tt fuv\_artifact} = 0 and {\tt fuv\_flags} = 0 or {\tt nuv\_artifact} = 0 and {\tt nuv\_flags} = 0, where {\tt GRANK}, {\tt f/nuv\_artifact}, and {\tt f/nuv\_flags} are the primary-source, artifact, and extraction flags, respectively.
A source with {\tt GRANK} = 0 has no other sources within 2\arcsec.5 while a source with {\tt GRANK} = 1 is the best source comprising more than one source within this radius.
In the GALEX pipeline, the program SExtractor \citep{Bertin} is used for the detection and photometry of sources in the GALEX.
Here, {\tt f/nuv\_flags} contains eight flag bits, with {\tt f/nuv\_flags} = 0 meaning that there are no warnings about the source extraction process.
{\tt f/nuv\_artifact} is the bitwise logical `OR', and {\tt f/nuv\_artifact} = 0 implies that a source is unaffected by any artifacts.
We refer the reader to Section 6.2 and Appendix A of a paper by \cite{Bianchi} and the SExtractor User Manual\footnote{\url{https://sextractor.readthedocs.io/en/latest/Flagging.html}} for more details.

%------------
%  u-band
%------------
\subsubsection{$u$-band Data}
\label{s_u}

Further, we utilized the $u$-band data from the SDSS and Kilo-Degree Survey \citep[KiDS;][]{deJong}.
In particular, we used the SDSS {\tt PhotoPrimary} table in DR17 \citep{Abdurrouf} and KiDS DR3 \citep{de_Jong}, which contain 469,053,874 and 48,736,590 sources, respectively.
The SDSS survey has a 5$\sigma$ limiting $u$-band magnitude of approximately 22.0, while that of the KiDS survey is approximately 24.3.
Because KiDS DR3 does not extend over the entire region covered by the HSC-SSP, we employed the SDSS $u$-band data for objects outside the KiDS footprint. 
To ensure reliable $u$-band fluxes for KiDS sources, we extracted sources with {\tt FLAGS\_U} = 0 where {\tt FLAGS\_U} is the extraction flag output by SExtractor (Section \ref{s_UV}).
If an object lies outside the KiDS footprint, we obtain its $u$-band flux densities based on the SDSS data; otherwise, we refer to the KiDS $u$-band flux.
We also checked the consistency between KiDS $u$-band mag ($u_{\rm KiDS}$) and SDSS $u$-band mag ($u_{\rm SDSS}$).
The weighted mean of $u_{\rm KiDS} - u_{\rm SDSS}$ is 0.05 mag, indicating that the $u$-band photometry is consistent with the KiDS and SDSS data.

%------------
% optical data
%------------
\subsubsection{Optical Data}
\label{s_op}

The HSC-SSP survey comprises three layers with different survey depths and areas: wide, deep, and ultradeep layers \citep[Tables 3 and 4 in][]{Aihara18a}.
We used the HSC-SSP s21a wide-layer data obtained between March 2014 and January 2021, which provide forced photometry in the $g$-, $r$-, $i$-, $z$-, and $y$-bands with 5$\sigma$ limiting magnitudes of 26.8, 26.4, 26.4, 25.5, and 24.7, respectively.
These photometry data are included in the CAMIRA catalog.

%------------
%   NIR
%------------
\subsubsection{Near-IR Data}
\label{s_NIR}

We compiled near-IR (NIR) data based on the VISTA Kilo-degree Infrared Galaxy Survey \citep[VIKING;][]{Edge} DR4\footnote{\url{https://www.eso.org/rm/api/v1/public/releaseDescriptions/135}}, containing 94,819,861 sources.
We used $J$-, $H$-, and $K_{\rm S}$-band data with 5$\sigma$ limiting magnitudes of approximately $J$ = 21 in Vega magnitude.
The VIKING catalog contains the Vega magnitude of each source. We converted these Vega magnitudes into AB magnitudes by employing the offset values $\Delta m$ ($m_{\rm AB} = m_{\rm Vega} + \Delta m$) for the $J$-, $H$-, and $K_{\rm S}$-bands of 0.916, 1.366, and 1.827, respectively, following \cite{Gonz}.
Before cross-matching, we extracted 80,580,274 objects with {\tt primary\_source} = 1 and ({\tt jpperrbits} $<$ 256 or {\tt hpperrbits} $<$ 256 or {\tt kspperrbits} $<$ 256) to ensure clean photometry for uniquely detected objects, similar to \cite{Toba19}.

Because the VIKING DR4 partially covers the HSC-SSP footprint, we employed data from the UKIRT Infrared Deep Sky Survey \citep[UKIDSS;][]{Lawrence} and Two Micron All Sky Survey \citep[2MASS;][]{Skrutskie}.
We utilized the UKIDSS Large Area Survey DR11plus obtained from the WSA--WFCAM Science Archive\footnote{\url{http://wsa.roe.ac.uk/index.html}}, containing 88,298,646 sources.
The limiting magnitudes of UKIDSS are 20.2, 19.6, 18.8, and 18.2 Vega magnitudes in the $Y$-, $J$-, $H$-, and $K$-bands, respectively.
The UKIDSS catalog lists the Vega magnitudes for each source; subsequently, they are converted into AB magnitudes using offset values $\Delta m$ ($m_{\rm AB} = m_{\rm Vega} + \Delta m$) for the $Y$-, $J$-, $H$-, and $K$-bands of 0.634, 0.938, 1.379, and 1.900, respectively, following \cite{Hewett}.
Before cross-matching, we selected 77,225,762 objects with ({\tt priOrSec} $\leq$ 0 or = {\tt frameSetID}) and ({\tt jpperrbits} $<$ 256 or {\tt hpperrbits} $<$ 256 or {\tt kpperrbits} $<$ 256) to ensure clean photometry for uniquely detected objects.
For the 2MASS data, we employed the AllWISE catalog (Section \ref{s_MIR}) that includes 2MASS photometry.
If an object lies outside the VIKING footprint but inside the UKIDSS footprint, we obtain its NIR flux densities based on UKIDSS.
If an object lies outside the VIKING footprint and even the UKIDSS footprint, we use 2MASS data.
Otherwise, we always refer to the VIKING NIR data.

We evaluated the consistency among the VIKING, UKIDSS, and 2MASS NIR photometry data.
For $J$-band, the weighted mean of $J_{\rm UKIDSS} - J_{\rm VIKING}$ and $J_{\rm UKIDSS} - J_{\rm 2MASS}$ is 0.06 and 0.08 mag, respectively.
For $H$-band, the weighted mean of $H_{\rm UKIDSS} - H_{\rm VIKING}$ and $H_{\rm UKIDSS} - H_{\rm 2MASS}$ is 0.03 and 0.08 mag, respectively.
For $K$-band, the weighted mean of $K_{\rm UKIDSS} - Ks_{\rm VIKING}$ and $K_{\rm UKIDSS} - Ks_{\rm 2MASS}$ is 0.02 and 0.11 mag, respectively.
These results indicate that the NIR photometry used herein is broadly consistent with the VIKING, UKIDSS, and 2MASS data.

%------------
%    MIR
%------------
\subsubsection{Mid-IR Data}
\label{s_MIR}

We obtained MIR data based on the Wide-field Infrared Survey Explorer \cite[WISE;][]{Wright}.
We utilized the unWISE \citep{Lang,Schlafly19} and AllWISE catalogs \citep{Cutri}, which provide 2,214,734,224 and 747,634,026 sources, respectively.
The 5$\sigma$ detection limits in the AllWISE catalog are approximately 0.054, 0.071, 0.73, and 5 mJy at 3.4, 4.6, 12, and 22 $\mu$m, respectively. 
The detection limit in the unWISE catalog is deeper by a factor of 2 at 3.4 and 4.6 $\mu$m.
Following \cite{Toba17}, we extracted 741,753,366 sources from the AllWISE catalog, with ({\tt w1sat} = 0 and {\tt w1cc\_map}=0), ({\tt w2sat} = 0 and {\tt w2cc\_map}=0), ({\tt w3sat} = 0 and {\tt w3cc\_map} = 0), or ({\tt w4sat} = 0 and {\tt w4cc\_map} = 0) to obtain reliable photometry in each band.
The unWISE and AllWISE catalogs contain the Vega magnitude of each source. We converted these Vega magnitudes into AB magnitudes using offset values $\Delta m$ ($m_{\rm AB} = m_{\rm Vega} + \Delta m$) of 2.699, 3.339, 5.174, and 6.620 for 3.4, 4.6, 12, and 22 $\micron$, respectively, following the Explanatory Supplement to the AllWISE Data Release Products\footnote{\url{https://wise2.ipac.caltech.edu/docs/release/allwise/expsup/}}.
For flux densities in unWISE, we corrected possible contamination from surrounding sources by using {\tt frac\_flux}.
Since unWISE 3.4 and 4.6 $\micron$ flux densities are deeper than those of ALLWISE, we basically used unWISE data for 3.4 and 4.6 $\micron$ data while ALLWISE was used for 12 and 22 $\micron$ data.

%--------------------
%  Cross-matching
%--------------------
\subsection{Cross-identification of Multiwavelength Data}
\label{s_cross}

Finally, multiwavelength data (taken from GALEX, SDSS, KiDS, VIKING, UKIDSS, unWISE, and ALLWISE) were added to the CAMIRA member galaxies by cross-matching each catalog with the CAMIRA member galaxy catalog.
We employed 1$\arcsec$ for KiDS, VIKING and SDSS and 3$\arcsec$ for GALEX, unWISE, and AllWISE as a search radius following \cite{Toba22}.
Consequently, among 877,642 CAMIRA member galaxies, 8,647 (1.0 \%), 263,891 (30.1\%), 113,415 (12.9\%), 293,360 (33.4\%), 347,653 (40.0\%), 594,405 (67.7\%), and 386,737 (44.1\%) objects from GALEX, SDSS, KiDS, VIKING, UKIDSS, unWISE, and AllWISE were identified, respectively.
Notably, 478/8,647 (5.5\%), 19,629/263,891 (7.4\%), 1,718/113,415 (1.5\%), 5,716/293,360 (1.9\%), 7,738/347,653 (2.2\%), 90.748/594,405 (15.2\%), and 63,546/386,737 (16.4\%) objects had more than two candidate counterparts within the search radius for the GALEX, SDSS, KiDS, VIKING, UKIDSS, unWISE, and AllWISE sources, respectively.
We chose the nearest object as the counterpart in these cases.
We confirmed that the following results remained unaffected even when the HSC sources with multiple candidate counterparts were discarded during the SED fitting.

%==================
%   SED fitting
%==================
\subsection{SED Fitting with {\tt CIGALE}}
\label{cigale}

To derive the AGN power fraction $f_{\rm AGN}$ = $L_{\rm IR}$ (AGNs)/$L_{\rm IR}$, we performed SED fitting. 
We employed the Code Investigating GALaxy Emission \citep[{\tt CIGALE\footnote{We use version 2022.1. See \url{https://cigale.lam.fr/2022/07/04/version-2022-1/}}};][]{Burgarella,Noll,Boquien,Yang,Yang22}.
This code allows us to include values for many parameters related to, e.g., the star formation history (SFH), single stellar population, attenuation law, AGN emissions, and dust emissions by considering the energy balance between the UV/optical and IR \cite[see e.g.,][]{Buat12,Ciesla_15,Boquien16,LoFaro,Toba20a,Toba20b,Setoguchi,Toba21b,Suleiman,Uematsu,Yamada,Setoguchi24,Uematsu24}. 
The parameter values used in the SED fitting are presented in Table \ref{Param}.

To find a best-fit SED and estimate physical properties with their uncertainties, {\tt CIGALE} adopted an analysis module, that is, the so-called {\tt pdf\_analysis}.
This module computes the likelihood for all the possible combinations of parameters and generates the probability distribution function for each parameter and each object.
Further, it scales the models by a factor ($\alpha$) to obtain physically meaningful values before computing the likelihood. 
$\alpha$ is described as follows:
\begin{equation}
\alpha = \frac{\sum_i \frac{f_i m_i}{\sigma_i^2}}{\sum_i \frac{m_i^2}{\sigma_i^2}} + \frac{\sum_j \frac{f_j m_j}{\sigma_j^2}}{\sum_j \frac{m_j^2}{\sigma_j^2}},
\end{equation}
where $f_{i}$ and $m_{i}$ are the observed and model flux densities, respectively; $f_{j}$ and $m_{j}$ are the observed and model extensive physical properties, respectively; $\sigma$ is the corresponding uncertainties.
The only parameter that is subjected to fitting is the scale factor $\alpha$.
Finally, {\tt pdf\_analysis} computes the probability-weighted mean and standard deviation corresponding to the resultant value and its uncertainty for each parameter.
An advantage of the methodology adopted in {\tt CIGALE} is that the models must be computed only once for all the objects because a fixed grid of models is used \cite[see Section 4.3 in ][for a detailed explanation of this module]{Boquien}.

%~~~~~~~~~~~~~~~~
\begin{table}
\renewcommand{\thetable}{\arabic{table}}
\centering
\caption{Parameter values used in SED fitting with {\tt CIGALE}.} 
\label{Param}
\begin{tabular}{ll}
\tablewidth{0pt}
\hline
\hline
Parameter & \multicolumn1c{Value} \\
\hline
\multicolumn2c{Delayed SFH}\\
\hline
$\tau_{\rm main}$ (Gyr) 	& 0.1, 0.5, 1.0, 2.0, \\  & 4.0, 6.0, 8.0, 10.0 \\
age (Gyr) 					& 0.1, 0.5, 1.0, 2.0, \\  & 4.0, 6.0, 8.0, 10.0, 12.0\\
\hline
\multicolumn2c{Single Stellar Population \citep{Bruzual}}\\
\hline
IMF				&	\cite{Chabrier} \\
Metallicity		&	0.02 \\
\hline
\multicolumn2c{Dust Attenuation \citep{Calzetti}}\\
\hline
$E(B-V)_{\rm line}$	& 0.05, 0.10, 0.15, 0.20, 0.25, \\ 
					& 0.40, 0.60, 0.80, 1.00, 2.00 \\ 
\hline
\multicolumn2c{AGN Disk +Torus Emission \citep{Stalevski}}\\
\hline
$\tau_{\rm 9.7}$ 			&  	3, 7, 11 		\\
$p$							&	0.5, 1.5	\\
$q$							&	0.5, 1.5	\\
$\Delta$ (\arcdeg)			&	30, 50, 70			\\
$R_{\rm max}/R_{\rm min}$ 	& 	30 				\\
Viewing angle $\theta$ (\arcdeg)			&	40			\\
AGN power fraction ($f_{\rm AGN}$) 	& 	0.00, 0.05, 0.10, 0.15, 0.20, 0.25, \\																						& 	0.30, 0.35, 0.40, 0.45, 0.50, 0.55, \\
									& 	0.60, 0.65, 0.70, 0.75, 0.80, 0.85, \\
									& 	0.90, 0.95, 0.99 \\
\hline
\multicolumn2c{AGN Polar Dust Emission \citep{Yang}}\\
\hline
extinction law					&	SMC		\\
$E(B-V)$						&	0.0, 0.1, 0.5, 1.0	\\
$T_{\rm dust}^{\rm polar}$ (K)	&	100	\\
Emissivity $\beta$				&	1.6\\
\hline
\multicolumn2c{Dust Emission \citep{Dale}} \\
\hline
IR power-law slope ($\alpha_{\rm dust}$) & 2.0 \\
\hline
\end{tabular}
\end{table}
%~~~~~~~~~~~~~~~~

We adopted a delayed-SFH model by assuming a single starburst with an exponential decay \citep[e.g.,][]{Ciesla_15,Ciesla_16}, where we parameterized the age of the main stellar population and $e$-folding time of the main stellar population ($\tau_{\rm main}$). 
We chose the single stellar population model of \citep{Bruzual} by assuming the initial mass function (IMF) of \cite{Chabrier} and employed the nebular emission model reported by \cite{Inoue} with the implementation of the new CLOUDY HII-region model \citep{Villa}.
For the attenuation of dust associated with the host galaxy, the models proposed by \cite{Calzetti} and \cite{Leitherer} were used.
We parameterized the color excess of the nebular emission lines $E(B-V)_{\rm lines}$, extended using the \cite{Leitherer} curve.
Additionally, we modeled the reprocessed IR emission of dust absorbed from UV/optical stellar emission considering the dust absorption and emission templates of \cite{Dale}. 

We modeled the AGN emission as emission from an accretion disk and dust torus using {\tt SKIRTOR}\footnote{\url{https://sites.google.com/site/skirtorus/sed-library}} \citep{Stalevski}, a clumpy two-phase torus model produced within the framework of the three-dimensional radiative-transfer code {\tt SKIRT} \citep{Baes,Camps}.
The AGN model in {\tt CIGALE} comprises seven parameters: the optical depth of the torus at 9.7 $\micron$ ($\tau_{\rm 9.7}$), a torus-density radial parameter ($p$), a torus-density angular parameter ($q$), the angle between the equatorial plane and torus edge ($\Delta$), the ratio between the maximum and minimum torus radii ($R_{\rm max}/R_{\rm min}$), the viewing angle ($\theta$), and the AGN power fraction ($f_{\rm AGN}$)\footnote{We note that $f_{\rm AGN}$ is not the quantity introduced into {\tt SKIRTOR} by \cite{Stalevski}. \cite{Yang} implemented this parameter as an AGN-related quantity in {\tt CIGALE}, wherein the relative normalization of dust emissions heated by AGN and SF components is achieved by $f_{\rm AGN}$.}.
$R_{\rm max}/R_{\rm min}$ and $\theta$ are fixed to be 30 and 40, respectively, to avoid degeneracy of AGN templates \citep[see][]{Yang}.
Because $f_{\rm AGN}$ is the primary topic of this study, we parameterized it using fine intervals.
For the polar dust emission, we assumed a gray body with a dust temperature $T_{\rm dust}^{\rm polar}$ = 100 K and emissivity $\beta$ = 1.6 \citep[e.g.,][]{Casey}.
We parametrized $E (B-V)$ for the polar dust considering the SMC extinction curve \citep{Prevot}.

\begin{figure}
\centering
 \includegraphics[width=0.45\textwidth]{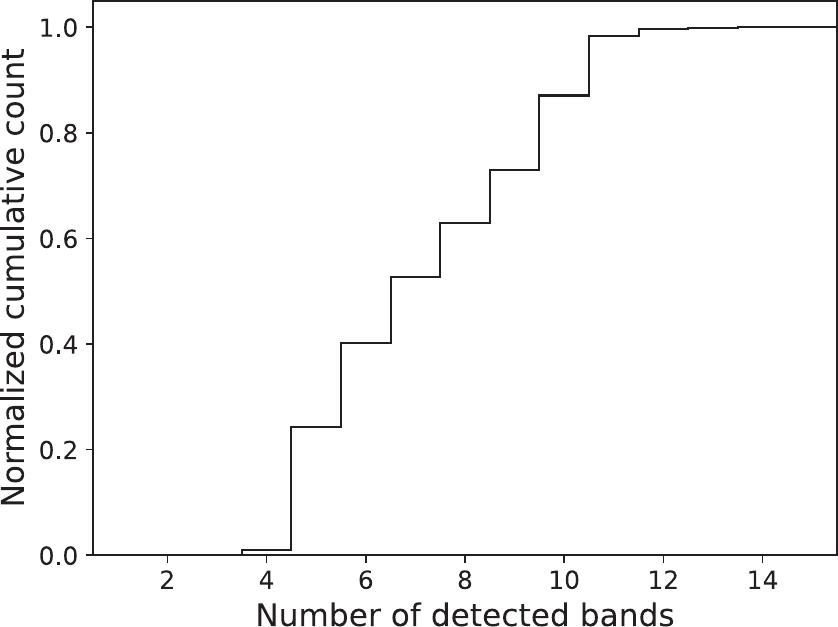}
\caption{Normalized cumulative histogram of the number of detected bands for CAMIRA member galaxies.}
\label{n_filter}
\end{figure}

Under the parameter settings summarized in Table \ref{Param}, we fit the stellar, nebular, AGN, and SF components to at most 15 photometric points (FUV, NUV, $u$, $g$, $r$, $i$, $z$, $y$, $J$, $H$, and $K/Ks$ bands, and 3.4, 4.6, 12, and 22 $\micron$) of 877,642 CAMIRA member galaxies.
Figure \ref{n_filter} shows the normalized cumulative histogram of the detected bands, demonstrating that roughly half of the sources have more than eight detected bands for the SED fitting.
Even if an object is not detected in a band, we utilize this information by setting 5$\sigma$ upper limits following \cite{Toba19} to further constrain the SEDs (Section \ref{SED}).
We will discuss how the number of detected bands affects the resultant $f_{\rm AGN}$ in Section \ref{d_mock}.

Notably, the photometry adopted in each catalog is different.
As the photometric flux densities are expected to trace the total flux densities, the influence of different photometry methods is likely to be small. 
Nevertheless, it is worth investigating if physical properties can actually be reliably estimated given the uncertainty of each photometry. This topic will be discussed in Section \ref{d_mock}.
We also note that the wavelength coverage of our data is limited to UV--MIR. 
This is because the energy output is expected to peak in the UV (if unprocessed) and IR (when dust-reprocessed); thus, the dataset we used will be sufficient to estimate the AGN power fraction.
However, since some CAMIRA member galaxies are detected by using other wavelengths such as X-ray and radio \citep{Hashiguchi}, it is worth investigating how the lack of these data, especially X-ray data, affects the resulting AGN power fraction. This topic will be discussed in Section \ref{d_X}.

%-------------
%   Figure
%-------------
%-------------

%%%%%%%%%%%%%%%%%%%%%%
%     Results 
%%%%%%%%%%%%%%%%%%%%%%
\section{Results} 
\label{Res}

%===========================
% Result of the SED fitting
%============================
%-------------
%   Figure
%-------------
\begin{figure*}
\centering
 \includegraphics[width=\textwidth]{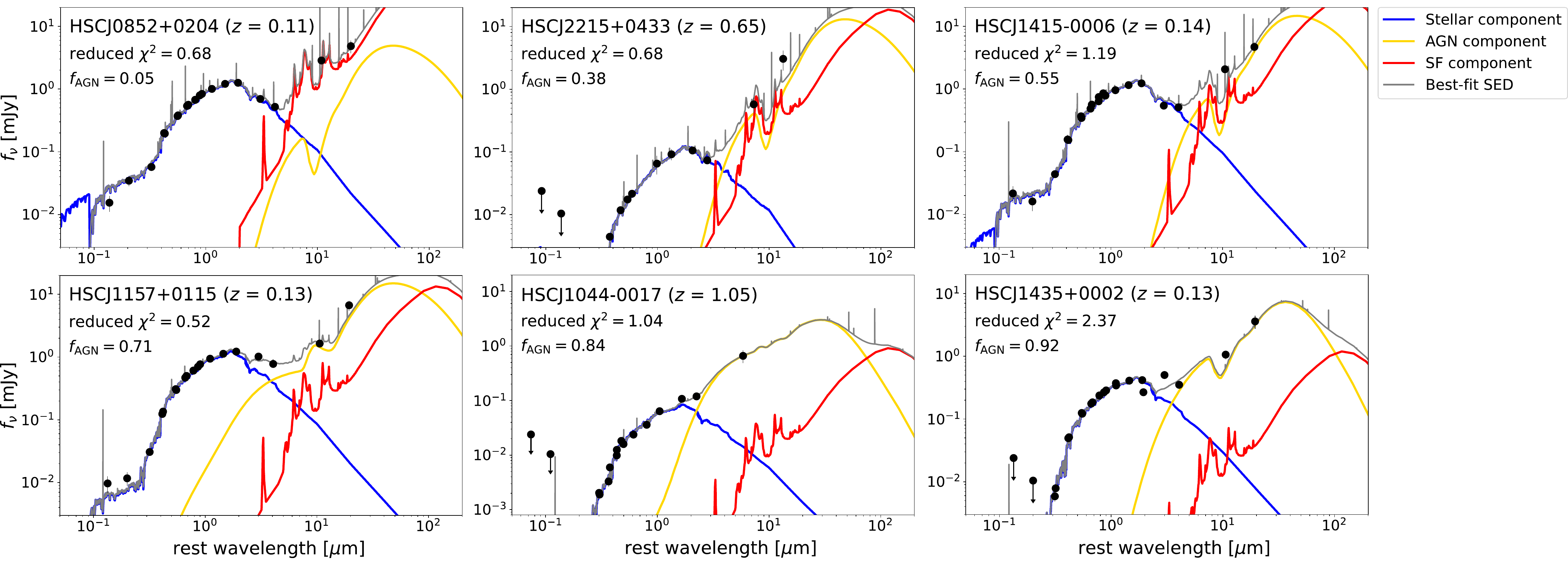}
\caption{Examples of SED fitting for CAMIRA member galaxies. The black points represent photometric data, while the gray solid line represents the best-fit SED; the contributions of the stellar, AGN, and SF components to the total SED are shown as blue, yellow, and red lines, respectively. The AGN power fraction increases from the top left to bottom right.}
\label{fig_SED}
\end{figure*}
%-------------

\subsection{SED Fitting Results}
\label{SED}

\begin{figure}
\centering
 \includegraphics[width=0.45\textwidth]{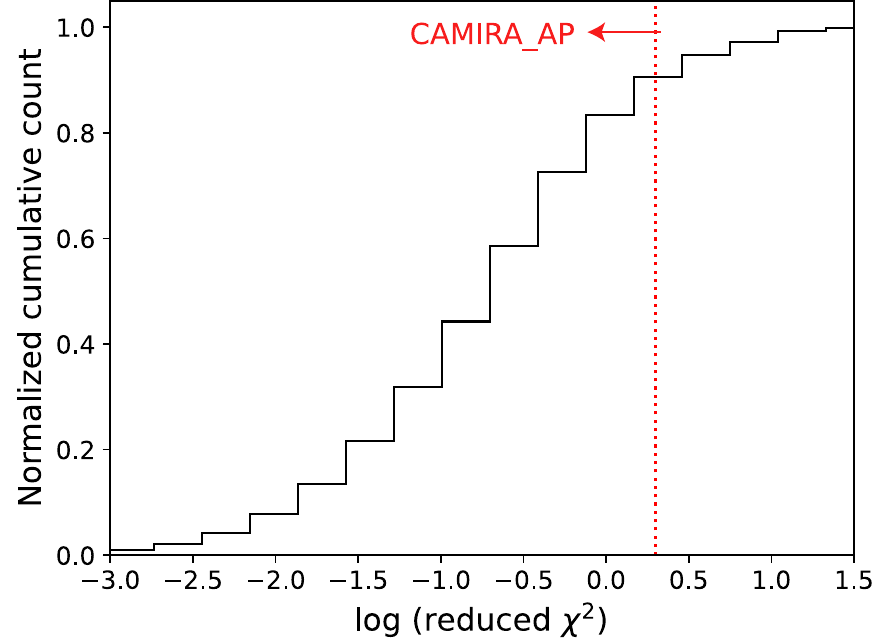}
\caption{Normalized cumulative histogram of reduced $\chi^2$ obtained through SED fitting for CAMIRA member galaxies. The vertical line indicates the reduced $\chi^2$ threshold to make a subsample.}
\label{chi2}
\end{figure}

Figure \ref{fig_SED} presents examples of SED fitting obtained with {\tt CIGALE}.
The normalized cumulative distribution of reduced $\chi^2$ is shown in Figure \ref{chi2}.
We confirmed that 756,140 out of 877,642 objects (approximately 86 \%) exhibited reduced $\chi^2 < 2.0$, while 789,329 out of 877,642 objects (approximately 90 \%) demonstrated reduced $\chi^2 < 3.0$.
Those results suggest that the data were moderately well fitted by {\tt CIGALE} using combination of the stellar, nebular, AGN, and SF components.
Meanwhile, notably, approximately 63\% of sources have reduced $\chi^2 < 0.5$, which may mean that the data are overfitted and that the estimated uncertainties may be underestimated.
In Section \ref{d_mock}, we will discuss how the AGN power fraction of sources with low reduced $\chi^2$ are affected.
Hereafter, we will focus on the subsample of 756,140 sources with reduced $\chi^2 < 2.0$.
We refer to these sources as ``CAMIRA AGN power sample (CAMIRA\_AP).''
In this work, we primarily focus on the AGN power fraction ($f_{\rm AGN}$) of CAMIRA\_AP member galaxies.
The other AGN properties estimated by {\tt CIGALE} (i.e., AGN torus-related values) are also provided in Appendix \ref{app3}.

%-------------
%   Figure
%-------------
%-------------

%===========================
% AGN fraction distribution
%============================
\subsection{AGN Power Fraction Distribution}
\label{r_fAGN}
 
In this study, we focused on galaxies that are expected to be member galaxies of clusters by comparing $z_{\rm cl}$ and $z_{\rm mem}$, as outlined in Section \ref{s_CAMIRA}.
Notably, the spectroscopic completeness of the CAMIRA\_AP member galaxies is considerably low (22,516/756,140 $\sim$ 3.0\%); therefore, we relied on $z_{\rm phot}$ for most galaxies. 
Hence, foreground and background galaxies could possibly contaminate the member galaxies. 
The CAMIRA catalog assigns a weight factor ($w_{\rm mem}$\footnote{Here, $w_{\rm mem}$ is defined as shown in Equation (20) of a paper by \cite{Oguri}. This equation is obtained by multiplying the cluster member galaxy ``number'' parameter, stellar-mass filter, and spatial filter, which are defined as shown in Equations (5), (8), and (9) of the paper by \cite{Oguri}, respectively.}) to each member galaxy corresponding to its membership probability. 
This factor is obtained by applying the fast Fourier transform to the two-dimensional richness map containing properties of galaxies (e.g., color and stellar mass).
A full description of $w_{\rm mem}$, ranging between 0 and 1, is provided by \citet{Oguri}.
A galaxy with $w_{\rm mem}$ closer to 1 is more likely to be a cluster member galaxy, as evident from its increased membership probability.

Therefore, we employed a membership-probability-weighted AGN power fraction to mitigate the influence of contamination from foreground and background galaxies, as introduced by \cite{Bufanda}.
Because AGN power fraction ($f_{\rm AGN} = L_{\rm IR}$ (AGNs)/$L_{\rm IR}$) can be calculated for each member galaxy and each galaxy is assigned a probability of being a cluster member ($w_{\rm mem}$), the membership-probability-weighted AGN power fraction is defined here in a CAMIRA\_AP cluster ($f^{\rm cl}_{\rm AGN}$) as
\begin{equation}
\label{df_fAGN_cl}
f^{\rm cl}_{\rm AGN} = \frac{\sum w^i_{\rm mem} ~ f^i_{\rm AGN}} {\sum w^i_{\rm mem}},
\end{equation}
where $f^i_{\rm AGN}$ and $w^i_{\rm mem}$ are the AGN power fractions of the $i$-th member galaxy in a CAMIRA\_AP cluster and its corresponding membership weight factor, respectively.
The uncertainty in $f^{\rm cl}_{\rm AGN}$ is calculated as its standard deviation.

As this work compares $f_{\rm AGN}$ for galaxy clusters with that for field galaxies (Section \ref{S_fAGN}), we define the AGN power fraction for field galaxies ($f^{\rm fd}_{\rm AGN}$).
For this purpose, we selected 503,595 field galaxies using the HSC-SSP with the same magnitude and color cuts as those used for red-sequence galaxies
\citep{Oguri,Oguri18}, although the field galaxies do not belong to a cluster.
We collected multiwavelength data from the UV--MIR range in exactly the same manner as that discussed in Section \ref{s_multi} and estimated $f_{\rm AGN}$ based on SED fitting using the parameter sets described in Section \ref{cigale}.
The range of the redshift of the field galaxies ($z_{\rm fd}$) is the same as that of the CAMIRA\_AP member galaxies.
For subsequent analysis, we extracted 371,993 field galaxies by adopting reduced $\chi^2 < 2.0$ from the SED fitting, in the same manner as that adopted for cluster member galaxies (Section \ref{SED}).
Because a field galaxy, by definition, does not have $w_{\rm mem}$, we used only field galaxies with $z_{\rm spec}$ and measured the AGN power fraction by weighting the uncertainties of $f_{\rm AGN}$ rather than $w_{\rm mem}$.
Hemce $f^{\rm fd}_{\rm AGN}$ is formulated as follows:
\begin{equation}
\label{df_fAGN_fd}
f^{\rm fd}_{\rm AGN} = \frac{\sum \sigma_{f^i_{\rm AGN}} ~ f^i_{\rm AGN}} {\sum \sigma_{f^i_{\rm AGN}} },
\end{equation}
where $f^i_{\rm AGN}$ and $\sigma_{f^i_{\rm AGN}}$ are the AGN power fractions of the $i$-th field galaxy and its uncertainty, respectively.
The uncertainty in $f^{\rm fd}_{\rm AGN}$ is calculated as its standard deviation.

\begin{figure}[h]
\centering
 \includegraphics[width=0.45\textwidth]{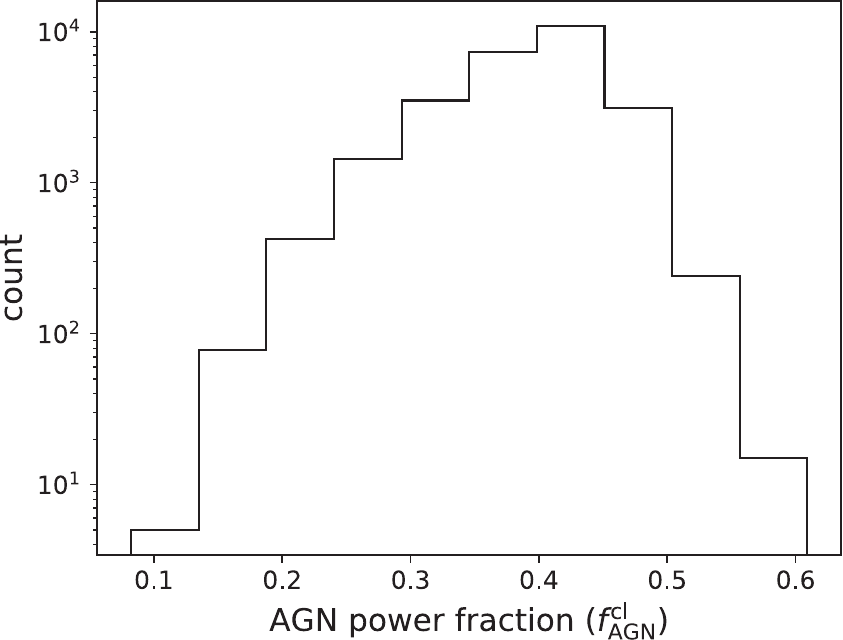}
\caption{Histogram of the membership-probability-weighted AGN power fraction ($f^{\rm cl}_{\rm AGN}$) for the CAMIRA\_AP clusters.}
\label{fAGN_hist}
\end{figure}

\begin{figure}[h]
\centering
 \includegraphics[width=0.45\textwidth]{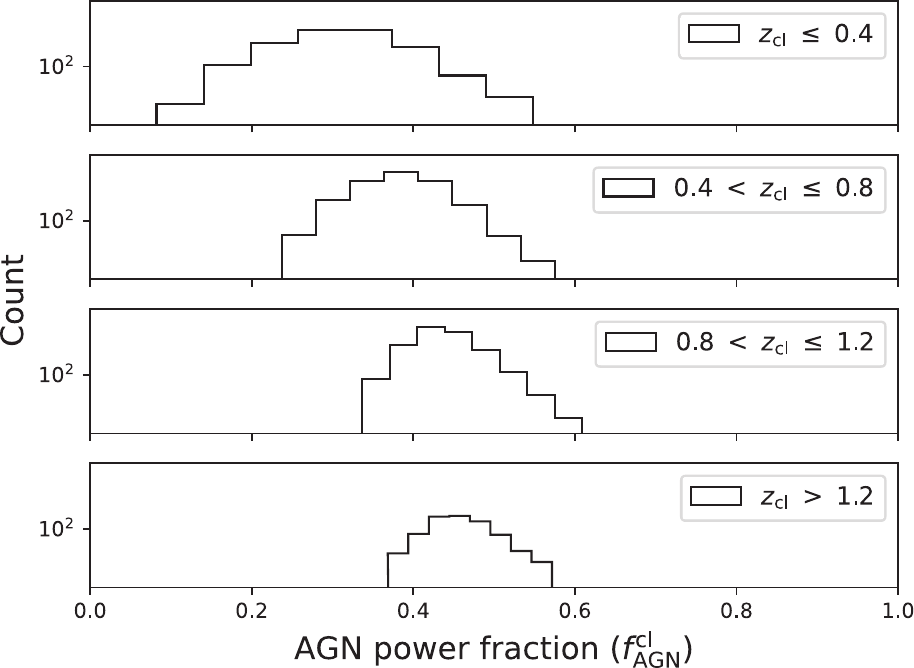}
\caption{Histograms of the AGN power fraction ($f^{\rm cl}_{\rm AGN}$) for the CAMIRA\_AP clusters in each redshift bin.}
\label{fAGN_hist_zbin}
\end{figure}

Figure \ref{fAGN_hist} presents the distribution of $f^{\rm cl}_{\rm AGN}$ in CAMIRA\_AP clusters.
The mean and standard deviations of $f^{\rm cl}_{\rm AGN}$ are 0.39 and 0.06, respectively.
We also divided the cluster sample into subsamples based on redshifts ($z_{\rm cl} ~\leq~ 0.4$, $0.4 < z_{\rm cl} ~\leq~ 0.8$, $0.8 < z_{\rm cl} ~\leq~ 1.2$, and $z_{\rm cl} > 1.2$) following \cite{Hashiguchi}.
The numbers of clusters in the redshift bins are 5,193, 10,819, 10,113, and 912, respectively.
We examined the $f^{\rm cl}_{\rm AGN}$ distribution of the clusters in each redshift bin (Figure \ref{fAGN_hist_zbin}).
The peak value of the histogram for each panel gradually increases with increasing $z_{\rm cl}$, providing tentative evidence for the redshift dependence of the AGN power fraction (Section \ref{S_z_fAGN} for a quantitative discussion). 
However, the range of the distribution is relatively large regardless of the redshift, and this should be kept in mind in the subsequent discussion.

%-------------
%   Figure
%-------------
%-------------

%-------------
%   Figure
%-------------
%-------------

\begin{figure}[h]
\centering
 \includegraphics[width=0.45\textwidth]{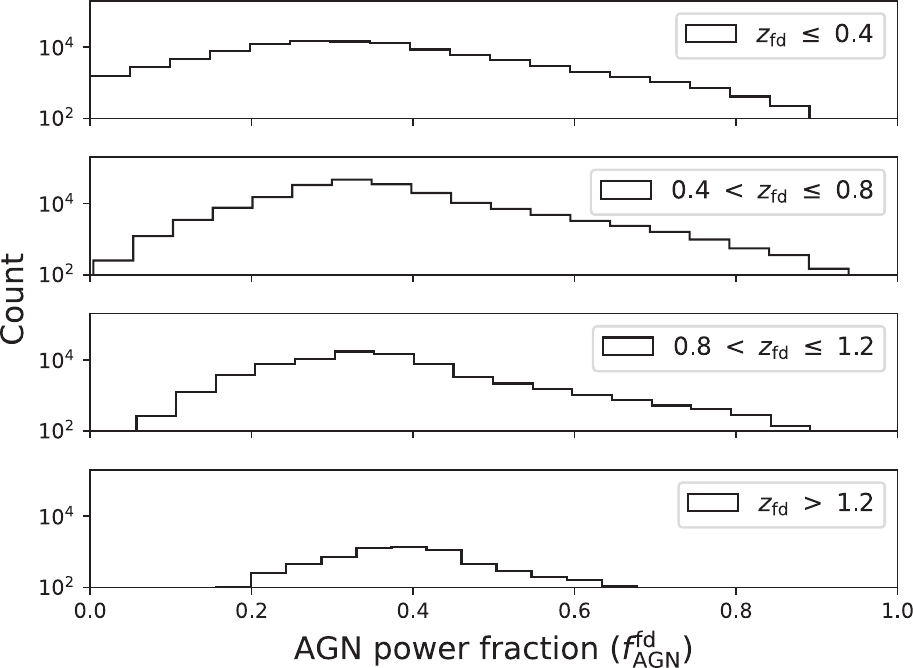}
\caption{Histograms of the AGN power fraction ($f^{\rm fd}_{\rm AGN}$) for the field galaxies in each redshift bin.}
\label{fAGN_hist_zbin_fd}
\end{figure}

Figure \ref{fAGN_hist_zbin_fd} shows the distribution of $f^{\rm fd}_{\rm AGN}$ of field galaxies in each redshift bin ($z_{\rm fd} \leq 0.4$, $0.4 < z_{\rm fd} \leq 0.8$, $0.8 < z_{\rm fd} \leq 1.2$, and $z_{\rm fd} > 1.2$), and these redshift bins have 98,584, 193,731, 72,849, and 6,829 field galaxies, respectively.
The peak value of the histogram for each panel slightly increases with increasing $z_{\rm fd}$; this increase also provides tentative evidence for the redshift dependence of the AGN power fraction even for field galaxies (see Section \ref{S_z_fAGN} for a quantitative discussion). The mean and standard deviations of $f^{\rm fd}_{\rm AGN}$ over the whole redshift range ($0.1 < z_{\rm fd} < 1.4$) are 0.35 and 0.13, respectively, which are lower than those for galaxy clusters, suggesting that the AGNs may tend to ignite in a dense environment.

%-------------
%   Figure
%-------------
%-------------

%=============================================
% Redshift and dist_cen dependence of the fAGN
%=============================================
\subsection{Redshift and Centric Radius Dependences of the AGN Power Fraction for Galaxy Clusters}
\label{S_fAGN}

We next consider the dependence of the AGN power fraction on (i) redshift and (ii) distance ($R$) from the cluster center (cluster centric radius) scaled using the virial radius ($R_{\rm 200}$), $R/R_{\rm 200}$.
Further, we compared the AGN power fraction for clusters with that for field galaxies.
Here, we stacked the redshift and radial distributions of member galaxies in redshift or $R/R_{\rm 200}$ bins.
We then calculated the AGN power fraction and its uncertainty in each bin by applying the $w_{\rm mem}$ weights, similar to that described in Section \ref{r_fAGN} (Equation \ref{df_fAGN_cl}).
Those enabled us to compare the AGN power fraction for clusters to that for field galaxies, as described in Sections \ref{S_z_fAGN} and \ref{S_dist_fAGN}.
Hereafter, we denote the AGN power fraction calculated in this manner as $f^{\rm mem}_{\rm AGN}$.

%-----------------
% z_cl vs. f_AGN 
%-----------------
\subsubsection{Redshift Dependence of the AGN Power Fraction}
\label{S_z_fAGN}

\begin{figure}[h]
\centering
 \includegraphics[width=0.45\textwidth]{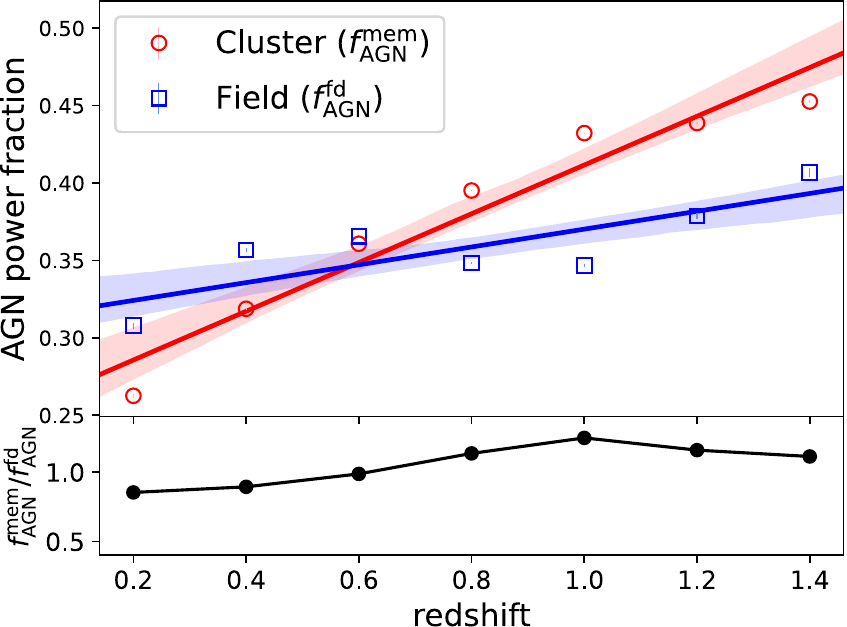}
\caption{Red circles and blue squares represent the AGN power fractions in the CAMIRA\_AP clusters and field, respectively. The vertical error bars are calculated based on the standard deviation of the weighted mean. The mean relative error of the AGN power fraction across the redshift bin is approximately 0.2\%. The solid lines with shaded regions represent the best-fit linear regressions with 1$\sigma$ confidence intervals. The bottom panel shows the ratios of $f_{\rm AGN}^{\rm mem}$ and $f_{\rm AGN}^{\rm fd}$.}
\label{z_fAGN}
\end{figure}

Figure \ref{z_fAGN} presents the AGN power fraction as a function of redshift.
The mean relative error of AGN power fraction across the redshift bin is approximately 0.2\%\footnote{If we employ weighted standard deviation (instead of the standard deviation of the weighted mean) as the uncertainty of AGN power fraction in each bin, the relative error gets larger.}.
We verified that $f^{\rm mem}_{\rm AGN}$ increases with increasing redshift, even from an AGN power fraction point of view, similarly to what is reported for the AGN {\it number} fraction by previous studies \citep[e.g.,][]{Galametz,Martini09,Haggard,Hashiguchi}.
We performed a linear regression to fit the $z_{\rm mem}$--$f^{\rm mem}_{\rm AGN}$ relation considering the uncertainties in each redshift bin.
Additionally, we calculated the correlation coefficient ($r$) based on a Bayesian regression method \citep{Kelly}, which yields the correlation coefficient with its corresponding uncertainty \cite[e.g.,][]{Toba19b,Toba21}.
The resulting value $r$ = 0.90 $\pm$ 0.14 indicates a strong positive correlation between redshift and AGN power fraction; in other words, a Butcher--Oemler \citep{Butcher}-like effect for AGNs in galaxy clusters was confirmed, even from the perspective of AGN power fraction.

%-------------
%   Figure
%-------------
%-------------

A positive correlation was found between redshift and AGN number fraction, as has been reported for ``field galaxies'' by many researchers \citep[e.g.,][]{Silverman,Haggard,Oi}.
\cite{Buat_15} also reported that the AGN power fraction increases with increasing redshift.
To determine if this observed trend is specific to the AGNs in clusters or if it merely reflects the trend observed in field galaxies, we calculated the AGN fraction for field galaxies detected using the HSC-SSP. 

Figure \ref{z_fAGN} shows the AGN power fraction for the field galaxies as a function of redshift. 
We here calculated the AGN power fraction and its uncertainty in each redshift bin by applying the $\sigma_{f_{\rm AGN}}$  weights, similar to that described in Equation \ref{df_fAGN_fd}.
An increasing trend of AGN fraction toward a high redshift is also confirmed for field galaxies, with a correlation coefficient of $r = 0.66 \pm 0.30$.
Moreover, we observe that the best-fit slope for clusters is steeper than that for the field, indicating accelerated AGN growth in cluster galaxies with increasing redshift compared to the field, as reported by previous studies \citep{Eastman,Hashiguchi}.
Average excess ($f^{\rm mem}_{\rm AGN}$/$f^{\rm fd}_{\rm AGN}$) is about 1.2 at $z > 0.6$ (bottom panel of Figure \ref{z_fAGN}).
Rapidly increasing AGN number fraction at high redshifts has also been reported for MIR-selected AGN \citep{Tomczak,Hashiguchi} \citep[see Section 4.2 in][for more details]{Hashiguchi}.
These results suggest that a denser environment (galaxy clusters), particularly in a high-$z$ universe, tends to boost AGN activity.

%=====================
% dist_cen vs. f_AGN 
%=====================
\subsubsection{Cluster Centric Radius Dependences on AGN Power Fraction}
\label{S_dist_fAGN}

Next, we examined the AGN power fraction as a function of the projected distance from the cluster center (cluster centric radius) scaled using $R_{\rm 200}$, $R/R_{\rm 200}$; we determined the cluster centers using centroids of the BCGs identified with the CAMIRA \citep{Oguri18}.
Here, $R_{\rm 200}$ is the radius within which the mass density is 200 times the mean universe mass density. 
We obtain $R_{\rm 200}$ by assuming the scaling relation between $N_{\rm mem}$ and cluster mass ($M_{\rm 200}$) reported for CAMIRA clusters by \citet{Okabe} \citep[see also][]{Murata,Chiu}, where $M_{\rm 200}$ is the total mass enclosed within a sphere of radius $R_{\rm 200}$.
$R_{\rm 200}$ ranges 0.5--2.5 Mpc for our cluster sample.

%-------------
%   Figure
%-------------
\begin{figure}[h]
\centering
 \includegraphics[width=0.45\textwidth]{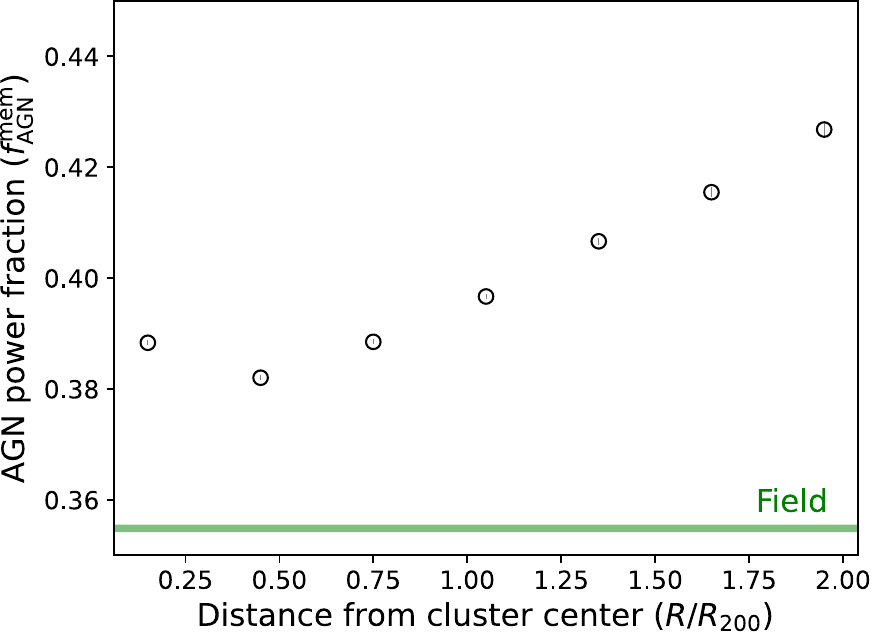}
\caption{AGN power fraction as a function of cluster centric radius scaled using the virial radius ($R/R_{\rm 200}$). The vertical error bars were calculated based on the standard deviation of the weighted mean. The mean relative error of AGN power fraction across the $R/R_{200}$ bin is about 0.2\%. The green-shaded region shows an average AGN power fraction for the field.}
\label{dist_fAGN}
\end{figure}
%-------------

Figure \ref{dist_fAGN} shows the resultant $f^{\rm mem}_{\rm AGN}$ as a function of $R/R_{\rm 200}$.
We found an excess of $f^{\rm mem}_{\rm AGN}$ at the outskirts of galaxy clusters, which is in good agreement with the results of previous studies \citep[e.g.,][]{Khabiboulline,Koulouridis}.
\cite{Hashiguchi} also reported that the MIR-selected AGNs are dominant at the outskirts of CAMIRA clusters, while the radio-selected AGNs are condensed to the cluster center.
Because the AGN power fraction in this research is based on the IR luminosity, the consistency of our result with that for the MIR-AGNs reported by \cite{Hashiguchi} (see also Section \ref{d_AH}) is reasonable.

%###################
%    Discussion 
%###################
\section{Discussion} 
\label{Dis}

%==================
% Possible Biases
%==================
\subsection{Possible Biases}

%------------------
%  Mock Analysis
%------------------
\subsubsection{Mock Analysis}
\label{d_mock}

Given the photometric uncertainties, we checked the reliability of the $f_{\rm AGN}$ derived using {\tt CIGALE}. 
We performed a mock analysis, which is a procedure provided by {\tt CIGALE}.
This mock analysis was performed by creating a mock catalog.
To build the mock catalog, the best-fit value for each object was considered; each quantity was then modified by adding a value taken from a Gaussian distribution with the same standard deviation as the observation uncertainty.
Finally, the same method used in the original estimation was applied to obtain mock estimations. 
Through this analysis, the reliability of the obtained estimations for the physical parameters was estimated.
A full description of this process can be found in a paper by \cite{Boquien} \citep[see also][]{Ciesla_15,Ciesla_16,Toba19,Toba20c,Pouliasis}.

\begin{figure}
    \centering
    \includegraphics[width=0.45\textwidth]{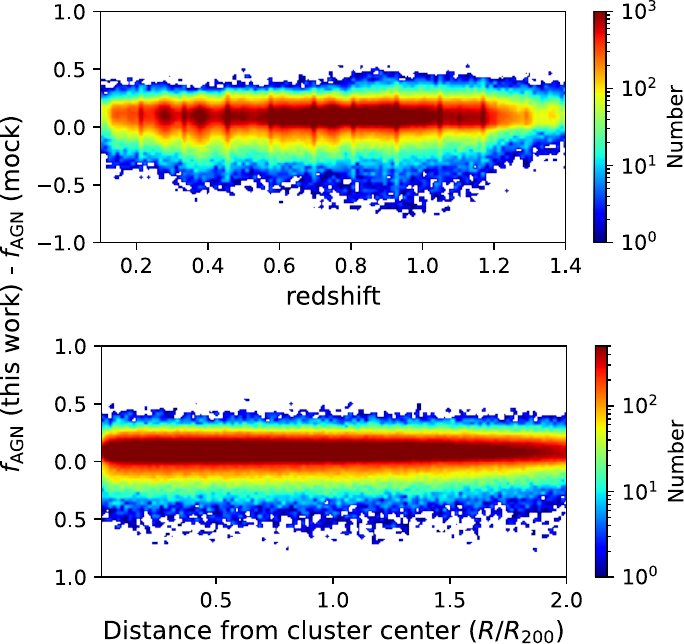}
\caption{Differences in the $f_{\rm AGN}$ derived herein using {\tt CIGALE} and those derived using the mock catalog as a function of redshift (top) and distance from the cluster center (bottom); both are color-coded according to the number of objects per pixel.}
\label{mock}
\end{figure}

Figure \ref{mock} shows the differences in the $f_{\rm AGN}$ derived herein using {\tt CIGALE} and those derived from the mock catalog as a function of redshift and cluster centric radius. 
The mean (with standard deviation) of $\Delta f_{\rm AGN}$ is $0.08 \pm 0.09$, which is acceptable for this research.
This result could also suggest that the derived AGN power fraction is not sensitive to the photometric uncertainty.
We also proved that $f_{\rm AGN}$ is not considerably dependent on the redshift and distance from the cluster center, indicating that $f_{\rm AGN}$ is well determined with negligible systematic uncertainty.

%-------------
%   Figure
%-------------
%-------------

\begin{figure}
\centering
    \includegraphics[width=0.45\textwidth]{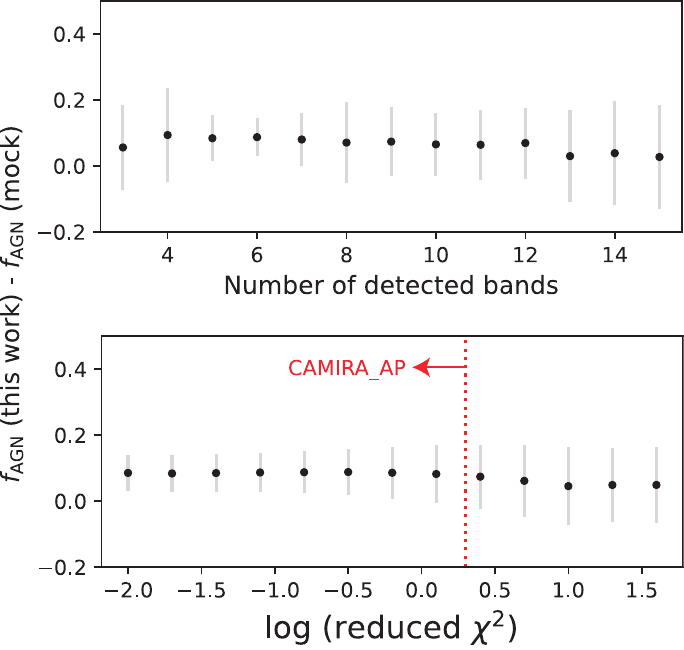}
\caption{Differences between the $f_{\rm AGN}$ derived herein using {\tt CIGALE} and the values derived using the mock catalog as a function of number of detected bands (top) and reduced $\chi^2$ (bottom).}
\label{mock2}
\end{figure}

In Sections \ref{cigale} and \ref{SED}, we reported that the number of detected bands and reduced $\chi^2$ obtained through SED fitting have a large variation (Figures \ref{n_filter} and \ref{chi2}).
We also tested how the number of detected bands and reduced $\chi^2$ depend on $\Delta f_{\rm AGN}$.
Figure \ref{mock2} shows $\Delta f_{\rm AGN}$ as a function of the numbers of detected bands and reduced $\chi^2$, demonstrating that $\Delta f_{\rm AGN}$ does not substantially depend on these values.
Therefore, we conclude that the limited number of detected bands and threshold of reduced $\chi^2$ in this research did not cause systematic uncertainty in the power fraction.
This result could also suggest that the influence of the difference in photometry for the SED fitting is likely to be small, as mentioned in Section \ref{cigale}.

%-------------
%   Figure
%-------------
%-------------

%---------------------------------
%  Lack of X-ray Information
%---------------------------------
\subsubsection{Lack of X-ray Information}
\label{d_X}

\begin{figure}
\centering
    \includegraphics[width=0.45\textwidth]{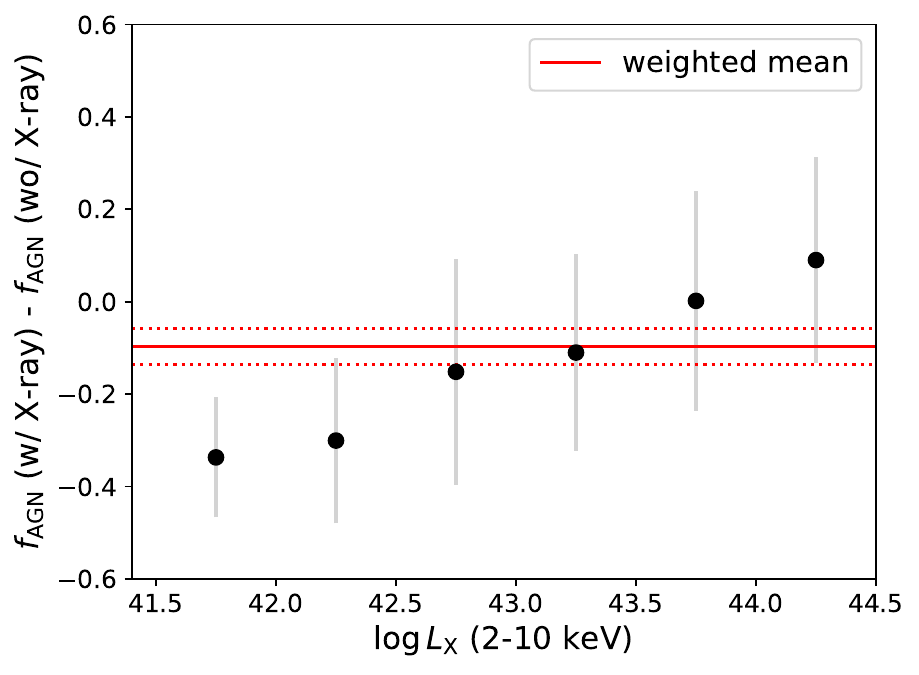}
\caption{Differences in the AGN power fraction derived using {\tt CIGALE} with and without X-ray information as a function of X-ray luminosity in the 2--10-keV range. The red solid and dotted lines indicate the weighted mean and standard deviation, respectively, over the X-ray luminosity range of the sample.}
\label{fAGN_X}
\end{figure}

Herein, UV--MIR data were employed to constrain the SED and estimate AGN power fraction, as described in Section \ref{DA}.
However, some CAMIRA member galaxies are detected using other wavelengths, such as those of X-rays and radio waves \citep{Hashiguchi}.
In particular, X-ray data may be crucial to constrain AGN properties such as $f_{\rm AGN}$, as reported by \cite{Yang}.
Since {\tt CIGALE} is capable of handling X-ray data, as demonstrated by \cite{Yang,Yang22}, we tested how the lack of X-ray data affects the resultant $f_{\rm AGN}$.
For this purpose, we used 263 X-ray-detected CAMIRA member galaxies reported by \cite{Hashiguchi}, who used XMM-Newton data to identify X-ray sources.
We conducted SED fitting to those 263 X-ray sources with and without X-ray flux and compared the calculated AGN power fractions.
Figure \ref{fAGN_X} shows the differences in the AGN power fraction with and without adding X-ray information to SED fitting with {\tt CIGALE} as a function of X-ray luminosity ($L_{\rm X}$) in the 2--10-keV range, where $L_{\rm X}$ was calculated as shown a study by \cite{Hashiguchi}. 
We found that weighted mean and its standard deviation of $f_{\rm AGN}$ (w/ X-ray) - $f_{\rm AGN}$ (wo/ X-ray) is $-0.10 \pm 0.04$.
We also found that this offset depends on the X-ray luminosity.
For a low-luminosity regime, $f_{\rm AGN}$ tends to be overestimated if X-ray information is not used.
Meanwhile, for a high-luminosity regime, $f_{\rm AGN}$ tends to be underestimated if X-ray information is not used.
These tendencies are in good agreement with those reported by \cite{Mountrichas}.
Thus, we should note the possible systematics of the AGN power fraction in this research.

%-------------
%   Figure
%-------------
%-------------

%---------------------------------
%  Comparison with Hashiguchi+23
%---------------------------------
\subsubsection{Comparisons with the AGN Number Fractions}
\label{d_AH}

\cite{Hashiguchi} have identified 2,688 AGNs based on multiwavelength data and determined the AGN number fraction for the same initial sample as in this study (i.e., CAMIRA clusters). 
They have also reported that 2,536 CAMIRA clusters have at least one member galaxy that hosts an AGN. 
We therefore expected that (i) the member galaxies hosting AGNs in a study by \cite{Hashiguchi} will have systematically larger values for $f_{\rm AGN}$ than those without AGNs and (ii) the AGN number fraction for each cluster will be correlated with $f^{\rm cl}_{\rm AGN}$. 

%-------------
%   Figure
%-------------
\begin{figure}[h]
    \centering
    \includegraphics[width=0.45\textwidth]{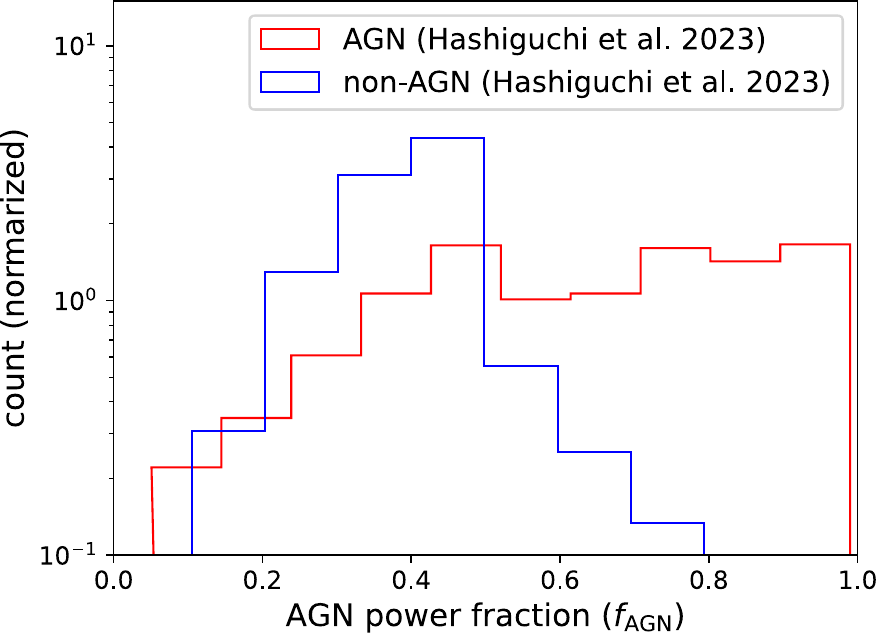}
	\caption{Comparison of $f_{\rm AGN}$ between member galaxies hosting AGNs (red) and those without AGNs (blue) reported by  \cite{Hashiguchi}.}
\label{comp_AH_mem}
\end{figure}
%-------------

%-------------
%   Figure
%-------------
\begin{figure}
 \centering
    \includegraphics[width=0.45\textwidth]{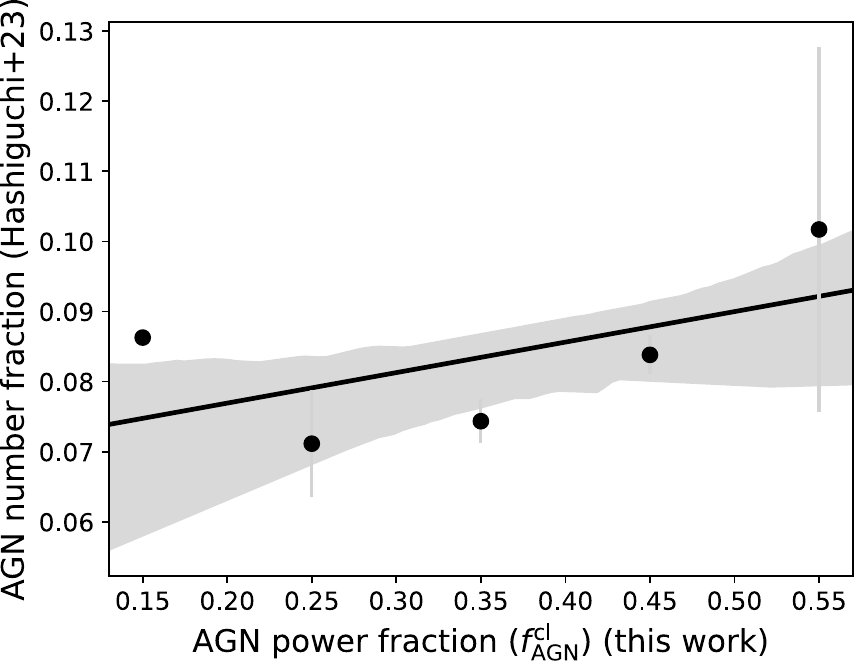}
	\caption{Comparison of the AGN power fraction ($f^{\rm cl}_{\rm AGN}$) obtained herein for each cluster with AGN number fraction reported by \cite{Hashiguchi}.}
\label{comp_AH_cl}
\end{figure}
%-------------

Figure \ref{comp_AH_mem} compares the distributions of the AGN power fraction (we obtained through SED fitting) for the member galaxies that do and do not host AGNs that were determined by \cite{Hashiguchi}.
We found that the $f_{\rm AGN}$ of AGNs identified by \cite{Hashiguchi} is systematically higher than that of objects not identified as AGNs by \cite{Hashiguchi}, which is supported with $>$99.9\% significance by a two-sided Kolmogorov--Smirnov (KS) test. 
The mean values of $f_{\rm AGN}$ for member galaxies with and without AGNs are 0.63 and 0.23, respectively.
Figure \ref{comp_AH_cl} compares the AGN number fraction \citep{Hashiguchi} with the AGN power fraction ($f^{\rm cl}_{\rm AGN}$) for each cluster.
We confirmed a positive correlation between the AGN number and power fractions, with a correlation coefficient of $r = 0.40 \pm 0.27$.
This implies that as expected, a cluster with high AGN number fraction tends to have high AGN power fraction.
We note that about 2\% of objects unclassified as AGN by \cite{Hashiguchi} have a large AGN power fraction ($f_{\rm AGN}$ $>$ 0.7).
Those objects may be good candidates for heavily obscured AGN that were missed by previous surveys.

%-------------
%   Figure
%-------------
\begin{figure}[h]
 \centering
 \includegraphics[width=0.45\textwidth]{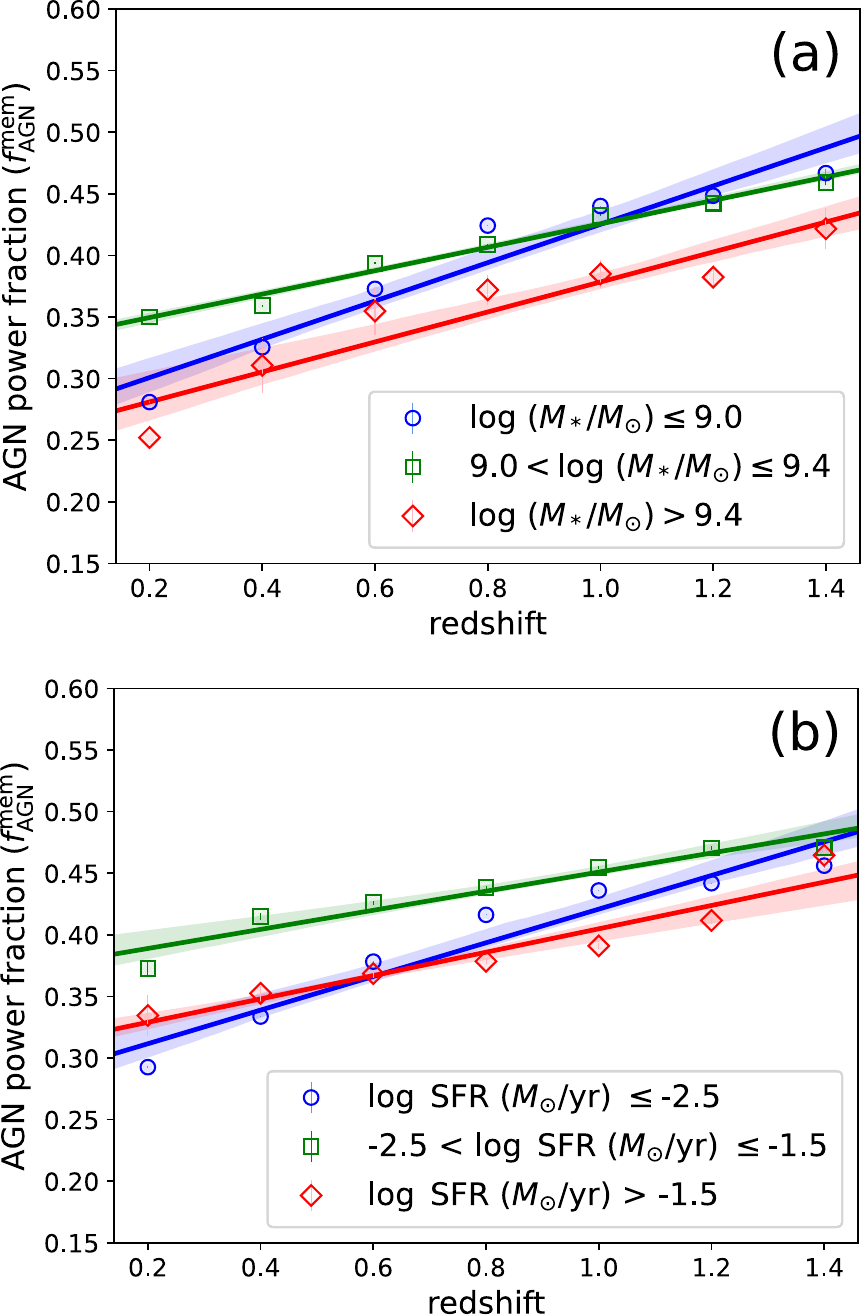}
\caption{AGN power fraction as a function of redshift for CAMIRA member galaxies with (a) different stellar masses (blue circles: $\log\, (M_*/M_{\odot}) ~\leq~ 10.5$, green squares: $10.5 < \log\, (M_*/M_{\odot}) ~\leq~ 11.5$, and red diamonds: $\log\, (M_*/M_{\odot}) > 11.5$) and (b) different SFR (blue circles: $\log\,{\rm SFR} (M_{\odot}/{\rm yr}) ~\leq~ -0.5$, green squares: $-0.5 < \log\,{\rm SFR} (M_{\odot}/{\rm yr}) ~\leq~ 0.0$, and red diamonds: $\log\,{\rm SFR} (M_{\odot}/{\rm yr}) > 0.0$). The vertical error bars represent the standard deviation of the weighted mean. The solid lines with shaded regions represent the best-fit linear regressions with 1$\sigma$ confidence intervals.}
\label{z_fAGN_stellar}
\end{figure}
%-------------

%--------------------------
%  Stellar mass dependence
%--------------------------
\subsubsection{Stellar Mass and SFR Dependence}

We observed a positive correlation between the AGN power fraction and redshift, both for member galaxies in galaxy clusters and those in the field, as discussed in Section \ref{S_z_fAGN}.
However, some studies have reported that the AGN number fraction depends on the stellar mass ($M_*$) of the host galaxy \citep[e.g.,][]{Kauffmann,Best05,Koss,Bitsakis,Miraghaei}.
Because our sample is flux-limited, more massive host galaxies at higher redshifts could exhibit more luminous AGNs, leading to a biased positive correlation between $f_{\rm AGN}$ and redshift.
This may also be the case for the SFR because the stellar mass and SFR are often related, and this relation would evolve along with the redshift reported for star-forming galaxies \citep[e.g.,][]{Noeske,Schreiber,Pearson}.

Following \cite{Hashiguchi}, we divided our sample into subsamples to investigate how the stellar mass and SFR of member galaxies influence the correlation between the AGN power fraction and redshift.
We defined the subsamples using the following ranges of stellar mass: $\log\, (M_*/M_{\odot}) ~\leq~ 10.5$, $10.5 < \log\, (M_*/M_{\odot}) ~\leq~ 11.5$, and $\log\, (M_*/M_{\odot}) > 11.5$; for SFR the range is as follows: $\log\,{\rm SFR} (M_{\odot}/{\rm yr}) ~\leq~ -0.5$, $-0.5 < \log\,{\rm SFR} (M_{\odot}/{\rm yr}) ~\leq~ 0.0$, and $\log\,{\rm SFR} (M_{\odot}/{\rm yr}) > 0.0$.
We obtained $M_*$ and SFR from {\tt CIGALE} outputs, as described in Section \ref{cigale}.
Figure \ref{z_fAGN_stellar} depicts $f_{\rm AGN}$ as a function of redshift for three subsamples with stellar mass and SFR.
We found no significant differences in the correlation coefficients among the subsamples for stellar mass or SFR that were within the respective errors.
Hence, we concluded that the observed positive correlation of $z_{\rm cl}$--$f^{\rm mem}_{\rm AGN}$ was not affected by such biases.

%=================================
%   Cluster mass vs. f_AGN
%=================================
\subsection{Cluster Mass Dependence on AGN Power Fraction}
\label{S_fAGN_nass}

\begin{figure}
 \centering
 \includegraphics[width=0.45\textwidth]{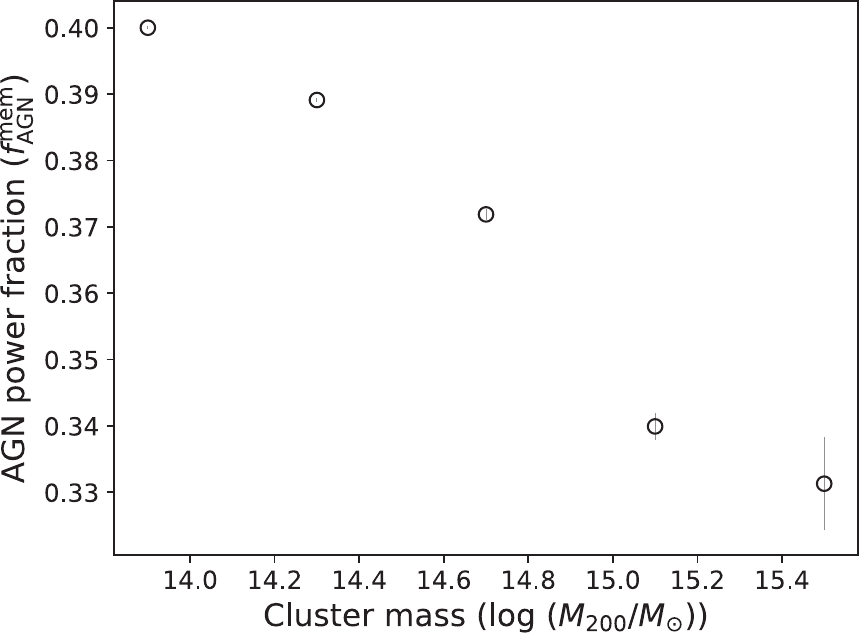}
\caption{AGN power fraction as a function of cluster mass ($M_{\rm 200}$). The vertical error bars represent the standard deviations of the weighted means.}
\label{Mcl_fAGN}
\end{figure}

We also investigated how the AGN power fraction depends on $M_{\rm 200}$ (or richness). 
As shown in Figure \ref{Mcl_fAGN}, the AGN power fraction decreases with increasing $M_{\rm 200}$, with the correlation coefficient being $r = -0.91 \pm 0.14$.
This trend is in good agreement with the results of previous studies \citep[e.g.,][]{Koulouridis18,Noordeh,Hashiguchi}. It was also observed in a study by \citet{Popesso}, who reported an anticorrelation between the AGN number fraction and velocity dispersion of galaxy clusters\footnote{Although our work considers galaxy groups and clusters together, \cite{Popesso} only focused on galaxy clusters.}.
Because the velocity dispersion of a galaxy cluster can be translated into $M_{\rm 200}$ \citep[e.g.,][]{Smith}; this provides additional support for our findings. 
These results suggest that a galaxy in a group environment is more likely to ignite an AGN than a cluster environment, which is consistent with previous studies by \cite{Li}, \cite{Pentericci}, and \cite{Hashiguchi}.

\begin{figure}
 \centering
 \includegraphics[width=0.45\textwidth]{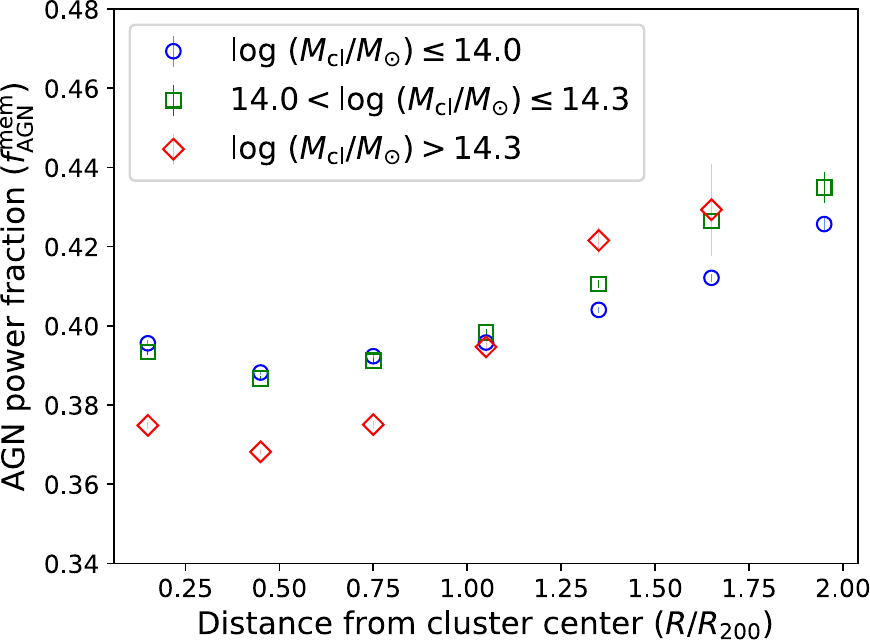}
\caption{AGN power fraction as a function of cluster centric radius scaled using the virial radius ($R/R_{\rm 200}$) for different cluster mass ranges (blue diamonds: $\log\,(M_{\rm cl}/M_{\odot})$ $\leq$ 14.0, green squares: $14.0 < \log\,(M_{\rm cl}/M_{\odot})$ $\leq$ 14.3, and red diamonds: $\log \, (M_{\rm cl}/M_{\odot}) > 14.3$). The vertical error bars represent the standard deviations of the weighted means.}
\label{dist_fAGN_Mcl}
\end{figure}

In addition, we examined how the AGN power fraction depends on the cluster centric radius in different $M_{\rm 200}$ ranges (Figure \ref{dist_fAGN_Mcl}).
Notably, massive clusters tend to be responsible for the rapid increase in the AGN power fraction at the cluster outskirts, as reported in Section \ref{S_dist_fAGN}.
We further discuss this point in Section \ref{S_fAGN_morph}, considering the morphologies of the galaxy clusters.

%-------------
%   Figure
%-------------
%-------------

%-------------
%   Figure
%-------------
%-------------

%=================================
%  Cluster morphology vs. f_AGN
%=================================
\subsection{Cluster Morphology Dependence on AGN Power Fraction}
\label{S_fAGN_morph}

We next consider how the emergence of AGNs depends on the substructure and morphology of galaxy clusters.
\citet{Okabe} identified merging-cluster candidates based on the CAMIRA cluster catalog using a peak-finding method. 
This method involves counting the number of peaks above a redshift-dependent threshold based on Gaussian-smoothed maps of the number densities of member galaxies. 
A galaxy cluster with a single peak was classified as a ``relaxed'' cluster, while a cluster with more than two peaks was considered a ``merging'' cluster. 
Accordingly, they classified 2,558 out of 27,037 CAMIRA clusters (approximately 9.5\%) as merging clusters, with 150,684 member galaxies associated with them.

%-------------
%   Figure
%-------------
\begin{figure}
 \centering
 \includegraphics[width=0.45\textwidth]{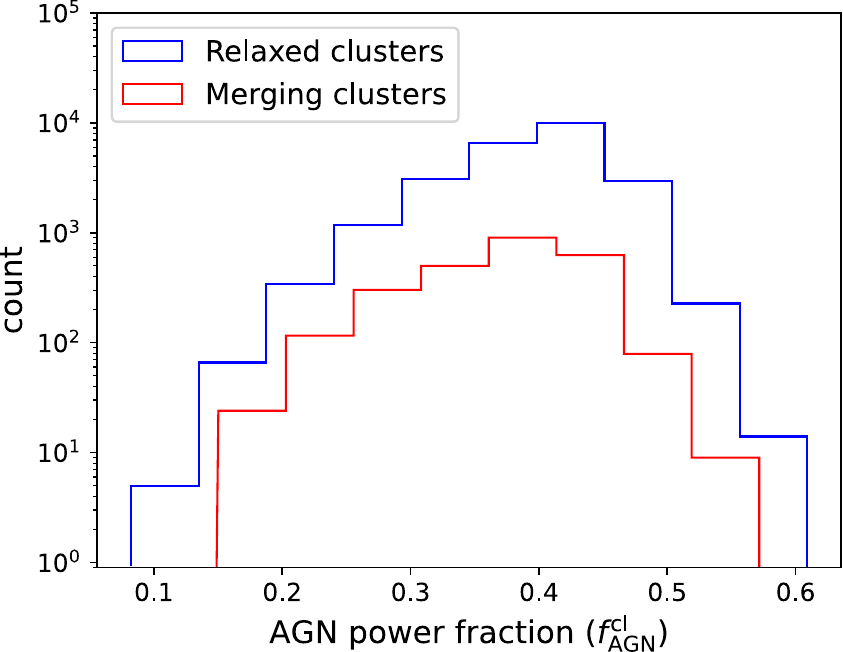}
\caption{$f^{\rm cl}_{\rm AGN}$ distributions for relaxed (blue) and merging (red) clusters.}
\label{fAGN_morph}
\end{figure}
%-------------

Figure \ref{fAGN_morph} presents the distribution of $f^{\rm cl}_{\rm AGN}$ for relaxed and merging clusters.
There was no significant difference between the two clusters, as we confirmed this with $>$99.9\% significance using a two-sided KS test.
This result suggests that cluster--cluster mergers may not necessarily trigger AGNs, as also reported by
 \cite{Silva,Hashiguchi}.
The dynamical activity within a cluster may also activate SF activity in member galaxies \citep[e.g.,][]{Miller,Sobral,Stroe15,Okabe}.
\cite{Stroe} recently reported that a large fraction of emission-line galaxies in merging clusters are powered by star formation rather than by AGNs. 
On the other hand, \cite{Noordeh} suggested that merging-cluster environments may contribute to the enhancement of AGN activity. 
Several enhancement mechanisms have been proposed for SF and AGN activity in merging clusters, including gas incorporation driven by ram pressure and galaxy--galaxy interactions \citep[][and references therein]{Treu}.
The relative strengths of SF and AGN may depend on the abovementioned dominant mechanism and the sequence of cluster--cluster mergers.
Future statistical work that considers the merger stage of a cluster--cluster merger may provide a way to resolve this issue.

%-------------
%   Figure
%-------------
\begin{figure}[h]
 \centering
 \includegraphics[width=0.45\textwidth]{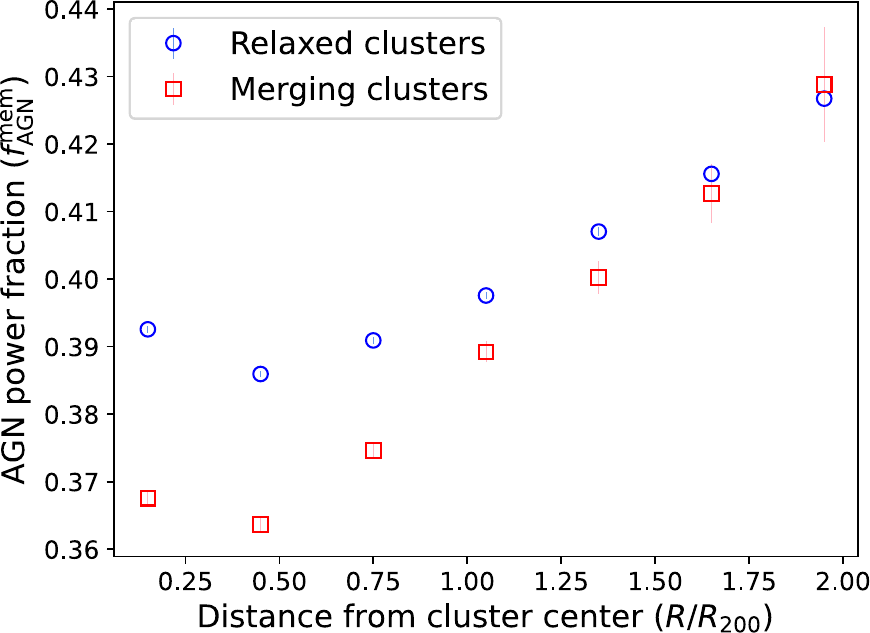}
\caption{AGN power fraction as a function of cluster centric radius scaled using the virial radius ($R/R_{\rm 200}$) for relaxed (blue circles) and merging (red squares) clusters.}
\label{dist_fAGN_morph}
\end{figure}
%-------------

%-------------
%   Figure
%-------------
\begin{figure}
 \centering
 \includegraphics[width=0.45\textwidth]{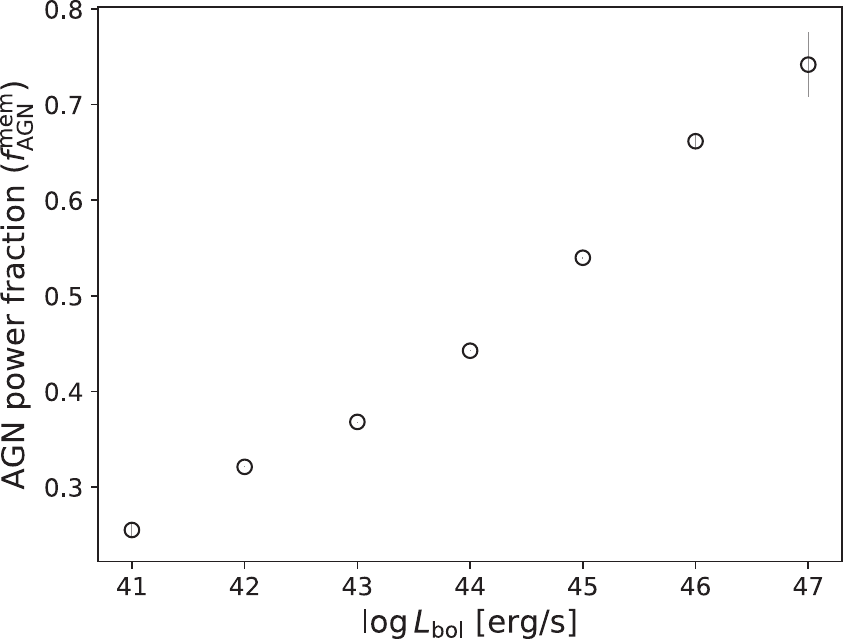}
\caption{AGN power fraction as a function of bolometric luminosity.}
\label{fAGN_Lbol}
\end{figure}
%-------------

%-------------
%   Figure
%-------------
\begin{figure*}
 \centering
 \includegraphics[width=\textwidth]{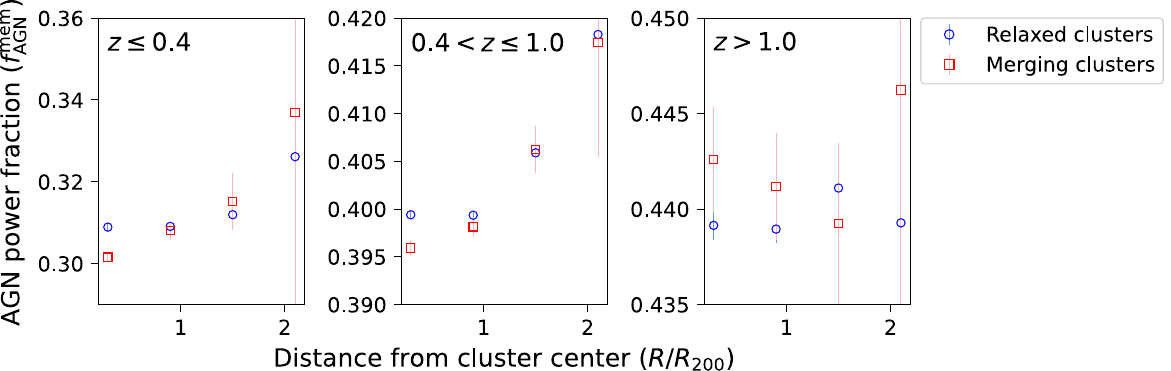}
\caption{AGN power fraction as a function of cluster centric radius scaled using the virial radius ($R/R_{\rm 200}$) for relaxed (blue circles) and merging (red squares) clusters in each redshift bin.}
\label{dist_fAGN_morph_zbin}
\end{figure*}
%-------------

Figure \ref{dist_fAGN_morph} illustrates the cluster centric radius dependence of the AGN fraction for relaxed and merging clusters. 
We found that AGNs were enhanced in relaxed clusters and that cluster--cluster mergers may not lead to increased AGN activity in the cluster center. 
The member galaxies may be moving too quickly to interact with each other, particularly in the cluster center, even if a cluster--cluster merger has occurred. 
We also found that AGNs are more likely to be enhanced at the outskirts of merging clusters rather than relaxed ones.
These findings are in good agreement with those obtained from the AGN number fraction \citep{Hashiguchi}.
The suppression of AGN activity at the cluster center of the merging clusters was also reported by a cosmological simulation \citep{Chadayammuri}.
Figure \ref{dist_fAGN_morph_zbin} illustrates the cluster centric radius dependence of the AGN fraction for relaxed and merging clusters in some redshift bins ($z \leq 0.4$, $0.4 < z \leq 1.0$, and $z > 1.0$).
The observed trend seen in Figure \ref{dist_fAGN_morph} seems to be established at $z \leq 1.0$.
Although the AGNs may be enhanced even in the cluster center in merging clusters at $z > 1.0$, we need more samples to confirm this possibility.

Because the AGN activity in massive clusters may be enhanced in the outskirts of clusters (as mentioned in Section \ref{S_fAGN_nass}), merging clusters with larger $M_{\rm 200}$ values are more likely to experience increased AGN activity in their outskirts.
Notably, the fraction of AGNs in galaxy--galaxy mergers increases with increasing luminosity \citep[e.g.,][]{Treister,Glikman,Dietrich,Weigel}.
We tested how the AGN power fraction depends on AGN bolometric luminosity ($L_{\rm bol}$).
To estimate $L_{\rm bol}$, we integrated the best-fit SED template of the AGN component output by {\tt CIGALE} over wavelengths longward of Ly$\alpha$ in the same manner as that performed by \cite{Toba17c}.
Figure \ref{fAGN_Lbol} shows the AGN power fraction as a function of AGN bolometric luminosity for CAMIRA\_AP member galaxies.
We found that $f_{\rm AGN}$ is strongly correlated with $L_{\rm bol}$ with a correlation coefficient of $r = 0.97 \pm 0.07$.
This result supports the idea that luminous AGNs can be enhanced in a dense environment, thereby merging clusters.

We note that the optical center of a cluster does not always correspond to the peak galaxy density, and it may also significantly differ from the X-ray center in certain instances, as reported by various researchers, including \cite{Mahdavi}, \cite{Oguri18}, and \cite{Ota23}.
This effect may be more severe when clusters merge. 
Although the fraction of merging clusters was small in the present study, this effect could impact the overall trend discussed in Section \ref{S_dist_fAGN}. 
It is therefore important to consider the potential uncertainty in $R$ for merging clusters.

%=================================
% AGN Feedback to SFR in Galaxy Clusters
%=================================
\subsection{AGN Feedback to SFR in Galaxy Clusters}
\label{S_AGN_feedback}

\begin{figure}
 \centering
 \includegraphics[width=0.45\textwidth]{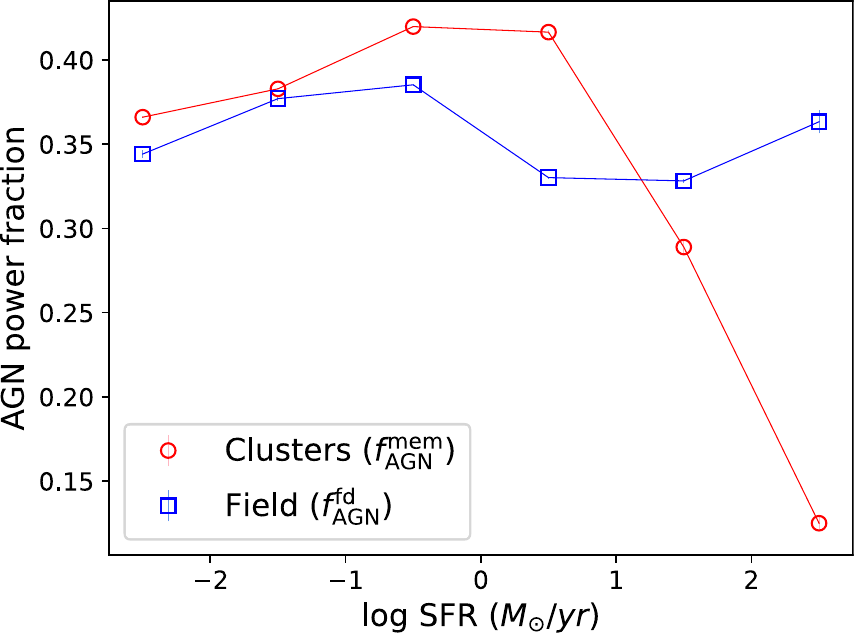}
\caption{AGN power fraction as a function of SFR for member galaxies in CAMIRA\_AP clusters (red circles) and field galaxies (blue squares).}
\label{fAGN_SFR}
\end{figure}

Finally, we discuss how the AGN power fraction depends on the SFR of CAMIRA\_AP member galaxies and field galaxies.
Figure \ref{fAGN_SFR} depicts the AGN power fraction as a function of SFR for member galaxies of clusters and field galaxies.
The AGN power fraction of field galaxies ($f^{\rm fd}_{\rm AGN}$) does not exhibit any remarkable SFR dependence.
In contrast, the AGN power fraction of CAMIRA\_AP member galaxies ($f^{\rm mem}_{\rm AGN}$) rapidly decreases when $\log\, {\rm SFR} > 0$.
In other words, when the AGN power fraction is high, the SFR tends to be small, possibly implying that the AGN feedback quenches the SF activity in member galaxies of the cluster.
This implication suggests that AGN feedback (such as gas-heating and outflow by AGN) is more effective in dense environments \citep[see also][]{Boselli,Maier,Peluso}.

%-------------
%   Figure
%-------------
%-------------

%##################
%    Summary
%##################
\section{Summary}
\label{S_summary}

We have investigated how AGN activity depends on the environment---in particular, on $z_{\rm cl}$ and the distance from the cluster center---from an AGN-energy-contribution point of view. 
Following \cite{Hashiguchi}, we utilized one of the largest optically selected galaxy cluster catalogs: the CAMIRA clusters selected with the Subaru HSC.
For approximately one million member galaxies of CAMIRA clusters in the redshift range $0.1 < z_{\rm cl} < 1.4$, we collected multiwavelength data from the UV-MIR range and have performed SED fitting to determine the AGN power fractions.
To mitigate against possible contamination from foreground and background galaxies, we introduced a membership-probability-weighted AGN power fraction and determined how this value depends on $z_{\rm cl}$ and cluster centric radius.
Our primary findings are as follows:
\begin{itemize}
\item In agreement with recent studies based on the AGN number fraction, we find that the AGN power fraction increases with increasing redshift for cluster members and field galaxies. 
In addition, the AGN power fraction for galaxy clusters increases more rapidly than for field galaxies (Section \ref{S_z_fAGN}).

\item The AGN power fraction increases toward the outskirts of galaxy clusters, which is consistent with the results reported by \cite{Hashiguchi} based on the number fraction of IR-selected AGNs. 
In contrast, the AGN power fraction decreases with increasing $M_{\rm 200}$, suggesting that AGN formation may be favored in galaxy group (Sections \ref{S_dist_fAGN} and \ref{S_fAGN_nass}).

\item Although the centers of merging clusters may be somewhat uncertain, we find that cluster--cluster mergers may not be the primary trigger for AGN activity in member galaxies.  
However, a cluster--cluster merger may enhance AGN formation at the outskirts of a cluster, especially for massive galaxy clusters (Section \ref{S_fAGN_morph}).

\item We have tentative evidence of AGN negative feedback in clusters, suggesting that AGN could suppress SF activity of member galaxies in denser environments (Section \ref{S_AGN_feedback}).
\end{itemize}

These results indicate that the emergence of an AGN population is influenced by its environment and redshift and that galaxy groups and clusters at high redshifts are perhaps crucial in AGN evolution. 
However, most CAMIRA cluster members have not yet been spectroscopically confirmed. 
Future spectroscopic studies using next-generation multiobject spectrographs---such as the Subaru Prime Focus Spectrograph \citep[PFS:][see also \citealt{Greene}]{Takada}---will help to resolve this issue and provide more-robust conclusions.

%\clearpage

%==================
% Acknowledgments
%==================
\begin{acknowledgments}
The authors gratefully acknowledge the anonymous referee for a careful reading of the manuscript and very helpful comments. 
We deeply thank Dr. I-Non Chiu for the fruitful discussion and comments.
We also thank Takuji Yamashita, Satoshi Yamada, Akatoki Noboriguchi, Mio Shibata, Nari Suzuki, Manami Furuse, and Yurika Matsuo for their support.

% HSC-SSP https://hscsurvey.pbworks.com/w/page/45370732/Survey%20Policies
The Hyper Suprime-Cam (HSC) collaboration includes the astronomical communities of Japan and Taiwan, and Princeton University.  The HSC instrumentation and software were developed by the National Astronomical Observatory of Japan (NAOJ), the Kavli Institute for the Physics and Mathematics of the Universe (Kavli IPMU), the University of Tokyo, the High Energy Accelerator Research Organization (KEK), the Academia Sinica Institute for Astronomy and Astrophysics in Taiwan (ASIAA), and Princeton University.  Funding was contributed by the FIRST program from the Japanese Cabinet Office, the Ministry of Education, Culture, Sports, Science and Technology (MEXT), the Japan Society for the Promotion of Science (JSPS), Japan Science and Technology Agency  (JST), the Toray Science  Foundation, NAOJ, Kavli IPMU, KEK, ASIAA, and Princeton University.
This paper makes use of software developed for the Large Synoptic Survey Telescope. We thank the LSST Project for making their code available as free software at  http://dm.lsst.org 
This paper is based [in part] on data collected at the Subaru Telescope and retrieved from the HSC data archive system, which is operated by Subaru Telescope and Astronomy Data Center (ADC) at NAOJ. Data analysis was in part carried out with the cooperation of Center for Computational Astrophysics (CfCA), NAOJ.

The Pan-STARRS1 Surveys (PS1) and the PS1 public science archive have been made possible through contributions by the Institute for Astronomy, the University of Hawaii, the Pan-STARRS Project Office, the Max Planck Society and its participating institutes, the Max Planck Institute for Astronomy, Heidelberg, and the Max Planck Institute for Extraterrestrial Physics, Garching, The Johns Hopkins University, Durham University, the University of Edinburgh, the Queen’s University Belfast, the Harvard-Smithsonian Center for Astrophysics, the Las Cumbres Observatory Global Telescope Network Incorporated, the National Central University of Taiwan, the Space Telescope Science Institute, the National Aeronautics and Space Administration under grant No. NNX08AR22G issued through the Planetary Science Division of the NASA Science Mission Directorate, the National Science Foundation grant No. AST-1238877, the University of Maryland, Eotvos Lorand University (ELTE), the Los Alamos National Laboratory, and the Gordon and Betty Moore Foundation.
\end{acknowledgments}

\begin{acknowledgments}
% KiDS DR3 https://cdsarc.u-strasbg.fr/viz-bin/getCatFile_Redirect/?-plus=-%2b&II/347/kids_dr3.intro.htx
Based on data products from observations made with ESO Telescopes at the La Silla Paranal Observatory under programme IDs 177.A-3016, 177.A-3017 and 177.A-3018, and on data products produced by Target/OmegaCEN, INAF-OACN, INAF-OAPD and the KiDS production team, on behalf of the KiDS consortium. 
OmegaCEN and the KiDS production team acknowledge support by NOVA and NWO-M grants. Members of INAF-OAPD and INAF-OACN also acknowledge the support from the Department of Physics \& Astronomy of the University of Padova, and of the Department of Physics of Univ. Federico II (Naples).

% SDSS IV
Funding for the Sloan Digital Sky Survey IV has been provided by the Alfred P. Sloan Foundation, the U.S. Department of Energy Office of Science, and the Participating Institutions. SDSS-IV acknowledges support and resources from the Center for High-Performance Computing at the University of Utah. The SDSS web site is www.sdss.org.

SDSS-IV is managed by the Astrophysical Research Consortium for the Participating Institutions of the SDSS Collaboration including the Brazilian Participation Group, the Carnegie Institution for Science, Carnegie Mellon University, the Chilean Participation Group, the French Participation Group, Harvard-Smithsonian Center for Astrophysics, Instituto de Astrof\'isica de Canarias, The Johns Hopkins University, Kavli Institute for the Physics and Mathematics of the Universe (IPMU) / University of Tokyo, the Korean Participation Group, Lawrence Berkeley National Laboratory, Leibniz Institut f\"ur Astrophysik Potsdam (AIP), Max-Planck-Institut f\"ur Astronomie (MPIA Heidelberg), Max-Planck-Institut f\"ur Astrophysik (MPA Garching), Max-Planck-Institut f\"ur Extraterrestrische Physik (MPE), National Astronomical Observatories of China, New Mexico State University, New York University, University of Notre Dame, Observat\'ario Nacional / MCTI, The Ohio State University, Pennsylvania State University, Shanghai Astronomical Observatory, United Kingdom Participation Group, Universidad Nacional Aut\'onoma de M\'exico, University of Arizona, University of Colorado Boulder, University of Oxford, University of Portsmouth, University of Utah, University of Virginia, University of Washington, University of Wisconsin, Vanderbilt University, and Yale University.

% VIKING https://www.eso.org/rm/api/v1/public/releaseDescriptions/135
This publication has made use of data from the VIKING survey from VISTA at the ESO Paranal Observatory, programme ID 179.A-2004. Data processing has been contributed by the VISTA Data Flow System at CASU, Cambridge and WFAU, Edinburgh.

% UKIDSS https://www.ukirt.hawaii.edu/surveys/UKIDSSdatapolicies.html
This work is based in part on data obtained as part of the UKIRT Infrared Deep Sky Survey.
\end{acknowledgments}

\begin{acknowledgments}
% 2MASS https://old.ipac.caltech.edu/2mass/releases/allsky/faq.html
This publication makes use of data products from the Two Micron All Sky Survey, which is a joint project of the University of Massachusetts and the Infrared Processing and Analysis Center/California Institute of Technology, funded by the National Aeronautics and Space Administration and the National Science Foundation.

% WISE
This publication makes use of data products from the Wide-field Infrared Survey Explorer, which is a joint project of the University of California, Los Angeles, and the Jet Propulsion Laboratory/California Institute of Technology, funded by the National Aeronautics and Space Administration.

% Ichiro
SED fitting was carried out using the SuMIRe cluster operated by the Extragalactic OIR group at ASIAA.
\end{acknowledgments}

\begin{acknowledgments}
% KAKENHI
This work is supported by JSPS KAKENHI Grant numbers JP18J01050, JP19K14759, and JP22H01266 (YT), JP20K04027 (NO), JP20H01946 (YU), JP21K03632 (MI) and JP20K22360, JP21H05449, and JP23K03460 (TO).
\end{acknowledgments}

\vspace{5mm}
\facilities{Subaru, GALEX, Sloan, CTIO:2MASS, FLWO:2MASS, UKIRT, ESO:VISTA, WISE}

\software{Astropy \citep{astropy:2013, astropy:2018, astropy:2022}, NumPy \citep{numpy}, SciPy \citep{scipy}, Matplotlib \citep{matplotlib}, Pandas \citep{pandas}, Statsmodels \citep{seabold2010statsmodels}, {\sf linmix} \citep{Kelly}, IDL, IDL Astronomy User's Library \citep{idl}, {\sf CIGALE} \citep{Boquien,Yang}, {\tt STILTS} \citep{stilts}.}

% https://github.com/leonoverweel/bibtex-python-package-citations
% https://www.astropy.org/acknowledging.html

%%%%%%%%%%%%%%%%%%%%%%%%%%
%       Appendix
%%%%%%%%%%%%%%%%%%%%%%%%%%
%===============================
% CAMIRA member galaxy catalog
%===============================
\appendix
\section{Value-added CAMIRA member galaxy catalog}
\label{app1}

We provide the physical properties of 877,642 CAMIRA member galaxies at $0 < z_{\rm cl} < 1.4$.
The catalog description is summarized in Table \ref{catalog}.

\startlongtable
\begin{deluxetable}{lccl}
\tablecaption{Physical properties of 877,642 CAMIRA member galaxies at $0 < z_{\rm cl} < 1.4$.\label{catalog}}
\tablehead{
\colhead{Column name} & \colhead{Format} & \colhead{Unit} & \colhead{Description}
}
%\colnumbers
\startdata
ID\_cl					& 	LONG64	&			&	Unique id \\
Name\_cl 				& 	STRING 	& 			& 	Object name \\
z\_cl					&	FLOAT  	&			&	Cluster redshift \citep{Oguri18} \\
N\_mem					&	FLOAT   &			&	Richness \citep{Oguri18} \\
log\_M200				&	FLOAT	& $M_{\sun}$&	Cluster mass \citep{Okabe} \\ 
Flag\_cl\_multi\_peaks	&	BOOLEAN &			&	True (multiple peaks), False (single peak) \citep{Okabe} \\ 
						&			&			&	In this work, a galaxy cluster with a single peak is classified as a relaxed cluster, \\ 		
						&			&			&	while a cluster with multiple peaks is considered a merging cluster (Section \ref{S_fAGN_morph}). \\
%-----------------
ID				& 	LONG64	&			&	Unique id for member galaxies \\
Name			& 	STRING 	& 			& 	Object name \\
R.A.			&	DOUBLE	&	degree	& 	Right Assignation (J2000.0) from the {\tt HSC s21a\_wide} \\
Decl.			&	DOUBLE	&	degree	& 	Declination (J2000.0) from the {\tt HSC s21a\_wide} \\
z\_mem			&	FLOAT 	& 			& 	Redshift (Section \ref{s_CAMIRA}) \\
R\_R200			&	FLOAT   &			&	Projected distance from the cluster center scaled using the virial radius \\
%-----------------
$u$mag			&	FLOAT & AB mag. & $u$-band magnitude \\
$u$mag\_err		&	FLOAT & AB mag. & $u$-band magnitude error \\
$g$mag			&	FLOAT & AB mag. & $g$-band magnitude \\
$g$mag\_err		&	FLOAT & AB mag. & $g$-band magnitude error \\
$r$mag			&	FLOAT & AB mag. & $r$-band magnitude \\
$r$mag\_err		&	FLOAT & AB mag. & $r$-band magnitude error\\
$i$mag			&	FLOAT & AB mag. & $i$-band magnitude \\
$i$mag\_err		&	FLOAT & AB mag. & $i$-band magnitude error\\
$z$mag			&	FLOAT & AB mag. & $z$-band magnitude \\
$z$mag\_err		&	FLOAT & AB mag. & $z$-band magnitude error\\
$y$mag			&	FLOAT & AB mag. & $y$-band magnitude \\
$y$mag\_err		&	FLOAT & AB mag. & $y$-band magnitude error \\
$j$mag			&	FLOAT & AB mag. & $J$-band magnitude \\
$j$mag\_err		&	FLOAT & AB mag. & $J$-band magnitude error \\
$h$mag			&	FLOAT & AB mag. & $H$-band magnitude \\
$h$mag\_err		&	FLOAT & AB mag. & $H$-band magnitude error \\
$k$mag			&	FLOAT & AB mag. & $K$-band magnitude \\
$k$mag\_err		&	FLOAT & AB mag. & $K$-band magnitude error \\
Flux\_34		&	FLOAT & mJy	& Flux density at 3.4 $\micron$ \\	
Flux\_34\_err	&	FLOAT & mJy	& Uncertainty of flux density at 3.4 $\micron$ \\
Flux\_46		&	FLOAT & mJy	& Flux density at 4.6 $\micron$ \\	
Flux\_46\_err	&	FLOAT & mJy	& Uncertainty of flux density at 4.6 $\micron$ \\
Flux\_12		&	FLOAT & mJy	& Flux density at 12 $\micron$ \\	
Flux\_12\_err	&	FLOAT & mJy	& Uncertainty of flux density at 12 $\micron$ \\
Flux\_22		&	FLOAT & mJy	& Flux density at 22 $\micron$ flux density \\	
Flux\_22\_err	&	FLOAT & mJy	& Uncertainty of flux density at 22 $\micron$ \\
Flag\_upper\_limit\_u	&	BOOLEAN	 & 		&	Upper limit flag for $u$mag. True (5$\sigma$ upper limit), False (otherwise) \\
Flag\_upper\_limit\_g	&	BOOLEAN	 & 		&	Upper limit flag for $g$mag. True (5$\sigma$ upper limit), False (otherwise) \\
Flag\_upper\_limit\_r	&	BOOLEAN	 & 		&	Upper limit flag for $r$mag. True (5$\sigma$ upper limit), False (otherwise) \\
Flag\_upper\_limit\_i	&	BOOLEAN	 & 		&	Upper limit flag for $i$mag. True (5$\sigma$ upper limit), False (otherwise) \\
Flag\_upper\_limit\_z	&	BOOLEAN	 & 		&	Upper limit flag for $z$mag. True (5$\sigma$ upper limit), False (otherwise) \\
Flag\_upper\_limit\_y	&	BOOLEAN	 & 		&	Upper limit flag for $y$mag. True (5$\sigma$ upper limit), False (otherwise) \\
Flag\_upper\_limit\_J	&	BOOLEAN	 & 		&	Upper limit flag for $J$mag. True (5$\sigma$ upper limit), False (otherwise) \\
Flag\_upper\_limit\_H	&	BOOLEAN	 & 		&	Upper limit flag for $H$mag. True (5$\sigma$ upper limit), False (otherwise) \\
Flag\_upper\_limit\_K	&	BOOLEAN	 & 		&	Upper limit flag for $K$mag. True (5$\sigma$ upper limit), False (otherwise) \\
Flag\_upper\_limit\_34	&	BOOLEAN	 & 		&	Upper limit flag for Flux\_34. True (5$\sigma$ upper limit), False (otherwise) \\
Flag\_upper\_limit\_46	&	BOOLEAN	 & 		&	Upper limit flag for Flux\_46. True (5$\sigma$ upper limit), False (otherwise) \\
Flag\_upper\_limit\_12	&	BOOLEAN	 & 		&	Upper limit flag for Flux\_12. True (5$\sigma$ upper limit), False (otherwise) \\
Flag\_upper\_limit\_22	&	BOOLEAN	 & 		&	Upper limit flag for Flux\_22. True (5$\sigma$ upper limit), False (otherwise) \\
%-------------------
rchi2				& FLOAT	&			&	Reduced $\chi^2$ obtained from {\tt CIGALE} \\
Flag\_CAMIRA\_AP	& BOOLEAN &			&	True (CAMIRA\_AP sample), False (otherwise) (Section \ref{SED}) \\	
E\_BV				& FLOAT	&			&	Color excess ($E(B-V)$) derived from {\tt CIGALE} \\
E\_BV\_err			& FLOAT	&			&	Uncertainty of color excess ($E(B-V)$) derived from {\tt CIGALE} \\
log\_M				& FLOAT & $M_{\sun}$ & Stellar mass derived from {\tt CIGALE} \\
log\_M\_err 		& FLOAT & $M_{\sun}$ & Uncertainty of stellar mass derived from {\tt CIGALE} \\
log\_SFR			& FLOAT & $M_{\sun}$ yr$^{-1}$ & SFR derived from {\tt CIGALE} \\
log\_SFR\_err 		& FLOAT & $M_{\sun}$ yr$^{-1}$ & Uncertainty of SFR derived from {\tt CIGALE} \\
log\_LIR			& FLOAT & $L_{\sun}$ & IR luminosity derived from {\tt CIGALE} \\
log\_LIR\_err		& FLOAT & $L_{\sun}$ & Uncertainty of IR luminosity derived from {\tt CIGALE} \\
log\_LIR\_AGN		& FLOAT & $L_{\sun}$ & IR luminosity contributed from AGNs derived from {\tt CIGALE} \\
log\_LIR\_AGN\_err	& FLOAT & $L_{\sun}$ & Uncertainty of log\_LIR\_AGN derived from {\tt CIGALE} \\
f\_AGN				& FLOAT &			 & AGN power fraction (i.e., log\_LIR\_AGN/log\_LIR) \\
f\_AGN\_err			& FLOAT &			 & Uncertainty of AGN power fraction \\
Flag\_AGN			& STRING &			 & IR (IR-AGNs), Radio (Radio-AGNs), X (X-ray AGNs) \citep{Hashiguchi} \\
\enddata
\tablecomments{The entire table is available in a machine-readable form in the online journal.}
\end{deluxetable}

%=============================================
% Best-fit SED for each CAMIRA member galaxy
%=============================================
\section{Best-fit SED for each CAMIRA member galaxy}
\label{app2}

The best-fit SED derived from {\tt CIGALE} is available in Table \ref{template}.
We encourage using a template of objects with reduced $\chi^{2} <$ 2.0 for science (Section \ref{SED}).

\startlongtable
\begin{deluxetable*}{lccl}
\tablecaption{Best-fit SED template of each CAMIRA member galaxy.\label{template}}
\tablehead{
\colhead{Column name} & \colhead{Format} & \colhead{Unit} & \colhead{Description}
}
%\colnumbers
\startdata
ID			&	LONG	&			& Unique id for member galaxies \\
Wavelength 	& 	DOUBLE 	& $\micron$ & Wavelength (observed frame) \\
FNU			&	DOUBLE	& mJy		& Flux density at each wavelength\\
LNU			&	DOUBLE	& W			& Luminosity density at each wavelength\\
\enddata
\tablecomments{The entire table is available in a machine-readable form in the online journal.}
\end{deluxetable*}

%=============================================
% Other AGN properties of CAMIRA member galaxies
%=============================================
\section{Other AGN properties of CAMIRA member galaxies}
\label{app3}

\begin{figure*}[h]
\centering
 \includegraphics[width=\textwidth]{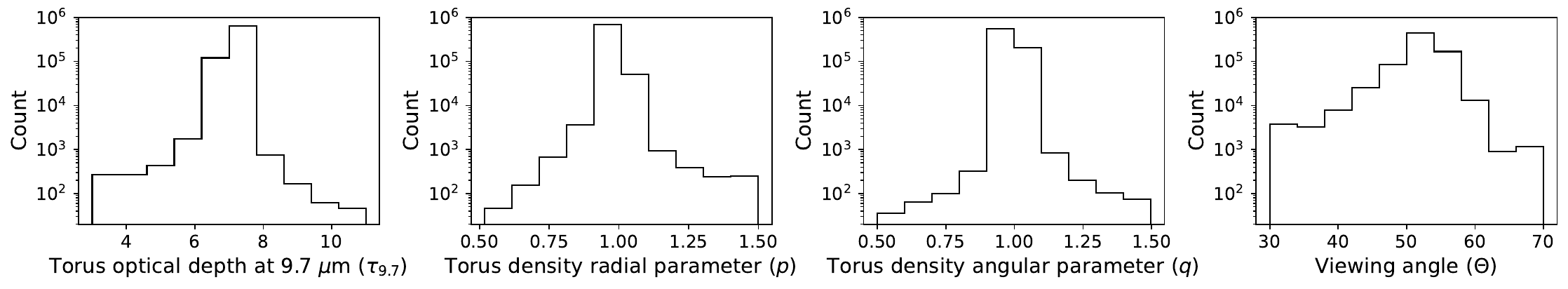}
\caption{Distributions of the AGN-related physical properties (except for the AGN power fraction) obtained through SED fitting using {\tt CIGALE}.}
\label{AGN_parm}
\end{figure*}

This work highlights AGN power fraction as a representative AGN property of the CAMIRA\_AP member galaxies.
We also derive other AGN properties primarily related to the torus AGN structure through SED fitting using {\tt CIGALE}, as described in Section \ref{SED}.
Herein, we present four quantities of the CAMIRA\_AP member galaxies: optical depth of the torus at 9.7 $\micron$ ($\tau_{\rm 9.7}$), torus-density radial parameter ($p$), torus-density angular parameter ($q$), and viewing angle ($\theta$).
Figure \ref{AGN_parm} shows the histograms of the four quantities, which are broadly distributed.
The weighted mean values of $\tau_{\rm 9.7}$, $p$, $q$, and $\theta$ are 7.10, 1.00, 0.99, and 52.2, respectively.
Notably, because the input parameter values of the quantities are sparser than those of the AGN power fraction (Table \ref{Param}), herein, the relative errors are considerably large (30\%--50\%).

%-------------
%   Figure
%-------------
%-------------

%%%%%%%%%%%%%%%%%%%%%%%%%%
%      References
%%%%%%%%%%%%%%%%%%%%%%%%%%
\bibliography{ref}{}

\begin{thebibliography}{}
\expandafter\ifx\csname natexlab\endcsname\relax\def\natexlab#1{#1}\fi
\providecommand{\url}[1]{\href{#1}{#1}}
\providecommand{\dodoi}[1]{doi:~\href{http://doi.org/#1}{\nolinkurl{#1}}}
\providecommand{\doeprint}[1]{\href{http://ascl.net/#1}{\nolinkurl{http://ascl.net/#1}}}
\providecommand{\doarXiv}[1]{\href{https://arxiv.org/abs/#1}{\nolinkurl{https://arxiv.org/abs/#1}}}

\bibitem[{{Abdurro'uf} {et~al.}(2022){Abdurro'uf}, {Accetta}, {Aerts}, {Silva
  Aguirre}, {Ahumada}, {Ajgaonkar}, {Filiz Ak}, {Alam}, {Allende Prieto},
  {Almeida}, {Anders}, {Anderson}, {Andrews}, {Anguiano}, {Aquino-Ort{\'\i}z},
  {Arag{\'o}n-Salamanca}, {Argudo-Fern{\'a}ndez}, {Ata}, {Aubert},
  {Avila-Reese}, {Badenes}, {Barb{\'a}}, {Barger}, {Barrera-Ballesteros},
  {Beaton}, {Beers}, {Belfiore}, {Bender}, {Bernardi}, {Bershady}, {Beutler},
  {Bidin}, {Bird}, {Bizyaev}, {Blanc}, {Blanton}, {Boardman}, {Bolton},
  {Boquien}, {Borissova}, {Bovy}, {Brandt}, {Brown}, {Brownstein}, {Brusa},
  {Buchner}, {Bundy}, {Burchett}, {Bureau}, {Burgasser}, {Cabang}, {Campbell},
  {Cappellari}, {Carlberg}, {Wanderley}, {Carrera}, {Cash}, {Chen}, {Chen},
  {Cherinka}, {Chiappini}, {Choi}, {Chojnowski}, {Chung}, {Clerc}, {Cohen},
  {Comerford}, {Comparat}, {da Costa}, {Covey}, {Crane}, {Cruz-Gonzalez},
  {Culhane}, {Cunha}, {Dai}, {Damke}, {Darling}, {Davidson}, {Davies},
  {Dawson}, {De Lee}, {Diamond-Stanic}, {Cano-D{\'\i}az}, {S{\'a}nchez},
  {Donor}, {Duckworth}, {Dwelly}, {Eisenstein}, {Elsworth}, {Emsellem},
  {Eracleous}, {Escoffier}, {Fan}, {Farr}, {Feng}, {Fern{\'a}ndez-Trincado},
  {Feuillet}, {Filipp}, {Fillingham}, {Frinchaboy}, {Fromenteau}, {Galbany},
  {Garc{\'\i}a}, {Garc{\'\i}a-Hern{\'a}ndez}, {Ge}, {Geisler}, {Gelfand},
  {G{\'e}ron}, {Gibson}, {Goddy}, {Godoy-Rivera}, {Grabowski}, {Green},
  {Greener}, {Grier}, {Griffith}, {Guo}, {Guy}, {Hadjara}, {Harding},
  {Hasselquist}, {Hayes}, {Hearty}, {Hern{\'a}ndez}, {Hill}, {Hogg},
  {Holtzman}, {Horta}, {Hsieh}, {Hsu}, {Hsu}, {Huber}, {Huertas-Company},
  {Hutchinson}, {Hwang}, {Ibarra-Medel}, {Chitham}, {Ilha}, {Imig}, {Jaekle},
  {Jayasinghe}, {Ji}, {Johnson}, {Jones}, {J{\"o}nsson}, {Katkov}, {Khalatyan},
  {Kinemuchi}, {Kisku}, {Knapen}, {Kneib}, {Kollmeier}, {Kong}, {Kounkel},
  {Kreckel}, {Krishnarao}, {Lacerna}, {Lane}, {Langgin}, {Lavender}, {Law},
  {Lazarz}, {Leung}, {Leung}, {Lewis}, {Li}, {Li}, {Lian}, {Liang}, {Lin},
  {Lin}, {Lin}, {Lintott}, {Long}, {Longa-Pe{\~n}a}, {L{\'o}pez-Cob{\'a}},
  {Lu}, {Lundgren}, {Luo}, {Mackereth}, {de la Macorra}, {Mahadevan},
  {Majewski}, {Manchado}, {Mandeville}, {Maraston}, {Margalef-Bentabol},
  {Masseron}, {Masters}, {Mathur}, {McDermid}, {Mckay}, {Merloni},
  {Merrifield}, {Meszaros}, {Miglio}, {Di Mille}, {Minniti}, {Minsley},
  {Monachesi}, {Moon}, {Mosser}, {Mulchaey}, {Muna}, {Mu{\~n}oz}, {Myers},
  {Myers}, {Nadathur}, {Nair}, {Nandra}, {Neumann}, {Newman}, {Nidever},
  {Nikakhtar}, {Nitschelm}, {O'Connell}, {Garma-Oehmichen}, {Luan Souza de
  Oliveira}, {Olney}, {Oravetz}, {Ortigoza-Urdaneta}, {Osorio}, {Otter},
  {Pace}, {Padilla}, {Pan}, {Pan}, {Parikh}, {Parker}, {Peirani}, {Pe{\~n}a
  Ram{\'\i}rez}, {Penny}, {Percival}, {Perez-Fournon}, {Pinsonneault},
  {Poidevin}, {Poovelil}, {Price-Whelan}, {B{\'a}rbara de Andrade Queiroz},
  {Raddick}, {Ray}, {Rembold}, {Riddle}, {Riffel}, {Riffel}, {Rix}, {Robin},
  {Rodr{\'\i}guez-Puebla}, {Roman-Lopes}, {Rom{\'a}n-Z{\'u}{\~n}iga}, {Rose},
  {Ross}, {Rossi}, {Rubin}, {Salvato}, {S{\'a}nchez}, {S{\'a}nchez-Gallego},
  {Sanderson}, {Santana Rojas}, {Sarceno}, {Sarmiento}, {Sayres}, {Sazonova},
  {Schaefer}, {Schiavon}, {Schlegel}, {Schneider}, {Schultheis}, {Schwope},
  {Serenelli}, {Serna}, {Shao}, {Shapiro}, {Sharma}, {Shen}, {Shetrone}, {Shu},
  {Simon}, {Skrutskie}, {Smethurst}, {Smith}, {Sobeck}, {Spoo}, {Sprague},
  {Stark}, {Stassun}, {Steinmetz}, {Stello}, {Stone-Martinez},
  {Storchi-Bergmann}, {Stringfellow}, {Stutz}, {Su}, {Taghizadeh-Popp},
  {Talbot}, {Tayar}, {Telles}, {Teske}, {Thakar}, {Theissen}, {Tkachenko},
  {Thomas}, {Tojeiro}, {Hernandez Toledo}, {Troup}, {Trump}, {Trussler},
  {Turner}, {Tuttle}, {Unda-Sanzana}, {V{\'a}zquez-Mata}, {Valentini},
  {Valenzuela}, {Vargas-Gonz{\'a}lez}, {Vargas-Maga{\~n}a}, {Alfaro},
  {Villanova}, {Vincenzo}, {Wake}, {Warfield}, {Washington}, {Weaver},
  {Weijmans}, {Weinberg}, {Weiss}, {Westfall}, {Wild}, {Wilde}, {Wilson},
  {Wilson}, {Wilson}, {Wolf}, {Wood-Vasey}, {Yan}, {Zamora}, {Zasowski},
  {Zhang}, {Zhao}, {Zheng}, {Zheng}, \& {Zhu}}]{Abdurrouf}
{Abdurro'uf}, {Accetta}, K., {Aerts}, C., {et~al.} 2022, \apjs, 259, 35,
  \dodoi{10.3847/1538-4365/ac4414}

\bibitem[{{Aguado} {et~al.}(2019){Aguado}, {Ahumada}, {Almeida}, {Anderson},
  {Andrews}, {Anguiano}, {Aquino Ort{\'\i}z}, {Arag{\'o}n-Salamanca},
  {Argudo-Fern{\'a}ndez}, {Aubert}, {Avila-Reese}, {Badenes}, {Barboza
  Rembold}, {Barger}, {Barrera-Ballesteros}, {Bates}, {Bautista}, {Beaton},
  {Beers}, {Belfiore}, {Bernardi}, {Bershady}, {Beutler}, {Bird}, {Bizyaev},
  {Blanc}, {Blanton}, {Blomqvist}, {Bolton}, {Boquien}, {Borissova}, {Bovy},
  {Brandt}, {Brinkmann}, {Brownstein}, {Bundy}, {Burgasser}, {Byler}, {Cano
  Diaz}, {Cappellari}, {Carrera}, {Cervantes Sodi}, {Chen}, {Cherinka}, {Choi},
  {Chung}, {Coffey}, {Comerford}, {Comparat}, {Covey}, {da Silva Ilha}, {da
  Costa}, {Dai}, {Damke}, {Darling}, {Davies}, {Dawson}, {de Sainte Agathe},
  {Deconto Machado}, {Del Moro}, {De Lee}, {Diamond-Stanic}, {Dom{\'\i}nguez
  S{\'a}nchez}, {Donor}, {Drory}, {du Mas des Bourboux}, {Duckworth}, {Dwelly},
  {Ebelke}, {Emsellem}, {Escoffier}, {Fern{\'a}ndez-Trincado}, {Feuillet},
  {Fischer}, {Fleming}, {Fraser-McKelvie}, {Freischlad}, {Frinchaboy}, {Fu},
  {Galbany}, {Garcia-Dias}, {Garc{\'\i}a-Hern{\'a}ndez}, {Garma Oehmichen},
  {Geimba Maia}, {Gil-Mar{\'\i}n}, {Grabowski}, {Gu}, {Guo}, {Ha},
  {Harrington}, {Hasselquist}, {Hayes}, {Hearty}, {Hernandez Toledo}, {Hicks},
  {Hogg}, {Holley-Bockelmann}, {Holtzman}, {Hsieh}, {Hunt}, {Hwang},
  {Ibarra-Medel}, {Jimenez Angel}, {Johnson}, {Jones}, {J{\"o}nsson},
  {Kinemuchi}, {Kollmeier}, {Krawczyk}, {Kreckel}, {Kruk}, {Lacerna}, {Lan},
  {Lane}, {Law}, {Lee}, {Li}, {Lian}, {Lin}, {Lin}, {Lintott}, {Long},
  {Longa-Pe{\~n}a}, {Mackereth}, {de la Macorra}, {Majewski}, {Malanushenko},
  {Manchado}, {Maraston}, {Mariappan}, {Marinelli}, {Marques-Chaves},
  {Masseron}, {Masters}, {McDermid}, {Medina Pe{\~n}a}, {Meneses-Goytia},
  {Merloni}, {Merrifield}, {Meszaros}, {Minniti}, {Minsley}, {Muna}, {Myers},
  {Nair}, {Correa do Nascimento}, {Newman}, {Nitschelm}, {Olmstead}, {Oravetz},
  {Oravetz}, {Ortega Minakata}, {Pace}, {Padilla}, {Palicio}, {Pan}, {Pan},
  {Parikh}, {Parker}, {Peirani}, {Penny}, {Percival}, {Perez-Fournon},
  {Peterken}, {Pinsonneault}, {Prakash}, {Raddick}, {Raichoor}, {Riffel},
  {Riffel}, {Rix}, {Robin}, {Roman-Lopes}, {Rose}, {Ross}, {Rossi}, {Rowlands},
  {Rubin}, {S{\'a}nchez}, {S{\'a}nchez-Gallego}, {Sayres}, {Schaefer},
  {Schiavon}, {Schimoia}, {Schlafly}, {Schlegel}, {Schneider}, {Schultheis},
  {Seo}, {Shamsi}, {Shao}, {Shen}, {Shetty}, {Simonian}, {Smethurst}, {Sobeck},
  {Souter}, {Spindler}, {Stark}, {Stassun}, {Steinmetz}, {Storchi-Bergmann},
  {Stringfellow}, {Su{\'a}rez}, {Sun}, {Taghizadeh-Popp}, {Talbot}, {Tayar},
  {Thakar}, {Thomas}, {Tissera}, {Tojeiro}, {Troup}, {Unda-Sanzana},
  {Valenzuela}, {Vargas-Maga{\~n}a}, {V{\'a}zquez-Mata}, {Wake}, {Weaver},
  {Weijmans}, {Westfall}, {Wild}, {Wilson}, {Woods}, {Yan}, {Yang}, {Zamora},
  {Zasowski}, {Zhang}, {Zheng}, {Zheng}, {Zhu}, {Zinn}, \& {Zou}}]{Aguado}
{Aguado}, D.~S., {Ahumada}, R., {Almeida}, A., {et~al.} 2019, \apjs, 240, 23,
  \dodoi{10.3847/1538-4365/aaf651}

\bibitem[{{Aihara} {et~al.}(2018{\natexlab{a}}){Aihara}, {Armstrong},
  {Bickerton}, {Bosch}, {Coupon}, {Furusawa}, {Hayashi}, {Ikeda}, {Kamata},
  {Karoji}, {Kawanomoto}, {Koike}, {Komiyama}, {Lang}, {Lupton}, {Mineo},
  {Miyatake}, {Miyazaki}, {Morokuma}, {Obuchi}, {Oishi}, {Okura}, {Price},
  {Takata}, {Tanaka}, {Tanaka}, {Tanaka}, {Uchida}, {Uraguchi}, {Utsumi},
  {Wang}, {Yamada}, {Yamanoi}, {Yasuda}, {Arimoto}, {Chiba}, {Finet},
  {Fujimori}, {Fujimoto}, {Furusawa}, {Goto}, {Goulding}, {Gunn}, {Harikane},
  {Hattori}, {Hayashi}, {He{\l}miniak}, {Higuchi}, {Hikage}, {Ho}, {Hsieh},
  {Huang}, {Huang}, {Imanishi}, {Iwata}, {Jaelani}, {Jian}, {Kashikawa},
  {Katayama}, {Kojima}, {Konno}, {Koshida}, {Kusakabe}, {Leauthaud}, {Lee},
  {Lin}, {Lin}, {Mandelbaum}, {Matsuoka}, {Medezinski}, {Miyama}, {Momose},
  {More}, {More}, {Mukae}, {Murata}, {Murayama}, {Nagao}, {Nakata}, {Niida},
  {Niikura}, {Nishizawa}, {Oguri}, {Okabe}, {Ono}, {Onodera}, {Onoue}, {Ouchi},
  {Pyo}, {Shibuya}, {Shimasaku}, {Simet}, {Speagle}, {Spergel}, {Strauss},
  {Sugahara}, {Sugiyama}, {Suto}, {Suzuki}, {Tait}, {Takada}, {Terai}, {Toba},
  {Turner}, {Uchiyama}, {Umetsu}, {Urata}, {Usuda}, {Yeh}, \&
  {Yuma}}]{Aihara18b}
{Aihara}, H., {Armstrong}, R., {Bickerton}, S., {et~al.} 2018{\natexlab{a}},
  \pasj, 70, S8, \dodoi{10.1093/pasj/psx081}

\bibitem[{{Aihara} {et~al.}(2018{\natexlab{b}}){Aihara}, {Arimoto},
  {Armstrong}, {Arnouts}, {Bahcall}, {Bickerton}, {Bosch}, {Bundy}, {Capak},
  {Chan}, {Chiba}, {Coupon}, {Egami}, {Enoki}, {Finet}, {Fujimori}, {Fujimoto},
  {Furusawa}, {Furusawa}, {Goto}, {Goulding}, {Greco}, {Greene}, {Gunn},
  {Hamana}, {Harikane}, {Hashimoto}, {Hattori}, {Hayashi}, {Hayashi},
  {He{\l}miniak}, {Higuchi}, {Hikage}, {Ho}, {Hsieh}, {Huang}, {Huang},
  {Ikeda}, {Imanishi}, {Inoue}, {Iwasawa}, {Iwata}, {Jaelani}, {Jian},
  {Kamata}, {Karoji}, {Kashikawa}, {Katayama}, {Kawanomoto}, {Kayo}, {Koda},
  {Koike}, {Kojima}, {Komiyama}, {Konno}, {Koshida}, {Koyama}, {Kusakabe},
  {Leauthaud}, {Lee}, {Lin}, {Lin}, {Lupton}, {Mandelbaum}, {Matsuoka},
  {Medezinski}, {Mineo}, {Miyama}, {Miyatake}, {Miyazaki}, {Momose}, {More},
  {More}, {Moritani}, {Moriya}, {Morokuma}, {Mukae}, {Murata}, {Murayama},
  {Nagao}, {Nakata}, {Niida}, {Niikura}, {Nishizawa}, {Obuchi}, {Oguri},
  {Oishi}, {Okabe}, {Okamoto}, {Okura}, {Ono}, {Onodera}, {Onoue}, {Osato},
  {Ouchi}, {Price}, {Pyo}, {Sako}, {Sawicki}, {Shibuya}, {Shimasaku},
  {Shimono}, {Shirasaki}, {Silverman}, {Simet}, {Speagle}, {Spergel},
  {Strauss}, {Sugahara}, {Sugiyama}, {Suto}, {Suyu}, {Suzuki}, {Tait},
  {Takada}, {Takata}, {Tamura}, {Tanaka}, {Tanaka}, {Tanaka}, {Tanaka},
  {Terai}, {Terashima}, {Toba}, {Tominaga}, {Toshikawa}, {Turner}, {Uchida},
  {Uchiyama}, {Umetsu}, {Uraguchi}, {Urata}, {Usuda}, {Utsumi}, {Wang}, {Wang},
  {Wong}, {Yabe}, {Yamada}, {Yamanoi}, {Yasuda}, {Yeh}, {Yonehara}, \&
  {Yuma}}]{Aihara18a}
{Aihara}, H., {Arimoto}, N., {Armstrong}, R., {et~al.} 2018{\natexlab{b}},
  \pasj, 70, S4, \dodoi{10.1093/pasj/psx066}

\bibitem[{{Aihara} {et~al.}(2019){Aihara}, {AlSayyad}, {Ando}, {Armstrong},
  {Bosch}, {Egami}, {Furusawa}, {Furusawa}, {Goulding}, {Harikane}, {Hikage},
  {Ho}, {Hsieh}, {Huang}, {Ikeda}, {Imanishi}, {Ito}, {Iwata}, {Jaelani},
  {Kakuma}, {Kawana}, {Kikuta}, {Kobayashi}, {Koike}, {Komiyama}, {Li},
  {Liang}, {Lin}, {Luo}, {Lupton}, {Lust}, {MacArthur}, {Matsuoka}, {Mineo},
  {Miyatake}, {Miyazaki}, {More}, {Murata}, {Namiki}, {Nishizawa}, {Oguri},
  {Okabe}, {Okamoto}, {Okura}, {Ono}, {Onodera}, {Onoue}, {Osato}, {Ouchi},
  {Shibuya}, {Strauss}, {Sugiyama}, {Suto}, {Takada}, {Takagi}, {Takata},
  {Takita}, {Tanaka}, {Terai}, {Toba}, {Uchiyama}, {Utsumi}, {Wang}, {Wang}, \&
  {Yamada}}]{Aihara19}
{Aihara}, H., {AlSayyad}, Y., {Ando}, M., {et~al.} 2019, \pasj, 71, 114,
  \dodoi{10.1093/pasj/psz103}

\bibitem[{{Aihara} {et~al.}(2022){Aihara}, {AlSayyad}, {Ando}, {Armstrong},
  {Bosch}, {Egami}, {Furusawa}, {Furusawa}, {Harasawa}, {Harikane}, {Hsieh},
  {Ikeda}, {Ito}, {Iwata}, {Kodama}, {Koike}, {Kokubo}, {Komiyama}, {Li},
  {Liang}, {Lin}, {Lupton}, {Lust}, {MacArthur}, {Mawatari}, {Mineo},
  {Miyatake}, {Miyazaki}, {More}, {Morishima}, {Murayama}, {Nakajima},
  {Nakata}, {Nishizawa}, {Oguri}, {Okabe}, {Okura}, {Ono}, {Osato}, {Ouchi},
  {Pan}, {Plazas Malag{\'o}n}, {Price}, {Reed}, {Rykoff}, {Shibuya},
  {Simunovic}, {Strauss}, {Sugimori}, {Suto}, {Suzuki}, {Takada}, {Takagi},
  {Takata}, {Takita}, {Tanaka}, {Tang}, {Taranu}, {Terai}, {Toba}, {Turner},
  {Uchiyama}, {Vijarnwannaluk}, {Waters}, {Yamada}, {Yamamoto}, \&
  {Yamashita}}]{Aihara22}
---. 2022, \pasj, 74, 247, \dodoi{10.1093/pasj/psab122}

\bibitem[{{Ando} {et~al.}(2023){Ando}, {Shimasaku}, \& {Ito}}]{Ando}
{Ando}, M., {Shimasaku}, K., \& {Ito}, K. 2023, \mnras, 519, 13,
  \dodoi{10.1093/mnras/stac3251}

\bibitem[{{Assef} {et~al.}(2018){Assef}, {Stern}, {Noirot}, {Jun}, {Cutri}, \&
  {Eisenhardt}}]{Assef}
{Assef}, R.~J., {Stern}, D., {Noirot}, G., {et~al.} 2018, \apjs, 234, 23,
  \dodoi{10.3847/1538-4365/aaa00a}

\bibitem[{{Astropy Collaboration} {et~al.}(2013){Astropy Collaboration},
  {Robitaille}, {Tollerud}, {Greenfield}, {Droettboom}, {Bray}, {Aldcroft},
  {Davis}, {Ginsburg}, {Price-Whelan}, {Kerzendorf}, {Conley}, {Crighton},
  {Barbary}, {Muna}, {Ferguson}, {Grollier}, {Parikh}, {Nair}, {Unther},
  {Deil}, {Woillez}, {Conseil}, {Kramer}, {Turner}, {Singer}, {Fox}, {Weaver},
  {Zabalza}, {Edwards}, {Azalee Bostroem}, {Burke}, {Casey}, {Crawford},
  {Dencheva}, {Ely}, {Jenness}, {Labrie}, {Lim}, {Pierfederici}, {Pontzen},
  {Ptak}, {Refsdal}, {Servillat}, \& {Streicher}}]{astropy:2013}
{Astropy Collaboration}, {Robitaille}, T.~P., {Tollerud}, E.~J., {et~al.} 2013,
  \aap, 558, A33, \dodoi{10.1051/0004-6361/201322068}

\bibitem[{{Astropy Collaboration} {et~al.}(2018){Astropy Collaboration},
  {Price-Whelan}, {Sip{\H{o}}cz}, {G{\"u}nther}, {Lim}, {Crawford}, {Conseil},
  {Shupe}, {Craig}, {Dencheva}, {Ginsburg}, {Vand erPlas}, {Bradley},
  {P{\'e}rez-Su{\'a}rez}, {de Val-Borro}, {Aldcroft}, {Cruz}, {Robitaille},
  {Tollerud}, {Ardelean}, {Babej}, {Bach}, {Bachetti}, {Bakanov}, {Bamford},
  {Barentsen}, {Barmby}, {Baumbach}, {Berry}, {Biscani}, {Boquien}, {Bostroem},
  {Bouma}, {Brammer}, {Bray}, {Breytenbach}, {Buddelmeijer}, {Burke},
  {Calderone}, {Cano Rodr{\'\i}guez}, {Cara}, {Cardoso}, {Cheedella}, {Copin},
  {Corrales}, {Crichton}, {D'Avella}, {Deil}, {Depagne}, {Dietrich}, {Donath},
  {Droettboom}, {Earl}, {Erben}, {Fabbro}, {Ferreira}, {Finethy}, {Fox},
  {Garrison}, {Gibbons}, {Goldstein}, {Gommers}, {Greco}, {Greenfield},
  {Groener}, {Grollier}, {Hagen}, {Hirst}, {Homeier}, {Horton}, {Hosseinzadeh},
  {Hu}, {Hunkeler}, {Ivezi{\'c}}, {Jain}, {Jenness}, {Kanarek}, {Kendrew},
  {Kern}, {Kerzendorf}, {Khvalko}, {King}, {Kirkby}, {Kulkarni}, {Kumar},
  {Lee}, {Lenz}, {Littlefair}, {Ma}, {Macleod}, {Mastropietro}, {McCully},
  {Montagnac}, {Morris}, {Mueller}, {Mumford}, {Muna}, {Murphy}, {Nelson},
  {Nguyen}, {Ninan}, {N{\"o}the}, {Ogaz}, {Oh}, {Parejko}, {Parley}, {Pascual},
  {Patil}, {Patil}, {Plunkett}, {Prochaska}, {Rastogi}, {Reddy Janga},
  {Sabater}, {Sakurikar}, {Seifert}, {Sherbert}, {Sherwood-Taylor}, {Shih},
  {Sick}, {Silbiger}, {Singanamalla}, {Singer}, {Sladen}, {Sooley},
  {Sornarajah}, {Streicher}, {Teuben}, {Thomas}, {Tremblay}, {Turner},
  {Terr{\'o}n}, {van Kerkwijk}, {de la Vega}, {Watkins}, {Weaver}, {Whitmore},
  {Woillez}, {Zabalza}, \& {Astropy Contributors}}]{astropy:2018}
{Astropy Collaboration}, {Price-Whelan}, A.~M., {Sip{\H{o}}cz}, B.~M., {et~al.}
  2018, \aj, 156, 123, \dodoi{10.3847/1538-3881/aabc4f}

\bibitem[{{Astropy Collaboration} {et~al.}(2022){Astropy Collaboration},
  {Price-Whelan}, {Lim}, {Earl}, {Starkman}, {Bradley}, {Shupe}, {Patil},
  {Corrales}, {Brasseur}, {N{"o}the}, {Donath}, {Tollerud}, {Morris},
  {Ginsburg}, {Vaher}, {Weaver}, {Tocknell}, {Jamieson}, {van Kerkwijk},
  {Robitaille}, {Merry}, {Bachetti}, {G{"u}nther}, {Aldcroft},
  {Alvarado-Montes}, {Archibald}, {B{'o}di}, {Bapat}, {Barentsen}, {Baz{'a}n},
  {Biswas}, {Boquien}, {Burke}, {Cara}, {Cara}, {Conroy}, {Conseil}, {Craig},
  {Cross}, {Cruz}, {D'Eugenio}, {Dencheva}, {Devillepoix}, {Dietrich},
  {Eigenbrot}, {Erben}, {Ferreira}, {Foreman-Mackey}, {Fox}, {Freij}, {Garg},
  {Geda}, {Glattly}, {Gondhalekar}, {Gordon}, {Grant}, {Greenfield}, {Groener},
  {Guest}, {Gurovich}, {Handberg}, {Hart}, {Hatfield-Dodds}, {Homeier},
  {Hosseinzadeh}, {Jenness}, {Jones}, {Joseph}, {Kalmbach}, {Karamehmetoglu},
  {Ka{l}uszy{'n}ski}, {Kelley}, {Kern}, {Kerzendorf}, {Koch}, {Kulumani},
  {Lee}, {Ly}, {Ma}, {MacBride}, {Maljaars}, {Muna}, {Murphy}, {Norman},
  {O'Steen}, {Oman}, {Pacifici}, {Pascual}, {Pascual-Granado}, {Patil},
  {Perren}, {Pickering}, {Rastogi}, {Roulston}, {Ryan}, {Rykoff}, {Sabater},
  {Sakurikar}, {Salgado}, {Sanghi}, {Saunders}, {Savchenko}, {Schwardt},
  {Seifert-Eckert}, {Shih}, {Jain}, {Shukla}, {Sick}, {Simpson},
  {Singanamalla}, {Singer}, {Singhal}, {Sinha}, {Sip{H{o}}cz}, {Spitler},
  {Stansby}, {Streicher}, {{ {S}}umak}, {Swinbank}, {Taranu}, {Tewary},
  {Tremblay}, {Val-Borro}, {Van Kooten}, {Vasovi{'c}}, {Verma}, {de Miranda
  Cardoso}, {Williams}, {Wilson}, {Winkel}, {Wood-Vasey}, {Xue}, {Yoachim},
  {Zhang}, {Zonca}, \& {Astropy Project Contributors}}]{astropy:2022}
{Astropy Collaboration}, {Price-Whelan}, A.~M., {Lim}, P.~L., {et~al.} 2022,
  apj, 935, 167, \dodoi{10.3847/1538-4357/ac7c74}

\bibitem[{{Baes} {et~al.}(2011){Baes}, {Verstappen}, {De Looze}, {Fritz},
  {Saftly}, {Vidal P{\'e}rez}, {Stalevski}, \& {Valcke}}]{Baes}
{Baes}, M., {Verstappen}, J., {De Looze}, I., {et~al.} 2011, \apjs, 196, 22,
  \dodoi{10.1088/0067-0049/196/2/22}

\bibitem[{{Baldry} {et~al.}(2018){Baldry}, {Liske}, {Brown}, {Robotham},
  {Driver}, {Dunne}, {Alpaslan}, {Brough}, {Cluver}, {Eardley}, {Farrow},
  {Heymans}, {Hildebrandt}, {Hopkins}, {Kelvin}, {Loveday}, {Moffett},
  {Norberg}, {Owers}, {Taylor}, {Wright}, {Bamford}, {Bland-Hawthorn},
  {Bourne}, {Bremer}, {Colless}, {Conselice}, {Croom}, {Davies}, {Foster},
  {Grootes}, {Holwerda}, {Jones}, {Kafle}, {Kuijken}, {Lara-Lopez},
  {L{\'o}pez-S{\'a}nchez}, {Meyer}, {Phillipps}, {Sutherland}, {van Kampen}, \&
  {Wilkins}}]{Baldry}
{Baldry}, I.~K., {Liske}, J., {Brown}, M.~J.~I., {et~al.} 2018, \mnras, 474,
  3875, \dodoi{10.1093/mnras/stx3042}

\bibitem[{{Bertin} \& {Arnouts}(1996)}]{Bertin}
{Bertin}, E., \& {Arnouts}, S. 1996, \aaps, 117, 393,
  \dodoi{10.1051/aas:1996164}

\bibitem[{{Best} {et~al.}(2005){Best}, {Kauffmann}, {Heckman}, {Brinchmann},
  {Charlot}, {Ivezi{\'c}}, \& {White}}]{Best05}
{Best}, P.~N., {Kauffmann}, G., {Heckman}, T.~M., {et~al.} 2005, \mnras, 362,
  25, \dodoi{10.1111/j.1365-2966.2005.09192.x}

\bibitem[{{Bhargava} {et~al.}(2023){Bhargava}, {Garrel}, {Koulouridis},
  {Pierre}, {Valtchanov}, {Cerardi}, {Maughan}, {Aguena}, {Benoist}, {Baguley},
  {Ramos-Ceja}, {Adami}, {Chiappetti}, {Vignali}, \& {Willis}}]{Bhargava}
{Bhargava}, S., {Garrel}, C., {Koulouridis}, E., {et~al.} 2023, \aap, 673, A92,
  \dodoi{10.1051/0004-6361/202244898}

\bibitem[{{Bianchi} {et~al.}(2017){Bianchi}, {Shiao}, \& {Thilker}}]{Bianchi}
{Bianchi}, L., {Shiao}, B., \& {Thilker}, D. 2017, \apjs, 230, 24,
  \dodoi{10.3847/1538-4365/aa7053}

\bibitem[{{Bitsakis} {et~al.}(2015){Bitsakis}, {Dultzin}, {Ciesla}, {Krongold},
  {Charmandaris}, \& {Zezas}}]{Bitsakis}
{Bitsakis}, T., {Dultzin}, D., {Ciesla}, L., {et~al.} 2015, \mnras, 450, 3114,
  \dodoi{10.1093/mnras/stv755}

\bibitem[{{Boquien} {et~al.}(2019){Boquien}, {Burgarella}, {Roehlly}, {Buat},
  {Ciesla}, {Corre}, {Inoue}, \& {Salas}}]{Boquien}
{Boquien}, M., {Burgarella}, D., {Roehlly}, Y., {et~al.} 2019, \aap, 622, A103,
  \dodoi{10.1051/0004-6361/201834156}

\bibitem[{{Boquien} {et~al.}(2016){Boquien}, {Kennicutt}, {Calzetti}, {Dale},
  {Galametz}, {Sauvage}, {Croxall}, {Draine}, {Kirkpatrick}, {Kumari}, {Hunt},
  {De Looze}, {Pellegrini}, {Rela{\~n}o}, {Smith}, \& {Tabatabaei}}]{Boquien16}
{Boquien}, M., {Kennicutt}, R., {Calzetti}, D., {et~al.} 2016, \aap, 591, A6,
  \dodoi{10.1051/0004-6361/201527759}

\bibitem[{{Bosch} {et~al.}(2018){Bosch}, {Armstrong}, {Bickerton}, {Furusawa},
  {Ikeda}, {Koike}, {Lupton}, {Mineo}, {Price}, {Takata}, {Tanaka}, {Yasuda},
  {AlSayyad}, {Becker}, {Coulton}, {Coupon}, {Garmilla}, {Huang}, {Krughoff},
  {Lang}, {Leauthaud}, {Lim}, {Lust}, {MacArthur}, {Mandelbaum}, {Miyatake},
  {Miyazaki}, {Murata}, {More}, {Okura}, {Owen}, {Swinbank}, {Strauss},
  {Yamada}, \& {Yamanoi}}]{Bosch}
{Bosch}, J., {Armstrong}, R., {Bickerton}, S., {et~al.} 2018, \pasj, 70, S5,
  \dodoi{10.1093/pasj/psx080}

\bibitem[{{Boselli} {et~al.}(2016){Boselli}, {Roehlly}, {Fossati}, {Buat},
  {Boissier}, {Boquien}, {Burgarella}, {Ciesla}, {Gavazzi}, \&
  {Serra}}]{Boselli}
{Boselli}, A., {Roehlly}, Y., {Fossati}, M., {et~al.} 2016, \aap, 596, A11,
  \dodoi{10.1051/0004-6361/201629221}

\bibitem[{{Bruzual} \& {Charlot}(2003)}]{Bruzual}
{Bruzual}, G., \& {Charlot}, S. 2003, \mnras, 344, 1000,
  \dodoi{10.1046/j.1365-8711.2003.06897.x}

\bibitem[{{Buat} {et~al.}(2012){Buat}, {Noll}, {Burgarella}, {Giovannoli},
  {Charmandaris}, {Pannella}, {Hwang}, {Elbaz}, {Dickinson}, {Magdis}, {Reddy},
  \& {Murphy}}]{Buat12}
{Buat}, V., {Noll}, S., {Burgarella}, D., {et~al.} 2012, \aap, 545, A141,
  \dodoi{10.1051/0004-6361/201219405}

\bibitem[{{Buat} {et~al.}(2015){Buat}, {Oi}, {Heinis}, {Ciesla}, {Burgarella},
  {Matsuhara}, {Malek}, {Goto}, {Malkan}, {Marchetti}, {Ohyama}, {Pearson},
  {Serjeant}, {Miyaji}, {Krumpe}, \& {Brunner}}]{Buat_15}
{Buat}, V., {Oi}, N., {Heinis}, S., {et~al.} 2015, \aap, 577, A141,
  \dodoi{10.1051/0004-6361/201425399}

\bibitem[{{Bufanda} {et~al.}(2017){Bufanda}, {Hollowood}, {Jeltema}, {Rykoff},
  {Rozo}, {Martini}, {Abbott}, {Abdalla}, {Allam}, {Banerji},
  {Benoit-L{\'e}vy}, {Bertin}, {Brooks}, {Carnero Rosell}, {Carrasco Kind},
  {Carretero}, {Cunha}, {da Costa}, {Desai}, {Diehl}, {Dietrich}, {Evrard},
  {Fausti Neto}, {Flaugher}, {Frieman}, {Gerdes}, {Goldstein}, {Gruen},
  {Gruendl}, {Gutierrez}, {Honscheid}, {James}, {Kuehn}, {Kuropatkin}, {Lima},
  {Maia}, {Marshall}, {Melchior}, {Miquel}, {Mohr}, {Ogando}, {Plazas},
  {Romer}, {Rooney}, {Sanchez}, {Santiago}, {Scarpine}, {Sevilla-Noarbe},
  {Smith}, {Soares-Santos}, {Sobreira}, {Suchyta}, {Tarle}, {Thomas}, {Tucker},
  {Walker}, \& {DES Collaboration}}]{Bufanda}
{Bufanda}, E., {Hollowood}, D., {Jeltema}, T.~E., {et~al.} 2017, \mnras, 465,
  2531, \dodoi{10.1093/mnras/stw2824}

\bibitem[{{Burgarella} {et~al.}(2005){Burgarella}, {Buat}, \&
  {Iglesias-P{\'a}ramo}}]{Burgarella}
{Burgarella}, D., {Buat}, V., \& {Iglesias-P{\'a}ramo}, J. 2005, \mnras, 360,
  1413, \dodoi{10.1111/j.1365-2966.2005.09131.x}

\bibitem[{{Butcher} \& {Oemler}(1984)}]{Butcher}
{Butcher}, H., \& {Oemler}, A., J. 1984, \apj, 285, 426, \dodoi{10.1086/162519}

\bibitem[{{Calzetti} {et~al.}(2000){Calzetti}, {Armus}, {Bohlin}, {Kinney},
  {Koornneef}, \& {Storchi-Bergmann}}]{Calzetti}
{Calzetti}, D., {Armus}, L., {Bohlin}, R.~C., {et~al.} 2000, \apj, 533, 682,
  \dodoi{10.1086/308692}

\bibitem[{{Camps} \& {Baes}(2015)}]{Camps}
{Camps}, P., \& {Baes}, M. 2015, Astronomy and Computing, 9, 20,
  \dodoi{10.1016/j.ascom.2014.10.004}

\bibitem[{{Casey}(2012)}]{Casey}
{Casey}, C.~M. 2012, \mnras, 425, 3094,
  \dodoi{10.1111/j.1365-2966.2012.21455.x}

\bibitem[{{Chabrier}(2003)}]{Chabrier}
{Chabrier}, G. 2003, \pasp, 115, 763, \dodoi{10.1086/376392}

\bibitem[{{Chadayammuri} {et~al.}(2021){Chadayammuri}, {Tremmel}, {Nagai},
  {Babul}, \& {Quinn}}]{Chadayammuri}
{Chadayammuri}, U., {Tremmel}, M., {Nagai}, D., {Babul}, A., \& {Quinn}, T.
  2021, \mnras, 504, 3922, \dodoi{10.1093/mnras/stab1010}

\bibitem[{{Chambers} {et~al.}(2016){Chambers}, {Magnier}, {Metcalfe},
  {Flewelling}, {Huber}, {Waters}, {Denneau}, {Draper}, {Farrow}, {Finkbeiner},
  {Holmberg}, {Koppenhoefer}, {Price}, {Saglia}, {Schlafly}, {Smartt},
  {Sweeney}, {Wainscoat}, {Burgett}, {Grav}, {Heasley}, {Hodapp}, {Jedicke},
  {Kaiser}, {Kudritzki}, {Luppino}, {Lupton}, {Monet}, {Morgan}, {Onaka},
  {Stubbs}, {Tonry}, {Banados}, {Bell}, {Bender}, {Bernard}, {Botticella},
  {Casertano}, {Chastel}, {Chen}, {Chen}, {Cole}, {Deacon}, {Frenk},
  {Fitzsimmons}, {Gezari}, {Goessl}, {Goggia}, {Goldman}, {Grebel}, {Hambly},
  {Hasinger}, {Heavens}, {Heckman}, {Henderson}, {Henning}, {Holman}, {Hopp},
  {Ip}, {Isani}, {Keyes}, {Koekemoer}, {Kotak}, {Long}, {Lucey}, {Liu},
  {Martin}, {McLean}, {Morganson}, {Murphy}, {Nieto-Santisteban}, {Norberg},
  {Peacock}, {Pier}, {Postman}, {Primak}, {Rae}, {Rest}, {Riess}, {Riffeser},
  {Rix}, {Roser}, {Schilbach}, {Schultz}, {Scolnic}, {Szalay}, {Seitz},
  {Shiao}, {Small}, {Smith}, {Soderblom}, {Taylor}, {Thakar}, {Thiel},
  {Thilker}, {Urata}, {Valenti}, {Walter}, {Watters}, {Werner}, {White},
  {Wood-Vasey}, \& {Wyse}}]{Chambers}
{Chambers}, K.~C., {Magnier}, E.~A., {Metcalfe}, N., {et~al.} 2016, ArXiv
  e-prints.
\newblock \doarXiv{1612.05560}

\bibitem[{{Chiu} {et~al.}(2020){Chiu}, {Umetsu}, {Murata}, {Medezinski}, \&
  {Oguri}}]{Chiu}
{Chiu}, I.~N., {Umetsu}, K., {Murata}, R., {Medezinski}, E., \& {Oguri}, M.
  2020, \mnras, 495, 428, \dodoi{10.1093/mnras/staa1158}

\bibitem[{{Ciesla} {et~al.}(2015){Ciesla}, {Charmandaris}, {Georgakakis},
  {Bernhard}, {Mitchell}, {Buat}, {Elbaz}, {LeFloc'h}, {Lacey}, {Magdis}, \&
  {Xilouris}}]{Ciesla_15}
{Ciesla}, L., {Charmandaris}, V., {Georgakakis}, A., {et~al.} 2015, \aap, 576,
  A10, \dodoi{10.1051/0004-6361/201425252}

\bibitem[{{Ciesla} {et~al.}(2016){Ciesla}, {Boselli}, {Elbaz}, {Boissier},
  {Buat}, {Charmandaris}, {Schreiber}, {B{\'e}thermin}, {Baes}, {Boquien}, {De
  Looze}, {Fern{\'a}ndez-Ontiveros}, {Pappalardo}, {Spinoglio}, \&
  {Viaene}}]{Ciesla_16}
{Ciesla}, L., {Boselli}, A., {Elbaz}, D., {et~al.} 2016, \aap, 585, A43,
  \dodoi{10.1051/0004-6361/201527107}

\bibitem[{{Coil} {et~al.}(2011){Coil}, {Blanton}, {Burles}, {Cool},
  {Eisenstein}, {Moustakas}, {Wong}, {Zhu}, {Aird}, {Bernstein}, {Bolton}, \&
  {Hogg}}]{Coil11}
{Coil}, A.~L., {Blanton}, M.~R., {Burles}, S.~M., {et~al.} 2011, \apj, 741, 8,
  \dodoi{10.1088/0004-637X/741/1/8}

\bibitem[{{Cool} {et~al.}(2013){Cool}, {Moustakas}, {Blanton}, {Burles},
  {Coil}, {Eisenstein}, {Wong}, {Zhu}, {Aird}, {Bernstein}, {Bolton}, {Hogg},
  \& {Mendez}}]{Cool13}
{Cool}, R.~J., {Moustakas}, J., {Blanton}, M.~R., {et~al.} 2013, \apj, 767,
  118, \dodoi{10.1088/0004-637X/767/2/118}

\bibitem[{{Cooper} {et~al.}(2012){Cooper}, {Griffith}, {Newman}, {Coil},
  {Davis}, {Dutton}, {Faber}, {Guhathakurta}, {Koo}, {Lotz}, {Weiner},
  {Willmer}, \& {Yan}}]{Cooper}
{Cooper}, M.~C., {Griffith}, R.~L., {Newman}, J.~A., {et~al.} 2012, \mnras,
  419, 3018, \dodoi{10.1111/j.1365-2966.2011.19938.x}

\bibitem[{{Coupon} {et~al.}(2018){Coupon}, {Czakon}, {Bosch}, {Komiyama},
  {Medezinski}, {Miyazaki}, \& {Oguri}}]{Coupon}
{Coupon}, J., {Czakon}, N., {Bosch}, J., {et~al.} 2018, \pasj, 70, S7,
  \dodoi{10.1093/pasj/psx047}

\bibitem[{{Cutri} {et~al.}(2021){Cutri}, {Wright}, {Conrow}, {Fowler},
  {Eisenhardt}, {Grillmair}, {Kirkpatrick}, {Masci}, {McCallon}, {Wheelock},
  {Fajardo-Acosta}, {Yan}, {Benford}, {Harbut}, {Jarrett}, {Lake}, {Leisawitz},
  {Ressler}, {Stanford}, {Tsai}, {Liu}, {Helou}, {Mainzer}, {Gettngs},
  {Gonzalez}, {Hoffman}, {Marsh}, {Padgett}, {Skrutskie}, {Beck}, {Papin}, \&
  {Wittman}}]{Cutri}
{Cutri}, R.~M., {Wright}, E.~L., {Conrow}, T., {et~al.} 2021, VizieR Online
  Data Catalog, II/328

\bibitem[{{Dale} {et~al.}(2014){Dale}, {Helou}, {Magdis}, {Armus},
  {D{\'\i}az-Santos}, \& {Shi}}]{Dale}
{Dale}, D.~A., {Helou}, G., {Magdis}, G.~E., {et~al.} 2014, \apj, 784, 83,
  \dodoi{10.1088/0004-637X/784/1/83}

\bibitem[{{de Jong} {et~al.}(2013){de Jong}, {Verdoes Kleijn}, {Kuijken}, \&
  {Valentijn}}]{deJong}
{de Jong}, J. T.~A., {Verdoes Kleijn}, G.~A., {Kuijken}, K.~H., \& {Valentijn},
  E.~A. 2013, Experimental Astronomy, 35, 25, \dodoi{10.1007/s10686-012-9306-1}

\bibitem[{{de Jong} {et~al.}(2017){de Jong}, {Verdoes Kleijn}, {Erben},
  {Hildebrandt}, {Kuijken}, {Sikkema}, {Brescia}, {Bilicki}, {Napolitano},
  {Amaro}, {Begeman}, {Boxhoorn}, {Buddelmeijer}, {Cavuoti}, {Getman}, {Grado},
  {Helmich}, {Huang}, {Irisarri}, {La Barbera}, {Longo}, {McFarland},
  {Nakajima}, {Paolillo}, {Puddu}, {Radovich}, {Rifatto}, {Tortora},
  {Valentijn}, {Vellucci}, {Vriend}, {Amon}, {Blake}, {Choi}, {Conti}, {Gwyn},
  {Herbonnet}, {Heymans}, {Hoekstra}, {Klaes}, {Merten}, {Miller}, {Schneider},
  \& {Viola}}]{de_Jong}
{de Jong}, J. T.~A., {Verdoes Kleijn}, G.~A., {Erben}, T., {et~al.} 2017, \aap,
  604, A134, \dodoi{10.1051/0004-6361/201730747}

\bibitem[{{Dietrich} {et~al.}(2018){Dietrich}, {Weiner}, {Ashby}, {Hayward},
  {Mart{\'\i}nez-Galarza}, {Ramos Padilla}, {Rosenthal}, {Smith}, {Willner}, \&
  {Zezas}}]{Dietrich}
{Dietrich}, J., {Weiner}, A.~S., {Ashby}, M. L.~N., {et~al.} 2018, \mnras, 480,
  3562, \dodoi{10.1093/mnras/sty2056}

\bibitem[{{Driver} {et~al.}(2011){Driver}, {Hill}, {Kelvin}, {Robotham},
  {Liske}, {Norberg}, {Baldry}, {Bamford}, {Hopkins}, {Loveday}, {Peacock},
  {Andrae}, {Bland-Hawthorn}, {Brough}, {Brown}, {Cameron}, {Ching}, {Colless},
  {Conselice}, {Croom}, {Cross}, {de Propris}, {Dye}, {Drinkwater}, {Ellis},
  {Graham}, {Grootes}, {Gunawardhana}, {Jones}, {van Kampen}, {Maraston},
  {Nichol}, {Parkinson}, {Phillipps}, {Pimbblet}, {Popescu}, {Prescott},
  {Roseboom}, {Sadler}, {Sansom}, {Sharp}, {Smith}, {Taylor}, {Thomas},
  {Tuffs}, {Wijesinghe}, {Dunne}, {Frenk}, {Jarvis}, {Madore}, {Meyer},
  {Seibert}, {Staveley-Smith}, {Sutherland}, \& {Warren}}]{Driver}
{Driver}, S.~P., {Hill}, D.~T., {Kelvin}, L.~S., {et~al.} 2011, \mnras, 413,
  971, \dodoi{10.1111/j.1365-2966.2010.18188.x}

\bibitem[{{Eastman} {et~al.}(2007){Eastman}, {Martini}, {Sivakoff}, {Kelson},
  {Mulchaey}, \& {Tran}}]{Eastman}
{Eastman}, J., {Martini}, P., {Sivakoff}, G., {et~al.} 2007, \apjl, 664, L9,
  \dodoi{10.1086/520577}

\bibitem[{{Edge} {et~al.}(2013){Edge}, {Sutherland}, {Kuijken}, {Driver},
  {McMahon}, {Eales}, \& {Emerson}}]{Edge}
{Edge}, A., {Sutherland}, W., {Kuijken}, K., {et~al.} 2013, The Messenger, 154,
  32

\bibitem[{{Ehlert} {et~al.}(2014){Ehlert}, {von der Linden}, {Allen}, {Brandt},
  {Xue}, {Luo}, {Mantz}, {Morris}, {Applegate}, \& {Kelly}}]{Ehlert14}
{Ehlert}, S., {von der Linden}, A., {Allen}, S.~W., {et~al.} 2014, \mnras, 437,
  1942, \dodoi{10.1093/mnras/stt2025}

\bibitem[{{Ellison} {et~al.}(2008){Ellison}, {Patton}, {Simard}, \&
  {McConnachie}}]{Ellison}
{Ellison}, S.~L., {Patton}, D.~R., {Simard}, L., \& {McConnachie}, A.~W. 2008,
  \aj, 135, 1877, \dodoi{10.1088/0004-6256/135/5/1877}

\bibitem[{{Fabian}(2012)}]{Fabian}
{Fabian}, A.~C. 2012, \araa, 50, 455,
  \dodoi{10.1146/annurev-astro-081811-125521}

\bibitem[{{Ferrarese} \& {Merritt}(2000)}]{Ferrarese}
{Ferrarese}, L., \& {Merritt}, D. 2000, \apjl, 539, L9, \dodoi{10.1086/312838}

\bibitem[{{Furusawa} {et~al.}(2018){Furusawa}, {Koike}, {Takata}, {Okura},
  {Miyatake}, {Lupton}, {Bickerton}, {Price}, {Bosch}, {Yasuda}, {Mineo},
  {Yamada}, {Miyazaki}, {Nakata}, {Koshida}, {Komiyama}, {Utsumi},
  {Kawanomoto}, {Jeschke}, {Noumaru}, {Schubert}, {Iwata}, {Finet},
  {Fujiyoshi}, {Tajitsu}, {Terai}, \& {Lee}}]{Furusawa}
{Furusawa}, H., {Koike}, M., {Takata}, T., {et~al.} 2018, \pasj, 70, S3,
  \dodoi{10.1093/pasj/psx079}

\bibitem[{{Galametz} {et~al.}(2009){Galametz}, {Stern}, {Eisenhardt},
  {Brodwin}, {Brown}, {Dey}, {Gonzalez}, {Jannuzi}, {Moustakas}, \&
  {Stanford}}]{Galametz}
{Galametz}, A., {Stern}, D., {Eisenhardt}, P. R.~M., {et~al.} 2009, \apj, 694,
  1309, \dodoi{10.1088/0004-637X/694/2/1309}

\bibitem[{{Glikman} {et~al.}(2015){Glikman}, {Simmons}, {Mailly}, {Schawinski},
  {Urry}, \& {Lacy}}]{Glikman}
{Glikman}, E., {Simmons}, B., {Mailly}, M., {et~al.} 2015, \apj, 806, 218,
  \dodoi{10.1088/0004-637X/806/2/218}

\bibitem[{{Gonz{\'a}lez-Fern{\'a}ndez}
  {et~al.}(2018){Gonz{\'a}lez-Fern{\'a}ndez}, {Hodgkin}, {Irwin},
  {Gonz{\'a}lez-Solares}, {Koposov}, {Lewis}, {Emerson}, {Hewett},
  {Yolda{\c{s}}}, \& {Riello}}]{Gonz}
{Gonz{\'a}lez-Fern{\'a}ndez}, C., {Hodgkin}, S.~T., {Irwin}, M.~J., {et~al.}
  2018, \mnras, 474, 5459, \dodoi{10.1093/mnras/stx3073}

\bibitem[{{Greene} {et~al.}(2022){Greene}, {Bezanson}, {Ouchi}, {Silverman}, \&
  {the PFS Galaxy Evolution Working Group}}]{Greene}
{Greene}, J., {Bezanson}, R., {Ouchi}, M., {Silverman}, J., \& {the PFS Galaxy
  Evolution Working Group}. 2022, arXiv e-prints, arXiv:2206.14908,
  \dodoi{10.48550/arXiv.2206.14908}

\bibitem[{{Guzzo} {et~al.}(2014){Guzzo}, {Scodeggio}, {Garilli}, {Granett},
  {Fritz}, {Abbas}, {Adami}, {Arnouts}, {Bel}, {Bolzonella}, {Bottini},
  {Branchini}, {Cappi}, {Coupon}, {Cucciati}, {Davidzon}, {De Lucia}, {de la
  Torre}, {Franzetti}, {Fumana}, {Hudelot}, {Ilbert}, {Iovino}, {Krywult}, {Le
  Brun}, {Le F{\`e}vre}, {Maccagni}, {Ma{\l}ek}, {Marulli}, {McCracken},
  {Paioro}, {Peacock}, {Polletta}, {Pollo}, {Schlagenhaufer}, {Tasca},
  {Tojeiro}, {Vergani}, {Zamorani}, {Zanichelli}, {Burden}, {Di Porto},
  {Marchetti}, {Marinoni}, {Mellier}, {Moscardini}, {Nichol}, {Percival},
  {Phleps}, \& {Wolk}}]{Guzzo}
{Guzzo}, L., {Scodeggio}, M., {Garilli}, B., {et~al.} 2014, \aap, 566, A108,
  \dodoi{10.1051/0004-6361/201321489}

\bibitem[{{Haggard} {et~al.}(2010){Haggard}, {Green}, {Anderson}, {Constantin},
  {Aldcroft}, {Kim}, \& {Barkhouse}}]{Haggard}
{Haggard}, D., {Green}, P.~J., {Anderson}, S.~F., {et~al.} 2010, \apj, 723,
  1447, \dodoi{10.1088/0004-637X/723/2/1447}

\bibitem[{Harris {et~al.}(2020)Harris, Millman, van~der Walt, Gommers,
  Virtanen, Cournapeau, Wieser, Taylor, Berg, Smith, Kern, Picus, Hoyer, van
  Kerkwijk, Brett, Haldane, del R{\'{i}}o, Wiebe, Peterson,
  G{\'{e}}rard-Marchant, Sheppard, Reddy, Weckesser, Abbasi, Gohlke, \&
  Oliphant}]{numpy}
Harris, C.~R., Millman, K.~J., van~der Walt, S.~J., {et~al.} 2020, Nature, 585,
  357, \dodoi{10.1038/s41586-020-2649-2}

\bibitem[{{Hashiguchi} {et~al.}(2023){Hashiguchi}, {Toba}, {Ota}, {Oguri},
  {Okabe}, {Ueda}, {Imanishi}, {Yamada}, {Goto}, {Koyama}, {Lee}, {Mitsuishi},
  {Nagao}, {Nishizawa}, {Noboriguchi}, {Oogi}, {Sakuta}, {Schramm}, {Shibata},
  {Terashima}, {Yamashita}, {Yanagawa}, \& {Yoshimoto}}]{Hashiguchi}
{Hashiguchi}, A., {Toba}, Y., {Ota}, N., {et~al.} 2023, \pasj, 75, 1246,
  \dodoi{10.1093/pasj/psad066}

\bibitem[{{Hewett} {et~al.}(2006){Hewett}, {Warren}, {Leggett}, \&
  {Hodgkin}}]{Hewett}
{Hewett}, P.~C., {Warren}, S.~J., {Leggett}, S.~K., \& {Hodgkin}, S.~T. 2006,
  \mnras, 367, 454, \dodoi{10.1111/j.1365-2966.2005.09969.x}

\bibitem[{{Hsieh} \& {Yee}(2014)}]{Hsieh}
{Hsieh}, B.~C., \& {Yee}, H.~K.~C. 2014, \apj, 792, 102,
  \dodoi{10.1088/0004-637X/792/2/102}

\bibitem[{{Hsieh} {et~al.}(2005){Hsieh}, {Yee}, {Lin}, \& {Gladders}}]{Hsieh05}
{Hsieh}, B.~C., {Yee}, H.~K.~C., {Lin}, H., \& {Gladders}, M.~D. 2005, \apjs,
  158, 161, \dodoi{10.1086/429293}

\bibitem[{{Huang} {et~al.}(2018){Huang}, {Leauthaud}, {Murata}, {Bosch},
  {Price}, {Lupton}, {Mandelbaum}, {Lackner}, {Bickerton}, {Miyazaki},
  {Coupon}, \& {Tanaka}}]{Huang}
{Huang}, S., {Leauthaud}, A., {Murata}, R., {et~al.} 2018, \pasj, 70, S6,
  \dodoi{10.1093/pasj/psx126}

\bibitem[{Hunter(2007)}]{matplotlib}
Hunter, J.~D. 2007, Computing in science \& engineering, 9, 90

\bibitem[{{Inoue}(2011)}]{Inoue}
{Inoue}, A.~K. 2011, \mnras, 415, 2920,
  \dodoi{10.1111/j.1365-2966.2011.18906.x}

\bibitem[{{Ivezi{\'c}} {et~al.}(2019){Ivezi{\'c}}, {Kahn}, {Tyson}, {Abel},
  {Acosta}, {Allsman}, {Alonso}, {AlSayyad}, {Anderson}, {Andrew}, \&
  et~al.}]{Ivezic}
{Ivezi{\'c}}, {\v Z}., {Kahn}, S.~M., {Tyson}, J.~A., {et~al.} 2019, \apj, 873,
  111, \dodoi{10.3847/1538-4357/ab042c}

\bibitem[{{Juri{\'c}} {et~al.}(2017){Juri{\'c}}, {Kantor}, {Lim}, {Lupton},
  {Dubois-Felsmann}, {Jenness}, {Axelrod}, {Aleksi{\'c}}, {Allsman},
  {AlSayyad}, {Alt}, {Armstrong}, {Basney}, {Becker}, {Becla}, {Biswas},
  {Bosch}, {Boutigny}, {Kind}, {Ciardi}, {Connolly}, {Daniel}, {Daues},
  {Economou}, {Chiang}, {Fausti}, {Fisher-Levine}, {Freemon}, {Gris},
  {Hernandez}, {Hoblitt}, {Ivezi{\'c}}, {Jammes}, {Jevremovi{\'c}}, {Jones},
  {Kalmbach}, {Kasliwal}, {Krughoff}, {Lurie}, {Lust}, {MacArthur}, {Melchior},
  {Moeyens}, {Nidever}, {Owen}, {Parejko}, {Peterson}, {Petravick},
  {Pietrowicz}, {Price}, {Reiss}, {Shaw}, {Sick}, {Slater}, {Strauss},
  {Sullivan}, {Swinbank}, {Van Dyk}, {Vuj{\v c}i{\'c}}, {Withers}, \&
  {Yoachim}}]{Juric}
{Juri{\'c}}, M., {Kantor}, J., {Lim}, K.-T., {et~al.} 2017, in Astronomical
  Society of the Pacific Conference Series, Vol. 512, Astronomical Data
  Analysis Software and Systems XXV, ed. N.~P.~F. {Lorente}, K.~{Shortridge},
  \& R.~{Wayth}, 279.
\newblock \doarXiv{1512.07914}

\bibitem[{{Kauffmann} {et~al.}(2003){Kauffmann}, {Heckman}, {Tremonti},
  {Brinchmann}, {Charlot}, {White}, {Ridgway}, {Brinkmann}, {Fukugita}, {Hall},
  {Ivezi{\'c}}, {Richards}, \& {Schneider}}]{Kauffmann}
{Kauffmann}, G., {Heckman}, T.~M., {Tremonti}, C., {et~al.} 2003, \mnras, 346,
  1055, \dodoi{10.1111/j.1365-2966.2003.07154.x}

\bibitem[{{Kawanomoto} {et~al.}(2018){Kawanomoto}, {Uraguchi}, {Komiyama},
  {Miyazaki}, {Furusawa}, {Finet}, {Hattori}, {Wang}, {Yasuda}, \&
  {Suzuki}}]{Kawanomoto}
{Kawanomoto}, S., {Uraguchi}, F., {Komiyama}, Y., {et~al.} 2018, \pasj, 70, 66,
  \dodoi{10.1093/pasj/psy056}

\bibitem[{{Kelly}(2007)}]{Kelly}
{Kelly}, B.~C. 2007, \apj, 665, 1489, \dodoi{10.1086/519947}

\bibitem[{{Khabiboulline} {et~al.}(2014){Khabiboulline}, {Steinhardt},
  {Silverman}, {Ellison}, {Mendel}, \& {Patton}}]{Khabiboulline}
{Khabiboulline}, E.~T., {Steinhardt}, C.~L., {Silverman}, J.~D., {et~al.} 2014,
  \apj, 795, 62, \dodoi{10.1088/0004-637X/795/1/62}

\bibitem[{{Komiyama} {et~al.}(2018){Komiyama}, {Obuchi}, {Nakaya}, {Kamata},
  {Kawanomoto}, {Utsumi}, {Miyazaki}, {Uraguchi}, {Furusawa}, {Morokuma},
  {Uchida}, {Miyatake}, {Mineo}, {Fujimori}, {Aihara}, {Karoji}, {Gunn}, \&
  {Wang}}]{Komiyama}
{Komiyama}, Y., {Obuchi}, Y., {Nakaya}, H., {et~al.} 2018, \pasj, 70, S2,
  \dodoi{10.1093/pasj/psx069}

\bibitem[{{Koss} {et~al.}(2011){Koss}, {Mushotzky}, {Veilleux}, {Winter},
  {Baumgartner}, {Tueller}, {Gehrels}, \& {Valencic}}]{Koss}
{Koss}, M., {Mushotzky}, R., {Veilleux}, S., {et~al.} 2011, \apj, 739, 57,
  \dodoi{10.1088/0004-637X/739/2/57}

\bibitem[{{Koulouridis} \& {Bartalucci}(2019)}]{Koulouridis}
{Koulouridis}, E., \& {Bartalucci}, I. 2019, \aap, 623, L10,
  \dodoi{10.1051/0004-6361/201935082}

\bibitem[{{Koulouridis} {et~al.}(2024){Koulouridis}, {Gkini}, \&
  {Drigga}}]{Koulouridis24}
{Koulouridis}, E., {Gkini}, A., \& {Drigga}, E. 2024, arXiv e-prints,
  arXiv:2401.05747, \dodoi{10.48550/arXiv.2401.05747}

\bibitem[{{Koulouridis} {et~al.}(2018){Koulouridis}, {Ricci}, {Giles}, {Adami},
  {Ramos-Ceja}, {Pierre}, {Plionis}, {Lidman}, {Georgantopoulos}, {Chiappetti},
  {Elyiv}, {Ettori}, {Faccioli}, {Fotopoulou}, {Gastaldello}, {Pacaud},
  {Paltani}, \& {Vignali}}]{Koulouridis18}
{Koulouridis}, E., {Ricci}, M., {Giles}, P., {et~al.} 2018, \aap, 620, A20,
  \dodoi{10.1051/0004-6361/201832974}

\bibitem[{{Krick} {et~al.}(2009){Krick}, {Surace}, {Thompson}, {Ashby}, {Hora},
  {Gorjian}, \& {Yan}}]{Krick}
{Krick}, J.~E., {Surace}, J.~A., {Thompson}, D., {et~al.} 2009, \apj, 700, 123,
  \dodoi{10.1088/0004-637X/700/1/123}

\bibitem[{{Landsman}(1993)}]{idl}
{Landsman}, W.~B. 1993, in Astronomical Society of the Pacific Conference
  Series, Vol.~52, Astronomical Data Analysis Software and Systems II, ed.
  R.~J. {Hanisch}, R.~J.~V. {Brissenden}, \& J.~{Barnes}, 246

\bibitem[{{Lang}(2014)}]{Lang}
{Lang}, D. 2014, \aj, 147, 108, \dodoi{10.1088/0004-6256/147/5/108}

\bibitem[{{Lawrence} {et~al.}(2007){Lawrence}, {Warren}, {Almaini}, {Edge},
  {Hambly}, {Jameson}, {Lucas}, {Casali}, {Adamson}, {Dye}, {Emerson},
  {Foucaud}, {Hewett}, {Hirst}, {Hodgkin}, {Irwin}, {Lodieu}, {McMahon},
  {Simpson}, {Smail}, {Mortlock}, \& {Folger}}]{Lawrence}
{Lawrence}, A., {Warren}, S.~J., {Almaini}, O., {et~al.} 2007, \mnras, 379,
  1599, \dodoi{10.1111/j.1365-2966.2007.12040.x}

\bibitem[{{Le F{\`e}vre} {et~al.}(2013){Le F{\`e}vre}, {Cassata}, {Cucciati},
  {Garilli}, {Ilbert}, {Le Brun}, {Maccagni}, {Moreau}, {Scodeggio}, {Tresse},
  {Zamorani}, {Adami}, {Arnouts}, {Bardelli}, {Bolzonella}, {Bondi},
  {Bongiorno}, {Bottini}, {Cappi}, {Charlot}, {Ciliegi}, {Contini}, {de la
  Torre}, {Foucaud}, {Franzetti}, {Gavignaud}, {Guzzo}, {Iovino}, {Lemaux},
  {L{\'o}pez-Sanjuan}, {McCracken}, {Marano}, {Marinoni}, {Mazure}, {Mellier},
  {Merighi}, {Merluzzi}, {Paltani}, {Pell{\`o}}, {Pollo}, {Pozzetti},
  {Scaramella}, {Tasca}, {Vergani}, {Vettolani}, {Zanichelli}, \&
  {Zucca}}]{LeFevre}
{Le F{\`e}vre}, O., {Cassata}, P., {Cucciati}, O., {et~al.} 2013, \aap, 559,
  A14, \dodoi{10.1051/0004-6361/201322179}

\bibitem[{{Leitherer} {et~al.}(2002){Leitherer}, {Li}, {Calzetti}, \&
  {Heckman}}]{Leitherer}
{Leitherer}, C., {Li}, I.~H., {Calzetti}, D., \& {Heckman}, T.~M. 2002, \apjs,
  140, 303, \dodoi{10.1086/342486}

\bibitem[{{Li} {et~al.}(2019){Li}, {Gu}, {Yuan}, {Bao}, {He}, \& {Bian}}]{Li}
{Li}, F., {Gu}, Y.-Z., {Yuan}, Q.-R., {et~al.} 2019, \mnras, 484, 3806,
  \dodoi{10.1093/mnras/stz267}

\bibitem[{{Lo Faro} {et~al.}(2017){Lo Faro}, {Buat}, {Roehlly},
  {Alvarez-Marquez}, {Burgarella}, {Silva}, \& {Efstathiou}}]{LoFaro}
{Lo Faro}, B., {Buat}, V., {Roehlly}, Y., {et~al.} 2017, \mnras, 472, 1372,
  \dodoi{10.1093/mnras/stx1901}

\bibitem[{{Magliocchetti} {et~al.}(2018){Magliocchetti}, {Popesso}, {Brusa}, \&
  {Salvato}}]{Magliocchetti}
{Magliocchetti}, M., {Popesso}, P., {Brusa}, M., \& {Salvato}, M. 2018, \mnras,
  478, 3848, \dodoi{10.1093/mnras/sty1309}

\bibitem[{{Magnier} {et~al.}(2013){Magnier}, {Schlafly}, {Finkbeiner}, {Juric},
  {Tonry}, {Burgett}, {Chambers}, {Flewelling}, {Kaiser}, {Kudritzki},
  {Morgan}, {Price}, {Sweeney}, \& {Stubbs}}]{Magnier}
{Magnier}, E.~A., {Schlafly}, E., {Finkbeiner}, D., {et~al.} 2013, \apjs, 205,
  20, \dodoi{10.1088/0067-0049/205/2/20}

\bibitem[{{Magorrian} {et~al.}(1998){Magorrian}, {Tremaine}, {Richstone},
  {Bender}, {Bower}, {Dressler}, {Faber}, {Gebhardt}, {Green}, {Grillmair},
  {Kormendy}, \& {Lauer}}]{Magorrian}
{Magorrian}, J., {Tremaine}, S., {Richstone}, D., {et~al.} 1998, \aj, 115,
  2285, \dodoi{10.1086/300353}

\bibitem[{{Mahdavi} {et~al.}(2013){Mahdavi}, {Hoekstra}, {Babul}, {Bildfell},
  {Jeltema}, \& {Henry}}]{Mahdavi}
{Mahdavi}, A., {Hoekstra}, H., {Babul}, A., {et~al.} 2013, \apj, 767, 116,
  \dodoi{10.1088/0004-637X/767/2/116}

\bibitem[{{Maier} {et~al.}(2022){Maier}, {Haines}, \& {Ziegler}}]{Maier}
{Maier}, C., {Haines}, C.~P., \& {Ziegler}, B.~L. 2022, \aap, 658, A190,
  \dodoi{10.1051/0004-6361/202141498}

\bibitem[{{Manzer} \& {De Robertis}(2014)}]{Manzer}
{Manzer}, L.~H., \& {De Robertis}, M.~M. 2014, \apj, 788, 140,
  \dodoi{10.1088/0004-637X/788/2/140}

\bibitem[{{Marshall} {et~al.}(2018){Marshall}, {Shabala}, {Krause}, {Pimbblet},
  {Croton}, \& {Owers}}]{Marshall}
{Marshall}, M.~A., {Shabala}, S.~S., {Krause}, M. G.~H., {et~al.} 2018, \mnras,
  474, 3615, \dodoi{10.1093/mnras/stx2996}

\bibitem[{{Martin} {et~al.}(2005){Martin}, {Fanson}, {Schiminovich},
  {Morrissey}, {Friedman}, {Barlow}, {Conrow}, {Grange}, {Jelinsky},
  {Milliard}, {Siegmund}, {Bianchi}, {Byun}, {Donas}, {Forster}, {Heckman},
  {Lee}, {Madore}, {Malina}, {Neff}, {Rich}, {Small}, {Surber}, {Szalay},
  {Welsh}, \& {Wyder}}]{Martin}
{Martin}, D.~C., {Fanson}, J., {Schiminovich}, D., {et~al.} 2005, \apjl, 619,
  L1, \dodoi{10.1086/426387}

\bibitem[{{Martini} {et~al.}(2009){Martini}, {Sivakoff}, \&
  {Mulchaey}}]{Martini09}
{Martini}, P., {Sivakoff}, G.~R., \& {Mulchaey}, J.~S. 2009, \apj, 701, 66,
  \dodoi{10.1088/0004-637X/701/1/66}

\bibitem[{McKinney(2010)}]{pandas}
McKinney, W. 2010, in Proceedings of the 9th Python in Science Conference, ed.
  S.~van~der Walt \& J.~Millman, 51 -- 56

\bibitem[{{Miller} \& {Owen}(2003)}]{Miller}
{Miller}, N.~A., \& {Owen}, F.~N. 2003, \aj, 125, 2427, \dodoi{10.1086/374767}

\bibitem[{{Miraghaei}(2020)}]{Miraghaei}
{Miraghaei}, H. 2020, \aj, 160, 227, \dodoi{10.3847/1538-3881/abafb1}

\bibitem[{{Mishra} \& {Dai}(2020)}]{Mishra}
{Mishra}, H.~D., \& {Dai}, X. 2020, \aj, 159, 69,
  \dodoi{10.3847/1538-3881/ab6225}

\bibitem[{{Miyazaki} {et~al.}(2018){Miyazaki}, {Komiyama}, {Kawanomoto}, {Doi},
  {Furusawa}, {Hamana}, {Hayashi}, {Ikeda}, {Kamata}, {Karoji}, {Koike},
  {Kurakami}, {Miyama}, {Morokuma}, {Nakata}, {Namikawa}, {Nakaya}, {Nariai},
  {Obuchi}, {Oishi}, {Okada}, {Okura}, {Tait}, {Takata}, {Tanaka}, {Tanaka},
  {Terai}, {Tomono}, {Uraguchi}, {Usuda}, {Utsumi}, {Yamada}, {Yamanoi},
  {Aihara}, {Fujimori}, {Mineo}, {Miyatake}, {Oguri}, {Uchida}, {Tanaka},
  {Yasuda}, {Takada}, {Murayama}, {Nishizawa}, {Sugiyama}, {Chiba}, {Futamase},
  {Wang}, {Chen}, {Ho}, {Liaw}, {Chiu}, {Ho}, {Lai}, {Lee}, {Jeng}, {Iwamura},
  {Armstrong}, {Bickerton}, {Bosch}, {Gunn}, {Lupton}, {Loomis}, {Price},
  {Smith}, {Strauss}, {Turner}, {Suzuki}, {Miyazaki}, {Muramatsu}, {Yamamoto},
  {Endo}, {Ezaki}, {Ito}, {Kawaguchi}, {Sofuku}, {Taniike}, {Akutsu}, {Dojo},
  {Kasumi}, {Matsuda}, {Imoto}, {Miwa}, {Suzuki}, {Takeshi}, \&
  {Yokota}}]{Miyazaki}
{Miyazaki}, S., {Komiyama}, Y., {Kawanomoto}, S., {et~al.} 2018, \pasj, 70, S1,
  \dodoi{10.1093/pasj/psx063}

\bibitem[{{Momcheva} {et~al.}(2016){Momcheva}, {Brammer}, {van Dokkum},
  {Skelton}, {Whitaker}, {Nelson}, {Fumagalli}, {Maseda}, {Leja}, {Franx},
  {Rix}, {Bezanson}, {Da Cunha}, {Dickey}, {F{\"o}rster Schreiber},
  {Illingworth}, {Kriek}, {Labb{\'e}}, {Ulf Lange}, {Lundgren}, {Magee},
  {Marchesini}, {Oesch}, {Pacifici}, {Patel}, {Price}, {Tal}, {Wake}, {van der
  Wel}, \& {Wuyts}}]{Momcheva}
{Momcheva}, I.~G., {Brammer}, G.~B., {van Dokkum}, P.~G., {et~al.} 2016, \apjs,
  225, 27, \dodoi{10.3847/0067-0049/225/2/27}

\bibitem[{{Mountrichas} {et~al.}(2021){Mountrichas}, {Buat}, {Yang}, {Boquien},
  {Burgarella}, \& {Ciesla}}]{Mountrichas}
{Mountrichas}, G., {Buat}, V., {Yang}, G., {et~al.} 2021, \aap, 646, A29,
  \dodoi{10.1051/0004-6361/202039401}

\bibitem[{{Mu{\~n}oz Rodr{\'\i}guez} {et~al.}(2023){Mu{\~n}oz Rodr{\'\i}guez},
  {Georgakakis}, {Shankar}, {Allevato}, {Bonoli}, {Brusa}, {Lapi}, \&
  {Viitanen}}]{Munoz}
{Mu{\~n}oz Rodr{\'\i}guez}, I., {Georgakakis}, A., {Shankar}, F., {et~al.}
  2023, \mnras, 518, 1041, \dodoi{10.1093/mnras/stac3114}

\bibitem[{{Murata} {et~al.}(2019){Murata}, {Oguri}, {Nishimichi}, {Takada},
  {Mandelbaum}, {More}, {Shirasaki}, {Nishizawa}, \& {Osato}}]{Murata}
{Murata}, R., {Oguri}, M., {Nishimichi}, T., {et~al.} 2019, \pasj, 71, 107,
  \dodoi{10.1093/pasj/psz092}

\bibitem[{{Newman} {et~al.}(2013){Newman}, {Cooper}, {Davis}, {Faber}, {Coil},
  {Guhathakurta}, {Koo}, {Phillips}, {Conroy}, {Dutton}, {Finkbeiner}, {Gerke},
  {Rosario}, {Weiner}, {Willmer}, {Yan}, {Harker}, {Kassin}, {Konidaris},
  {Lai}, {Madgwick}, {Noeske}, {Wirth}, {Connolly}, {Kaiser}, {Kirby},
  {Lemaux}, {Lin}, {Lotz}, {Luppino}, {Marinoni}, {Matthews}, {Metevier}, \&
  {Schiavon}}]{Newman}
{Newman}, J.~A., {Cooper}, M.~C., {Davis}, M., {et~al.} 2013, \apjs, 208, 5,
  \dodoi{10.1088/0067-0049/208/1/5}

\bibitem[{{Nishizawa} {et~al.}(2020){Nishizawa}, {Hsieh}, {Tanaka}, \&
  {Takata}}]{Nishizawa}
{Nishizawa}, A.~J., {Hsieh}, B.-C., {Tanaka}, M., \& {Takata}, T. 2020, arXiv
  e-prints, arXiv:2003.01511, \dodoi{10.48550/arXiv.2003.01511}

\bibitem[{{Noeske} {et~al.}(2007){Noeske}, {Weiner}, {Faber}, {Papovich},
  {Koo}, {Somerville}, {Bundy}, {Conselice}, {Newman}, {Schiminovich}, {Le
  Floc'h}, {Coil}, {Rieke}, {Lotz}, {Primack}, {Barmby}, {Cooper}, {Davis},
  {Ellis}, {Fazio}, {Guhathakurta}, {Huang}, {Kassin}, {Martin}, {Phillips},
  {Rich}, {Small}, {Willmer}, \& {Wilson}}]{Noeske}
{Noeske}, K.~G., {Weiner}, B.~J., {Faber}, S.~M., {et~al.} 2007, \apjl, 660,
  L43, \dodoi{10.1086/517926}

\bibitem[{{Noll} {et~al.}(2009){Noll}, {Burgarella}, {Giovannoli}, {Buat},
  {Marcillac}, \& {Mu{\~n}oz-Mateos}}]{Noll}
{Noll}, S., {Burgarella}, D., {Giovannoli}, E., {et~al.} 2009, \aap, 507, 1793,
  \dodoi{10.1051/0004-6361/200912497}

\bibitem[{{Noordeh} {et~al.}(2020){Noordeh}, {Canning}, {King}, {Allen},
  {Mantz}, {Morris}, {Ehlert}, {von der Linden}, {Brandt}, {Luo}, {Xue}, \&
  {Kelly}}]{Noordeh}
{Noordeh}, E., {Canning}, R.~E.~A., {King}, A., {et~al.} 2020, \mnras, 498,
  4095, \dodoi{10.1093/mnras/staa2682}

\bibitem[{{Oguri}(2014)}]{Oguri}
{Oguri}, M. 2014, \mnras, 444, 147, \dodoi{10.1093/mnras/stu1446}

\bibitem[{{Oguri} {et~al.}(2018){Oguri}, {Lin}, {Lin}, {Nishizawa}, {More},
  {More}, {Hsieh}, {Medezinski}, {Miyatake}, {Jian}, {Lin}, {Takada}, {Okabe},
  {Speagle}, {Coupon}, {Leauthaud}, {Lupton}, {Miyazaki}, {Price}, {Tanaka},
  {Chiu}, {Komiyama}, {Okura}, {Tanaka}, \& {Usuda}}]{Oguri18}
{Oguri}, M., {Lin}, Y.-T., {Lin}, S.-C., {et~al.} 2018, \pasj, 70, S20,
  \dodoi{10.1093/pasj/psx042}

\bibitem[{{Oi} {et~al.}(2021){Oi}, {Goto}, {Matsuhara}, {Utsumi}, {Momose},
  {Toba}, {Malkan}, {Takagi}, {Huang}, {Kim}, \& {Ohyama}}]{Oi}
{Oi}, N., {Goto}, T., {Matsuhara}, H., {et~al.} 2021, \mnras, 500, 5024,
  \dodoi{10.1093/mnras/staa3080}

\bibitem[{{Okabe} {et~al.}(2019){Okabe}, {Oguri}, {Akamatsu}, {Hamabata},
  {Nishizawa}, {Medezinski}, {Koyama}, {Hayashi}, {Okabe}, {Ueda}, {Mitsuishi},
  \& {Ota}}]{Okabe}
{Okabe}, N., {Oguri}, M., {Akamatsu}, H., {et~al.} 2019, \pasj, 71, 79,
  \dodoi{10.1093/pasj/psz059}

\bibitem[{{Ota} {et~al.}(2023){Ota}, {Nguyen-Dang}, {Mitsuishi}, {Oguri},
  {Klein}, {Okabe}, {Ramos-Ceja}, {Reiprich}, {Pacaud}, {Bulbul},
  {Br{\"u}ggen}, {Liu}, {Migkas}, {Chiu}, {Ghirardini}, {Grandis}, {Lin},
  {Miyatake}, {Miyazaki}, \& {Sanders}}]{Ota23}
{Ota}, N., {Nguyen-Dang}, N.~T., {Mitsuishi}, I., {et~al.} 2023, \aap, 669,
  A110, \dodoi{10.1051/0004-6361/202244260}

\bibitem[{{Pearson} {et~al.}(2018){Pearson}, {Wang}, {Hurley}, {Ma{\l}ek},
  {Buat}, {Burgarella}, {Farrah}, {Oliver}, {Smith}, \& {van der
  Tak}}]{Pearson}
{Pearson}, W.~J., {Wang}, L., {Hurley}, P.~D., {et~al.} 2018, \aap, 615, A146,
  \dodoi{10.1051/0004-6361/201832821}

\bibitem[{{Peluso} {et~al.}(2022){Peluso}, {Vulcani}, {Poggianti}, {Moretti},
  {Radovich}, {Smith}, {Jaff{\'e}}, {Crossett}, {Gullieuszik}, {Fritz}, \&
  {Ignesti}}]{Peluso}
{Peluso}, G., {Vulcani}, B., {Poggianti}, B.~M., {et~al.} 2022, \apj, 927, 130,
  \dodoi{10.3847/1538-4357/ac4225}

\bibitem[{{Pentericci} {et~al.}(2013){Pentericci}, {Castellano}, {Menci},
  {Salimbeni}, {Dahlen}, {Galametz}, {Santini}, {Grazian}, \&
  {Fontana}}]{Pentericci}
{Pentericci}, L., {Castellano}, M., {Menci}, N., {et~al.} 2013, \aap, 552,
  A111, \dodoi{10.1051/0004-6361/201219759}

\bibitem[{{Popesso} \& {Biviano}(2006)}]{Popesso}
{Popesso}, P., \& {Biviano}, A. 2006, \aap, 460, L23,
  \dodoi{10.1051/0004-6361:20066269}

\bibitem[{{Pouliasis} {et~al.}(2020){Pouliasis}, {Mountrichas},
  {Georgantopoulos}, {Ruiz}, {Yang}, \& {Bonanos}}]{Pouliasis20}
{Pouliasis}, E., {Mountrichas}, G., {Georgantopoulos}, I., {et~al.} 2020,
  \mnras, 495, 1853, \dodoi{10.1093/mnras/staa1263}

\bibitem[{{Pouliasis} {et~al.}(2022){Pouliasis}, {Mountrichas},
  {Georgantopoulos}, {Ruiz}, {Gilli}, {Koulouridis}, {Akiyama}, {Ueda},
  {Garrel}, {Nagao}, {Paltani}, {Pierre}, {Toba}, \& {Vignali}}]{Pouliasis}
---. 2022, \aap, 667, A56, \dodoi{10.1051/0004-6361/202243502}

\bibitem[{{Prevot} {et~al.}(1984){Prevot}, {Lequeux}, {Maurice}, {Prevot}, \&
  {Rocca-Volmerange}}]{Prevot}
{Prevot}, M.~L., {Lequeux}, J., {Maurice}, E., {Prevot}, L., \&
  {Rocca-Volmerange}, B. 1984, \aap, 132, 389

\bibitem[{{Santos} {et~al.}(2021){Santos}, {Goto}, {Kim}, {Wang}, {Ho},
  {Hashimoto}, {Huang}, {Lu}, {On}, {Wong}, {Hsiao}, {Pollo}, {Malkan},
  {Miyaji}, {Toba}, {Kilerci-Eser}, {Ma{\l}ek}, {Hwang}, {Jeong}, {Shim},
  {Pearson}, {Poliszczuk}, \& {Chen}}]{Santos}
{Santos}, D. J.~D., {Goto}, T., {Kim}, S.~J., {et~al.} 2021, \mnras, 507, 3070,
  \dodoi{10.1093/mnras/stab2352}

\bibitem[{{Schlafly} {et~al.}(2019){Schlafly}, {Meisner}, \&
  {Green}}]{Schlafly19}
{Schlafly}, E.~F., {Meisner}, A.~M., \& {Green}, G.~M. 2019, \apjs, 240, 30,
  \dodoi{10.3847/1538-4365/aafbea}

\bibitem[{{Schlafly} {et~al.}(2012){Schlafly}, {Finkbeiner}, {Juri{\'c}},
  {Magnier}, {Burgett}, {Chambers}, {Grav}, {Hodapp}, {Kaiser}, {Kudritzki},
  {Martin}, {Morgan}, {Price}, {Rix}, {Stubbs}, {Tonry}, \&
  {Wainscoat}}]{Schlafly}
{Schlafly}, E.~F., {Finkbeiner}, D.~P., {Juri{\'c}}, M., {et~al.} 2012, \apj,
  756, 158, \dodoi{10.1088/0004-637X/756/2/158}

\bibitem[{{Schreiber} {et~al.}(2015){Schreiber}, {Pannella}, {Elbaz},
  {B{\'e}thermin}, {Inami}, {Dickinson}, {Magnelli}, {Wang}, {Aussel}, {Daddi},
  {Juneau}, {Shu}, {Sargent}, {Buat}, {Faber}, {Ferguson}, {Giavalisco},
  {Koekemoer}, {Magdis}, {Morrison}, {Papovich}, {Santini}, \&
  {Scott}}]{Schreiber}
{Schreiber}, C., {Pannella}, M., {Elbaz}, D., {et~al.} 2015, \aap, 575, A74,
  \dodoi{10.1051/0004-6361/201425017}

\bibitem[{{Scodeggio} {et~al.}(2018){Scodeggio}, {Guzzo}, {Garilli}, {Granett},
  {Bolzonella}, {de la Torre}, {Abbas}, {Adami}, {Arnouts}, {Bottini}, {Cappi},
  {Coupon}, {Cucciati}, {Davidzon}, {Franzetti}, {Fritz}, {Iovino}, {Krywult},
  {Le Brun}, {Le F{\`e}vre}, {Maccagni}, {Ma{\l}ek}, {Marchetti}, {Marulli},
  {Polletta}, {Pollo}, {Tasca}, {Tojeiro}, {Vergani}, {Zanichelli}, {Bel},
  {Branchini}, {De Lucia}, {Ilbert}, {McCracken}, {Moutard}, {Peacock},
  {Zamorani}, {Burden}, {Fumana}, {Jullo}, {Marinoni}, {Mellier}, {Moscardini},
  \& {Percival}}]{Scodeggio}
{Scodeggio}, M., {Guzzo}, L., {Garilli}, B., {et~al.} 2018, \aap, 609, A84,
  \dodoi{10.1051/0004-6361/201630114}

\bibitem[{Seabold \& Perktold(2010)}]{seabold2010statsmodels}
Seabold, S., \& Perktold, J. 2010, in 9th Python in Science Conference

\bibitem[{{Setoguchi} {et~al.}(2021){Setoguchi}, {Ueda}, {Toba}, \&
  {Akiyama}}]{Setoguchi}
{Setoguchi}, K., {Ueda}, Y., {Toba}, Y., \& {Akiyama}, M. 2021, \apj, 909, 188,
  \dodoi{10.3847/1538-4357/abdf55}

\bibitem[{{Setoguchi} {et~al.}(2024){Setoguchi}, {Ueda}, {Toba}, {Li},
  {Silverman}, \& {Uematsu}}]{Setoguchi24}
{Setoguchi}, K., {Ueda}, Y., {Toba}, Y., {et~al.} 2024, \apj, 961, 246,
  \dodoi{10.3847/1538-4357/ad1186}

\bibitem[{{Silva} {et~al.}(2021){Silva}, {Marchesini}, {Silverman}, {Martis},
  {Iono}, {Espada}, \& {Skelton}}]{Silva}
{Silva}, A., {Marchesini}, D., {Silverman}, J.~D., {et~al.} 2021, \apj, 909,
  124, \dodoi{10.3847/1538-4357/abdbb1}

\bibitem[{{Silverman} {et~al.}(2009){Silverman}, {Kova{\v{c}}}, {Knobel},
  {Lilly}, {Bolzonella}, {Lamareille}, {Mainieri}, {Brusa}, {Cappelluti},
  {Peng}, {Hasinger}, {Zamorani}, {Scodeggio}, {Contini}, {Carollo}, {Jahnke},
  {Kneib}, {Le Fevre}, {Bardelli}, {Bongiorno}, {Brunner}, {Caputi}, {Civano},
  {Comastri}, {Coppa}, {Cucciati}, {de la Torre}, {de Ravel}, {Elvis},
  {Finoguenov}, {Fiore}, {Franzetti}, {Garilli}, {Gilli}, {Griffiths},
  {Iovino}, {Kampczyk}, {Koekemoer}, {Le Borgne}, {Le Brun}, {Maier},
  {Mignoli}, {Pello}, {Perez Montero}, {Ricciardelli}, {Tanaka}, {Tasca},
  {Tresse}, {Vergani}, {Vignali}, {Zucca}, {Bottini}, {Cappi}, {Cassata},
  {Marinoni}, {McCracken}, {Memeo}, {Meneux}, {Oesch}, {Porciani}, \&
  {Salvato}}]{Silverman}
{Silverman}, J.~D., {Kova{\v{c}}}, K., {Knobel}, C., {et~al.} 2009, \apj, 695,
  171, \dodoi{10.1088/0004-637X/695/1/171}

\bibitem[{{Skelton} {et~al.}(2014){Skelton}, {Whitaker}, {Momcheva}, {Brammer},
  {van Dokkum}, {Labb{\'e}}, {Franx}, {van der Wel}, {Bezanson}, {Da Cunha},
  {Fumagalli}, {F{\"o}rster Schreiber}, {Kriek}, {Leja}, {Lundgren}, {Magee},
  {Marchesini}, {Maseda}, {Nelson}, {Oesch}, {Pacifici}, {Patel}, {Price},
  {Rix}, {Tal}, {Wake}, \& {Wuyts}}]{Skelton}
{Skelton}, R.~E., {Whitaker}, K.~E., {Momcheva}, I.~G., {et~al.} 2014, \apjs,
  214, 24, \dodoi{10.1088/0067-0049/214/2/24}

\bibitem[{{Skrutskie} {et~al.}(2006){Skrutskie}, {Cutri}, {Stiening},
  {Weinberg}, {Schneider}, {Carpenter}, {Beichman}, {Capps}, {Chester},
  {Elias}, {Huchra}, {Liebert}, {Lonsdale}, {Monet}, {Price}, {Seitzer},
  {Jarrett}, {Kirkpatrick}, {Gizis}, {Howard}, {Evans}, {Fowler}, {Fullmer},
  {Hurt}, {Light}, {Kopan}, {Marsh}, {McCallon}, {Tam}, {Van Dyk}, \&
  {Wheelock}}]{Skrutskie}
{Skrutskie}, M.~F., {Cutri}, R.~M., {Stiening}, R., {et~al.} 2006, \aj, 131,
  1163, \dodoi{10.1086/498708}

\bibitem[{{Smith}(1936)}]{Smith}
{Smith}, S. 1936, \apj, 83, 23, \dodoi{10.1086/143697}

\bibitem[{{Sobral} {et~al.}(2015){Sobral}, {Stroe}, {Dawson}, {Wittman}, {Jee},
  {R{\"o}ttgering}, {van Weeren}, \& {Br{\"u}ggen}}]{Sobral}
{Sobral}, D., {Stroe}, A., {Dawson}, W.~A., {et~al.} 2015, \mnras, 450, 630,
  \dodoi{10.1093/mnras/stv521}

\bibitem[{{Stalevski} {et~al.}(2016){Stalevski}, {Ricci}, {Ueda}, {Lira},
  {Fritz}, \& {Baes}}]{Stalevski}
{Stalevski}, M., {Ricci}, C., {Ueda}, Y., {et~al.} 2016, \mnras, 458, 2288,
  \dodoi{10.1093/mnras/stw444}

\bibitem[{{Stroe} \& {Sobral}(2021)}]{Stroe}
{Stroe}, A., \& {Sobral}, D. 2021, \apj, 912, 55,
  \dodoi{10.3847/1538-4357/abe7f8}

\bibitem[{{Stroe} {et~al.}(2015){Stroe}, {Sobral}, {Dawson}, {Jee}, {Hoekstra},
  {Wittman}, {van Weeren}, {Br{\"u}ggen}, \& {R{\"o}ttgering}}]{Stroe15}
{Stroe}, A., {Sobral}, D., {Dawson}, W., {et~al.} 2015, \mnras, 450, 646,
  \dodoi{10.1093/mnras/stu2519}

\bibitem[{{Suleiman} {et~al.}(2022){Suleiman}, {Noboriguchi}, {Toba},
  {Bal{\'a}zs}, {Burgarella}, {Kov{\'a}cs}, {Marton}, {Talafha}, {Frey}, \&
  {T{\'o}th}}]{Suleiman}
{Suleiman}, N., {Noboriguchi}, A., {Toba}, Y., {et~al.} 2022, \pasj, 74, 1157,
  \dodoi{10.1093/pasj/psac061}

\bibitem[{{Takada} {et~al.}(2014){Takada}, {Ellis}, {Chiba}, {Greene},
  {Aihara}, {Arimoto}, {Bundy}, {Cohen}, {Dor{\'e}}, {Graves}, {Gunn},
  {Heckman}, {Hirata}, {Ho}, {Kneib}, {Le F{\`e}vre}, {Lin}, {More},
  {Murayama}, {Nagao}, {Ouchi}, {Seiffert}, {Silverman}, {Sodr{\'e}},
  {Spergel}, {Strauss}, {Sugai}, {Suto}, {Takami}, \& {Wyse}}]{Takada}
{Takada}, M., {Ellis}, R.~S., {Chiba}, M., {et~al.} 2014, \pasj, 66, R1,
  \dodoi{10.1093/pasj/pst019}

\bibitem[{{Tanaka} {et~al.}(2018){Tanaka}, {Coupon}, {Hsieh}, {Mineo},
  {Nishizawa}, {Speagle}, {Furusawa}, {Miyazaki}, \& {Murayama}}]{Tanaka}
{Tanaka}, M., {Coupon}, J., {Hsieh}, B.-C., {et~al.} 2018, \pasj, 70, S9,
  \dodoi{10.1093/pasj/psx077}

\bibitem[{{Taylor}(2006)}]{stilts}
{Taylor}, M.~B. 2006, in Astronomical Society of the Pacific Conference Series,
  Vol. 351, Astronomical Data Analysis Software and Systems XV, ed.
  C.~{Gabriel}, C.~{Arviset}, D.~{Ponz}, \& S.~{Enrique}, 666

\bibitem[{{Toba} {et~al.}(2017{\natexlab{a}}){Toba}, {Bae}, {Nagao}, {Woo},
  {Wang}, {Wagner}, {Sun}, \& {Chang}}]{Toba17c}
{Toba}, Y., {Bae}, H.-J., {Nagao}, T., {et~al.} 2017{\natexlab{a}}, \apj, 850,
  140, \dodoi{10.3847/1538-4357/aa918a}

\bibitem[{{Toba} {et~al.}(2019{\natexlab{a}}){Toba}, {Ueda}, {Matsuoka},
  {Shidatsu}, {Nagao}, {Terashima}, {Wang}, \& {Chang}}]{Toba19b}
{Toba}, Y., {Ueda}, Y., {Matsuoka}, K., {et~al.} 2019{\natexlab{a}}, \mnras,
  484, 196, \dodoi{10.1093/mnras/sty3523}

\bibitem[{{Toba} {et~al.}(2015){Toba}, {Nagao}, {Strauss}, {Aoki}, {Goto},
  {Imanishi}, {Kawaguchi}, {Terashima}, {Ueda}, {Bosch}, {Bundy}, {Doi},
  {Inami}, {Komiyama}, {Lupton}, {Matsuhara}, {Matsuoka}, {Miyazaki},
  {Morokuma}, {Nakata}, {Oi}, {Onoue}, {Oyabu}, {Price}, {Tait}, {Takata},
  {Tanaka}, {Terai}, {Turner}, {Uchida}, {Usuda}, {Utsumi}, {Yamada}, \&
  {Wang}}]{Toba15}
{Toba}, Y., {Nagao}, T., {Strauss}, M.~A., {et~al.} 2015, \pasj, 67, 86,
  \dodoi{10.1093/pasj/psv057}

\bibitem[{{Toba} {et~al.}(2017{\natexlab{b}}){Toba}, {Nagao}, {Wang},
  {Matsuhara}, {Akiyama}, {Goto}, {Koyama}, {Ohyama}, \& {Yamamura}}]{Toba17}
{Toba}, Y., {Nagao}, T., {Wang}, W.-H., {et~al.} 2017{\natexlab{b}}, \apj, 840,
  21, \dodoi{10.3847/1538-4357/aa6d0a}

\bibitem[{{Toba} {et~al.}(2019{\natexlab{b}}){Toba}, {Yamashita}, {Nagao},
  {Wang}, {Ueda}, {Ichikawa}, {Kawaguchi}, {Akiyama}, {Hsieh}, {Kajisawa},
  {Lee}, {Matsuoka}, {Noboriguchi}, {Onoue}, {Schramm}, {Tanaka}, \&
  {Komiyama}}]{Toba19}
{Toba}, Y., {Yamashita}, T., {Nagao}, T., {et~al.} 2019{\natexlab{b}}, \apjs,
  243, 15, \dodoi{10.3847/1538-4365/ab238d}

\bibitem[{{Toba} {et~al.}(2020{\natexlab{a}}){Toba}, {Yamada}, {Ueda}, {Ricci},
  {Terashima}, {Nagao}, {Wang}, {Tanimoto}, \& {Kawamuro}}]{Toba20a}
{Toba}, Y., {Yamada}, S., {Ueda}, Y., {et~al.} 2020{\natexlab{a}}, \apj, 888,
  8, \dodoi{10.3847/1538-4357/ab5718}

\bibitem[{{Toba} {et~al.}(2020{\natexlab{b}}){Toba}, {Wang}, {Nagao}, {Ueda},
  {Ueda}, {Lim}, {Chang}, {Saito}, \& {Kawabe}}]{Toba20b}
{Toba}, Y., {Wang}, W.-H., {Nagao}, T., {et~al.} 2020{\natexlab{b}}, \apj, 889,
  76, \dodoi{10.3847/1538-4357/ab616d}

\bibitem[{{Toba} {et~al.}(2020{\natexlab{c}}){Toba}, {Goto}, {Oi}, {Wang},
  {Kim}, {Ho}, {Burgarella}, {Hashimoto}, {Hsieh}, {Huang}, {Hwang}, {Ikeda},
  {Kim}, {Kim}, {Lee}, {Malkan}, {Matsuhara}, {Miyaji}, {Momose}, {Ohyama},
  {Oyabu}, {Pearson}, {Santos}, {Shim}, {Takagi}, {Ueda}, {Utsumi}, \&
  {Wada}}]{Toba20c}
{Toba}, Y., {Goto}, T., {Oi}, N., {et~al.} 2020{\natexlab{c}}, \apj, 899, 35,
  \dodoi{10.3847/1538-4357/ab9cb7}

\bibitem[{{Toba} {et~al.}(2021{\natexlab{a}}){Toba}, {Brusa}, {Liu}, {Buchner},
  {Terashima}, {Urrutia}, {Salvato}, {Akiyama}, {Arcodia}, {Goulding},
  {Higuchi}, {Inoue}, {Kawaguchi}, {Lamer}, {Merloni}, {Nagao}, {Ueda}, \&
  {Nandra}}]{Toba21b}
{Toba}, Y., {Brusa}, M., {Liu}, T., {et~al.} 2021{\natexlab{a}}, \aap, 649,
  L11, \dodoi{10.1051/0004-6361/202140317}

\bibitem[{{Toba} {et~al.}(2021{\natexlab{b}}){Toba}, {Ueda}, {Gandhi}, {Ricci},
  {Burgarella}, {Buat}, {Nagao}, {Oyabu}, {Matsuhara}, \& {Hsieh}}]{Toba21}
{Toba}, Y., {Ueda}, Y., {Gandhi}, P., {et~al.} 2021{\natexlab{b}}, \apj, 912,
  91, \dodoi{10.3847/1538-4357/abe94a}

\bibitem[{{Toba} {et~al.}(2022){Toba}, {Liu}, {Urrutia}, {Salvato}, {Li},
  {Ueda}, {Brusa}, {Yutani}, {Wada}, {Nishizawa}, {Buchner}, {Nagao},
  {Merloni}, {Akiyama}, {Arcodia}, {Hsieh}, {Ichikawa}, {Imanishi}, {Inoue},
  {Kawaguchi}, {Lamer}, {Nandra}, {Silverman}, \& {Terashima}}]{Toba22}
{Toba}, Y., {Liu}, T., {Urrutia}, T., {et~al.} 2022, \aap, 661, A15,
  \dodoi{10.1051/0004-6361/202141547}

\bibitem[{{Tomczak} {et~al.}(2011){Tomczak}, {Tran}, \& {Saintonge}}]{Tomczak}
{Tomczak}, A.~R., {Tran}, K.-V.~H., \& {Saintonge}, A. 2011, \apj, 738, 65,
  \dodoi{10.1088/0004-637X/738/1/65}

\bibitem[{{Tonry} {et~al.}(2012){Tonry}, {Stubbs}, {Lykke}, {Doherty},
  {Shivvers}, {Burgett}, {Chambers}, {Hodapp}, {Kaiser}, {Kudritzki},
  {Magnier}, {Morgan}, {Price}, \& {Wainscoat}}]{Tonry}
{Tonry}, J.~L., {Stubbs}, C.~W., {Lykke}, K.~R., {et~al.} 2012, \apj, 750, 99,
  \dodoi{10.1088/0004-637X/750/2/99}

\bibitem[{{Treister} {et~al.}(2012){Treister}, {Schawinski}, {Urry}, \&
  {Simmons}}]{Treister}
{Treister}, E., {Schawinski}, K., {Urry}, C.~M., \& {Simmons}, B.~D. 2012,
  \apjl, 758, L39, \dodoi{10.1088/2041-8205/758/2/L39}

\bibitem[{{Treu} {et~al.}(2003){Treu}, {Ellis}, {Kneib}, {Dressler}, {Smail},
  {Czoske}, {Oemler}, \& {Natarajan}}]{Treu}
{Treu}, T., {Ellis}, R.~S., {Kneib}, J.-P., {et~al.} 2003, \apj, 591, 53,
  \dodoi{10.1086/375314}

\bibitem[{{Tzanavaris} {et~al.}(2014){Tzanavaris}, {Gallagher},
  {Hornschemeier}, {Fedotov}, {Eracleous}, {Brandt}, {Desjardins}, {Charlton},
  \& {Gronwall}}]{Tzanavaris}
{Tzanavaris}, P., {Gallagher}, S.~C., {Hornschemeier}, A.~E., {et~al.} 2014,
  \apjs, 212, 9, \dodoi{10.1088/0067-0049/212/1/9}

\bibitem[{{Uematsu} {et~al.}(2023){Uematsu}, {Ueda}, {Kohno}, {Yamada}, {Toba},
  {Fujimoto}, {Hatsukade}, {Umehata}, {Espada}, {Sun}, {Magdis}, {Kokorev}, \&
  {Ao}}]{Uematsu}
{Uematsu}, R., {Ueda}, Y., {Kohno}, K., {et~al.} 2023, \apj, 945, 121,
  \dodoi{10.3847/1538-4357/acb4e9}

\bibitem[{{Uematsu} {et~al.}(2024){Uematsu}, {Ueda}, {Kohno}, {Toba}, {Yamada},
  {Smail}, {Umehata}, {Fujimoto}, {Hatsukade}, {Ao}, {Bauer}, {Brammer},
  {Dessauges-Zavadsky}, {Espada}, {Jolly}, {Koekemoer}, {Kokorev}, {Magdis},
  {Oguri}, \& {Sun}}]{Uematsu24}
---. 2024, arXiv e-prints, arXiv:2402.05849, \dodoi{10.48550/arXiv.2402.05849}

\bibitem[{{Villa-V{\'e}lez} {et~al.}(2021){Villa-V{\'e}lez}, {Buat},
  {Theul{\'e}}, {Boquien}, \& {Burgarella}}]{Villa}
{Villa-V{\'e}lez}, J.~A., {Buat}, V., {Theul{\'e}}, P., {Boquien}, M., \&
  {Burgarella}, D. 2021, \aap, 654, A153, \dodoi{10.1051/0004-6361/202140890}

\bibitem[{{Virtanen} {et~al.}(2020){Virtanen}, {Gommers}, {Oliphant},
  {Haberland}, {Reddy}, {Cournapeau}, {Burovski}, {Peterson}, {Weckesser},
  {Bright}, {van der Walt}, {Brett}, {Wilson}, {Jarrod Millman}, {Mayorov},
  {Nelson}, {Jones}, {Kern}, {Larson}, {Carey}, {Polat}, {Feng}, {Moore},
  {VanderPlas}, {Laxalde}, {Perktold}, {Cimrman}, {Henriksen}, {Quintero},
  {Harris}, {Archibald}, {Ribeiro}, {Pedregosa}, {van Mulbregt}, \&
  {Contributors}}]{scipy}
{Virtanen}, P., {Gommers}, R., {Oliphant}, T.~E., {et~al.} 2020, Nature
  Methods, \dodoi{https://doi.org/10.1038/s41592-019-0686-2}

\bibitem[{{Weigel} {et~al.}(2018){Weigel}, {Schawinski}, {Treister},
  {Trakhtenbrot}, \& {Sanders}}]{Weigel}
{Weigel}, A.~K., {Schawinski}, K., {Treister}, E., {Trakhtenbrot}, B., \&
  {Sanders}, D.~B. 2018, \mnras, 476, 2308, \dodoi{10.1093/mnras/sty383}

\bibitem[{{Woo} {et~al.}(2013){Woo}, {Schulze}, {Park}, {Kang}, {Kim}, \&
  {Riechers}}]{Woo}
{Woo}, J.-H., {Schulze}, A., {Park}, D., {et~al.} 2013, \apj, 772, 49,
  \dodoi{10.1088/0004-637X/772/1/49}

\bibitem[{{Wright} {et~al.}(2010){Wright}, {Eisenhardt}, {Mainzer}, {Ressler},
  {Cutri}, {Jarrett}, {Kirkpatrick}, {Padgett}, {McMillan}, {Skrutskie},
  {Stanford}, {Cohen}, {Walker}, {Mather}, {Leisawitz}, {Gautier}, {McLean},
  {Benford}, {Lonsdale}, {Blain}, {Mendez}, {Irace}, {Duval}, {Liu}, {Royer},
  {Heinrichsen}, {Howard}, {Shannon}, {Kendall}, {Walsh}, {Larsen}, {Cardon},
  {Schick}, {Schwalm}, {Abid}, {Fabinsky}, {Naes}, \& {Tsai}}]{Wright}
{Wright}, E.~L., {Eisenhardt}, P. R.~M., {Mainzer}, A.~K., {et~al.} 2010, \aj,
  140, 1868, \dodoi{10.1088/0004-6256/140/6/1868}

\bibitem[{{Yamada} {et~al.}(2023){Yamada}, {Ueda}, {Herrera-Endoqui}, {Toba},
  {Miyaji}, {Ogawa}, {Uematsu}, {Tanimoto}, {Imanishi}, \& {Ricci}}]{Yamada}
{Yamada}, S., {Ueda}, Y., {Herrera-Endoqui}, M., {et~al.} 2023, \apjs, 265, 37,
  \dodoi{10.3847/1538-4365/acb349}

\bibitem[{{Yang} {et~al.}(2020){Yang}, {Boquien}, {Buat}, {Burgarella},
  {Ciesla}, {Duras}, {Stalevski}, {Brandt}, \& {Papovich}}]{Yang}
{Yang}, G., {Boquien}, M., {Buat}, V., {et~al.} 2020, \mnras, 491, 740,
  \dodoi{10.1093/mnras/stz3001}

\bibitem[{{Yang} {et~al.}(2022){Yang}, {Boquien}, {Brandt}, {Buat},
  {Burgarella}, {Ciesla}, {Lehmer}, {Ma{\l}ek}, {Mountrichas}, {Papovich},
  {Pons}, {Stalevski}, {Theul{\'e}}, \& {Zhu}}]{Yang22}
{Yang}, G., {Boquien}, M., {Brandt}, W.~N., {et~al.} 2022, \apj, 927, 192,
  \dodoi{10.3847/1538-4357/ac4971}

\bibitem[{{York} {et~al.}(2000){York}, {Adelman}, {Anderson}, {Anderson},
  {Annis}, {Bahcall}, {Bakken}, {Barkhouser}, {Bastian}, {Berman}, {Boroski},
  {Bracker}, {Briegel}, {Briggs}, {Brinkmann}, {Brunner}, {Burles}, {Carey},
  {Carr}, {Castander}, {Chen}, {Colestock}, {Connolly}, {Crocker}, {Csabai},
  {Czarapata}, {Davis}, {Doi}, {Dombeck}, {Eisenstein}, {Ellman}, {Elms},
  {Evans}, {Fan}, {Federwitz}, {Fiscelli}, {Friedman}, {Frieman}, {Fukugita},
  {Gillespie}, {Gunn}, {Gurbani}, {de Haas}, {Haldeman}, {Harris}, {Hayes},
  {Heckman}, {Hennessy}, {Hindsley}, {Holm}, {Holmgren}, {Huang}, {Hull},
  {Husby}, {Ichikawa}, {Ichikawa}, {Ivezi{\'c}}, {Kent}, {Kim}, {Kinney},
  {Klaene}, {Kleinman}, {Kleinman}, {Knapp}, {Korienek}, {Kron}, {Kunszt},
  {Lamb}, {Lee}, {Leger}, {Limmongkol}, {Lindenmeyer}, {Long}, {Loomis},
  {Loveday}, {Lucinio}, {Lupton}, {MacKinnon}, {Mannery}, {Mantsch}, {Margon},
  {McGehee}, {McKay}, {Meiksin}, {Merelli}, {Monet}, {Munn}, {Narayanan},
  {Nash}, {Neilsen}, {Neswold}, {Newberg}, {Nichol}, {Nicinski}, {Nonino},
  {Okada}, {Okamura}, {Ostriker}, {Owen}, {Pauls}, {Peoples}, {Peterson},
  {Petravick}, {Pier}, {Pope}, {Pordes}, {Prosapio}, {Rechenmacher}, {Quinn},
  {Richards}, {Richmond}, {Rivetta}, {Rockosi}, {Ruthmansdorfer}, {Sandford},
  {Schlegel}, {Schneider}, {Sekiguchi}, {Sergey}, {Shimasaku}, {Siegmund},
  {Smee}, {Smith}, {Snedden}, {Stone}, {Stoughton}, {Strauss}, {Stubbs},
  {SubbaRao}, {Szalay}, {Szapudi}, {Szokoly}, {Thakar}, {Tremonti}, {Tucker},
  {Uomoto}, {Vanden Berk}, {Vogeley}, {Waddell}, {Wang}, {Watanabe},
  {Weinberg}, {Yanny}, {Yasuda}, \& {SDSS Collaboration}}]{York}
{York}, D.~G., {Adelman}, J., {Anderson}, John~E., J., {et~al.} 2000, \aj, 120,
  1579, \dodoi{10.1086/301513}

\bibitem[{{Zucker} {et~al.}(2016){Zucker}, {Walker}, {Johnson}, {Gallagher},
  {Alatalo}, \& {Tzanavaris}}]{Zucker}
{Zucker}, C., {Walker}, L.~M., {Johnson}, K., {et~al.} 2016, \apj, 821, 113,
  \dodoi{10.3847/0004-637X/821/2/113}

\end{thebibliography}
\bibliographystyle{aasjournal}

\end{document}